\documentclass[12pt,a4paper]{article}

\usepackage{lmodern}[lmr]
\usepackage{amsfonts, amsmath, amssymb, amsthm, color, epstopdf, enumitem, float, graphicx, lipsum, longtable, lscape, mathrsfs, mathtools, multirow, nccmath, pdflscape, rotating, tocloft}
\usepackage[top=1.4in, bottom=1.4in, left=1.35in, right=1.35in]{geometry}
\usepackage{natbib}
\usepackage{url}
\usepackage[pdftex,colorlinks=true,hypertexnames=false]{hyperref}
\usepackage[doublespacing]{setspace}
\usepackage[table]{xcolor}
\usepackage{tikz}
\usetikzlibrary{arrows,decorations.pathmorphing,backgrounds,positioning,fit,matrix}
\usepackage{subcaption}
\usepackage{palatino,eulervm}
\usepackage[T1]{fontenc}

\definecolor{darkblue}{rgb}{0,0,.6}
\hypersetup{citecolor=darkblue,linkcolor=darkblue,urlcolor=darkblue}
\def\be{\begin{equation}}
\def\ee{\end{equation}}
\def\bea{\begin{eqnarray}}
\def\eea{\end{eqnarray}}

\providecommand{\U}[1]{\protect\rule{.1in}{.1in}}

\graphicspath{{plots/}}

\DeclareMathOperator*{\argmin}{\arg\!\min}
\DeclareMathOperator*{\argmax}{\arg\!\max}

\begin{document}

\newtheorem{corollary}{Corollary}
\newtheorem{definition}{Definition}
\newtheorem{lemma}{Lemma}
\newtheorem{proposition}{Proposition}
\newtheorem{remark}{Remark}
\newtheorem{theorem}{Theorem}
\newtheorem{example}{Example}
\newtheorem{assumption}{Assumption}
\newtheorem{prop}{Proposition}

\numberwithin{corollary}{section}
\numberwithin{definition}{section}
\numberwithin{equation}{section}
\numberwithin{lemma}{section}
\numberwithin{proposition}{section}
\numberwithin{remark}{section}
\numberwithin{theorem}{section}

\allowdisplaybreaks[4]

\title{Estimation of Grouped Time-Varying Network Vector Autoregression Models}
\author{{\normalsize Degui Li\thanks{%
Department of Mathematics, University of York, UK. },\ \ \ Bin Peng\thanks{Department of Econometrics and Business Statistics, Monash University, Australia. },\ \ \ Songqiao Tang\thanks{%
School of Mathematical Sciences, Zhejiang University, China.},\ \ \ Weibiao Wu\thanks{Department of Statistics, University of Chicago, US.}}\\
{\normalsize\em University of York,\ \ Monash University,\ \ Zhejiang University,\ \ University of Chicago}}
\date{{\normalsize Version: \today}}
\maketitle

\begin{abstract}

This paper introduces a flexible time-varying network vector autoregressive model framework for large-scale time series. A latent group structure is imposed on the heterogeneous and node-specific time-varying momentum and network spillover effects so that the number of unknown time-varying coefficients to be estimated can be reduced considerably. A classic agglomerative clustering algorithm with nonparametrically estimated distance matrix is combined with a ratio criterion to consistently estimate the latent group number and membership. A post-grouping local linear smoothing method is proposed to estimate the group-specific time-varying momentum and network effects, substantially improving the convergence rates of the preliminary estimates which ignore the latent structure. We further modify the methodology and theory to allow for structural breaks in either the group membership, group number or group-specific coefficient functions. Numerical studies including Monte-Carlo simulation and an empirical application are presented to examine the finite-sample performance of the developed model and methodology.

\bigskip 


\end{abstract}

\newpage

%%%%%%%%%%%%%%%%%%%%%%

\section{Introduction}\label{sec1} 
\renewcommand{\theequation}{1.\arabic{equation}} \setcounter{equation}{0}

Modeling large-scale time series has been the main frontier of recent advances of time series analysis and is of fundamental importance in various fields of applications such as climatology, economics, finance and social networks. Since \cite{Si80}'s seminal work, the vector autoregressive (VAR) model has become a commonly-used statistical tool to tackle multivariate time series, see \cite{Lu06} and \cite{KL17} for a comprehensive review of classic estimation and forecasting techniques. However, the VAR-based estimation and forecasting are challenging when the number of time series sequences $N$ diverges to infinity. In this case, the number of unknown parameters in VAR transition matrices is of order $O(N^2)$, which may be substantially larger than the time series length $T$. In order to construct sensible estimate and forecast, two dimension-reduction approaches are often employed: VAR with sparse transition matrices and regularised estimation \citep[e.g.,][]{SB11, BM15, KC15, DZZ16, MPS23} and factor-augmented VAR \citep[e.g.,][]{BBE05, BN06, BLL16}. Although some sound asymptotic properties have been developed for the sparse or factor-augmented VAR estimates, they often neglect possible network structures in large-scale time series and cannot directly capture dynamic network effects.   

\smallskip

Consider time series observation vectors $X_{t}=\left(x_{1,t},\cdots,x_{N,t}\right)^{^\intercal}$ with $N$ being the number of nodes in the large-scale network, and denote an adjacency matrix by ${\mathbf W}=(w_{ij})_{N\times N}$, where $w_{ii}=0$, $w_{ij}=1$ for $i\neq j$ if there exists a directed edge from $i$ to $j$ and $w_{ij}=0$ otherwise. The matrix ${\mathbf W}$ is assumed to be observable and can be either directed (${\mathbf W}^{^\intercal}\neq {\mathbf W}$) or undirected (${\mathbf W}^{^\intercal}={\mathbf W}$). The classic network VAR model is defined by 
\begin{equation}\label{nVAR}
x_{i,t}=\beta_1 \sum_{j\neq i}\widetilde{w}_{ij}x_{j,t-1}+\beta_2x_{i,t-1}+\varepsilon_{i,t},\ \ i=1,\cdots,N,
\end{equation}
where $\widetilde{w}_{ij}=w_{ij}/n_i$ with $n_i=\sum_{j\neq i} w_{ij}$,  $\beta_1$ and $\beta_2$ are unknown parameters, and $\varepsilon_t=(\varepsilon_{1,t},\cdots,\varepsilon_{N,t})^{^\intercal}$ is a sequence of independent and identically distributed (i.i.d.) random vectors. The above network VAR model formulation contains two regression components: $\beta_1 \sum_{j\neq i}\widetilde{w}_{ij}x_{j,t-1}$ and $\beta_2x_{i,t-1}$, corresponding to the network (cross-lag) and momentum (own-lag) effects, respectively. \cite{ZPLLW17} discuss stationarity conditions for an extended network VAR model with extra nodal effects, propose the least squares estimation method and derive the relevant asymptotic theory. Although the classic linear network VAR model is easy to interpret and implement, it may be invalid in empirical applications. In particular, there exist two practical issues: (i) the stable network VAR model cannot capture smooth structural changes in the underlying data generating process over a long time span; and (ii) it is often too restrictive to impose the homogeneity assumption on the autoregressive coefficients over $N$ nodes. Consequently, the homogenous linear network VAR (\ref{nVAR}) may suffer from the model misspecification problem, resulting in biased estimates, inaccurate forecast and misleading inference. There have been some attempts in recent years to address one of the aforementioned two issues. To incorporate structural changes in the autoregressive structures, \cite{Su16}, \cite{SM18}, \cite{Wu19}, \cite{CLLL23} and \cite{YSM24} extend the linear network VAR models, allowing the coefficients to vary smoothly over time or with a stationary index variable. On the other hand, to relax the homogeneity restriction in linear VAR models, \cite{YSM23, YSM24} introduce fully heterogenous network VAR models with the momentum and network spillover effects varying over nodes. However, the number of unknown coefficients in the latter models grow with the number of nodes, resulting in slow estimation convergence when the time span is not sufficiently long. 

\smallskip

This paper aims to jointly tackle the aforementioned two issues by introducing a general time-varying network VAR model framework satisfying a latent group structure, i.e., the time-varying network autoregressive relationships are invariant within a group of nodes, but change over different groups. The grouped time-varying network VAR model achieves a good balance between model flexibility and parsimony. It not only covers the homogenous network VAR models \citep{ZPLLW17, Wu19} as a special case, but also provides a more parsimonious model formulation than the fully heterogenous network VAR models \citep{YSM23, YSM24}, achieving dimension reduction in estimation and improving the subsequent out-of-sample forecasting performance. The main methodological and theoretical contributions of our paper with connection to the existing literature are summarized as follows.
 
\begin{itemize}

\item {\em General network VAR model framework with a latent group structure on time-varying momentum and network spillover effects}. There has been increasing interest in recent years to explore a group structure under the classic stable VAR or network VAR model framework. For example, \cite{ZP20} introduce a grouped linear network VAR model via a mixture Gaussian distribution and use an EM estimation algorithm; \cite{GB21} propose a stochastic block VAR model and detect a latent group structure on the network spillover effects; \cite{CFZ23} study a community network VAR model and allow network effects to vary over different communities; and \cite{ZXF23} introduce a least squares algorithm to simultaneously estimate the parameters and identify the group structure for heterogeneous network VAR models. In this paper, we relax the somehow restricted stable model assumption in the aforementioned literature, allowing for structural changes in the underlying data generating process, a typical dynamic feature for large-scale network time series collected over a long time span. With the latent group structure, we substantially reduce the number of unknown coefficient functions for momentum and network spillover effects, which is appealing when the model is applied to the out-of-sample prediction. As in \cite{ZXF23}, we allow the time-varying network effects to depend on both the sender and receiver's group information, resulting in a further homogeneity structure over the network effects.

\smallskip

\item {\em Easy-to-implement clustering algorithm and post-grouping nonparametric estimation}. Since neither the group number nor membership is known a priori, we combine a classic agglomerative clustering algorithm and a simple ratio criterion to consistently estimate the group structure. With the nonparametrically estimated distance matrix, we may directly adopt a standard package in the computing software to implement the cluster analysis. The three-stage procedure introduced in Section \ref{sec3.1} does not require iterative computation \citep[e.g.,][]{BM15, ZXF23} to obtain consistent group membership estimates. The developed clustering methodology complements recent developments on latent group estimation in the context of large panel data \citep[e.g.,][]{BM15, KLZ16, SSP16, AB17, VL17, VL20, Ch19}. However, since the underlying high-dimensional time series process is locally stationary, it is technically more challenging to derive the asymptotic property of the developed methodology. To improve the convergence rates of the fully heterogenous time-varying coefficient estimation which only uses the sample information from one node and its direct neighbors, we propose a post-grouping local linear smoothing method in Section \ref{sec4} to estimate the group-specific time-varying momentum and network effects by making use of the consistently estimated group structure. The asymptotic normal distribution theory is derived with the convergence rates comparable to those for homogenous time-varying coefficient estimation. 

\smallskip

\item {\em Structural breaks in the group structure}. The existing literature on grouped network VAR models requires the assumption that the latent group structure is time-invariant. This assumption may be restrictive for some empirical case studies. For example, a macroeconomic shock such as the global financial crisis may not only lead to a structural break in the group structure among a large number of countries, but also result in abrupt changes in the vector autoregressive structure of macroeconomic time series. This paper extends the model, methodology and theory, allowing for structural breaks in either the group membership, group number or group-specific coefficient functions. With the two-stage estimation procedure introduced in Section \ref{sec5}, we consistently estimate the scaled break location and the group structures over the two time periods separated by the break point. The online supplement \citep{LPTW24} further introduces a refined break point estimation using the consistently estimated group structure to improve the break point estimation accuracy. Our model and methodology can be seen as an extension of the linear panel model framework (with break in the group structure) considered by \cite{LOW23} and \cite{WPS23}, taking into account the network structure and allowing for structural changes over time.

\end{itemize}

The finite-sample Monte-Carlo simulation study shows that the proposed clustering algorithm and ratio criterion can consistently estimate the latent group membership and number as long as the time series length $T$ is moderately large; the post-grouping local linear estimates perform significantly better than the naive heterogeneous local linear estimates; and the developed two-stage estimation procedure can precisely locate the break point and estimate the group structure before and after the break. The developed model and methodology are further applied to analyze the monthly temperature data collected in $37$ UK weather stations over the time period between January 1950 and February 2023. The empirical result reveals that there exist two groups over the node-specific time-varying coefficients: weather stations in Northern Ireland and Wales tend to form one group and those in England and Scotland form the other. In addition, our proposed method outperforms other competing methods in terms of out-of-sample prediction.

\smallskip

The rest of the paper is organized as follows. Section \ref{sec2} introduces the time-varying network VAR model together with the latent group structure and provides some fundamental model assumptions. Section \ref{sec3} describes the estimation methodology for the group structure and establishes the consistency property for the resulting estimation. Section \ref{sec4} proposes the post-grouping nonparametric estimation and derives the relevant asymptotic theory. Section \ref{sec5} extends the model, method and theory to the case with breaks in either the group structure or group-specific coefficient functions. Section \ref{sec6} reports both the simulation and empirical studies. Section \ref{sec7} concludes the paper. Appendix A introduces the clustering algorithm to identify the homogeneity structure on the network effects. The online supplement \citep{LPTW24} contains proofs of the main asymptotic theorems, some technical lemmas with their proofs, discussion on the refined break point estimation, and extra numerical results. Throughout the paper, we let $\lfloor\cdot\rfloor$ and $\lceil\cdot\rceil$ be the floor and ceiling functions, respectively. For a vector $x=(x_1,\cdots,x_p)^{^\intercal}\in\mathscr{R}^p$, we write $\vert x\vert_q=(\sum_{i=1}^p|x_i|^q)^{1/q}$ for $q\geq1$ and $\vert x\vert_\infty=\max_{1\leq i\leq p}|x_i|$; for a matrix $\boldsymbol{\Sigma}=(\sigma_{ij})_{p\times p}$, write $\vert\boldsymbol{\Sigma}\vert_{\sf F}=(\sum_{i=1}^{p}\sum_{j=1}^{p}\sigma_{ij}^2)^{1/2}$, $\vert\boldsymbol{\Sigma}\vert_\infty=\max_{1\leq i,j\leq p}\vert\sigma_{ij}\vert$ and $\vert\boldsymbol{\Sigma}\vert_{\sf O}=\max_{x\in\mathscr{R}^p: \vert x\vert_2=1}\vert\boldsymbol{\Sigma} x\vert$; and for a $p$-dimensional random vector $Z$, write $Z\in\mathscr{L}^\kappa$ if $\Vert Z\Vert_\kappa:=[{\sf E}(\vert Z\vert_2^\kappa)]^{1/\kappa}<\infty, \kappa\ge1$. Let $\mathbf{I}_k$ be a $k\times k$ identity matrix and $\mathbf{O}_{k\times l}$ a $k\times l$ zero matrix. For a square matrix, $\lambda_{\min}(\cdot)$ and $\lambda_{\max}(\cdot)$ denote the minimum and maximum eigenvalues and ${\sf det}(\cdot)$ denotes its determinant. Let $a_n=o(b_n)$, $a_n=o_P(b_n)$ and $a_n\propto b_n$ denote that $a_n/b_n\to 0$ as $n\to\infty$, $a_n/b_n\to 0$ with probability approaching one ({\em w.p.a.1}), and $0<\underline{c}\leq a_n/b_n\leq \overline{c}<\infty$, respectively.

%%%%%%%%%%%%%%%%%%%

\section{Time-varying network VAR and latent groups}\label{sec2} 
\renewcommand{\theequation}{2.\arabic{equation}} \setcounter{equation}{0}

In this section, we introduce the main model framework, i.e., the time-varying network VAR model with a latent group structure, and impose some fundamental assumptions, ensuring the network time series are locally stable. 

\subsection{Grouped time-varying network VAR}\label{sec2.1}

Suppose that there exists a partition of the node index set $\{1,2,\cdots,N\}$, denoted by ${\mathscr{G}}=\{{\mathscr G}_1,{\mathscr G}_2,\cdots,{\mathscr G}_{K_0}\}$, such that ${\mathscr G}_i\cap {\mathscr G}_j=\emptyset$ for $1\leq i\neq j\leq K_0$. Let $g_i\in\{1,\cdots,K_0\}$ be the group membership of the $i$-th node, i.e., $g_i=k$ is equivalent to $i\in{\mathscr G}_k$. Neither the group membership nor the group number is known a priori. Consider the following grouped time-varying network VAR model:
\begin{equation}\label{eq2.1}
x_{i,t}=\sum_{j\neq i}\alpha_{g_ig_j}(\tau_t)\widetilde{w}_{ij}x_{j,t-1}+\alpha_{g_i}(\tau_t)x_{i,t-1}+\varepsilon_{i,t},\ \ i=1,\cdots,N,\ \ t=1,\cdots,T,
\end{equation}
where $\tau_t=t/T$ denotes the scaled time point, $\alpha_{g_ig_j}(\cdot)$ and $\alpha_{g_i}(\cdot)$ are smooth coefficient functions, and the remaining elements are defined as those in (\ref{nVAR}). In contrast to the linear network VAR model (\ref{nVAR}), the grouped time-varying network VAR model (\ref{eq2.1}) provides a much more flexible framework, allowing the network and momentum effects to change over time and nodes. In particular, the interaction between nodes from two groups share the same time-varying network effects which is appealing for modeling social networks with smooth structural changes. Although we only consider one lag in model (\ref{eq2.1}) for simplicity of exposition, the method and theory developed in Sections \ref{sec3} and \ref{sec4} below can be easily extended to the model setting with finite lags. It is worth pointing out that our network structure is deterministic \citep{ZPLLW17} and the group membership is determined by node-specific time-varying momentum and network spillover effects. This is substantially different from the community network VAR model studied by \cite{CFZ23}, where the network structure is random and the community structure is used for the network generating mechanism. 

\smallskip

Let $\widetilde{\mathbf W}$ be the row-normalized adjacency matrix with the $(i,j)$-entry being $\widetilde{w}_{ij}$, ${\mathbf B}_1(\cdot)$ be an $N\times N$ matrix being the diagonal entries being zeros and the off-diagonal $(i,j)$-entry being $\alpha_{g_ig_j}(\cdot)$ and ${\mathbf B}_2(\cdot)={\sf diag}\{\alpha_{g_1}(\cdot),\cdots,\alpha_{g_N}(\cdot)\}$. Then, we may rewrite model (\ref{eq2.1}) as
\begin{equation}\label{eq2.2}
X_t={\mathbf B}(\tau_t)X_{t-1}+\varepsilon_t,\ \ {\mathbf B}(\tau_t)={\mathbf B}_1(\tau_t)\odot\widetilde{\mathbf W}+{\mathbf B}_2(\tau_t),\ \ t=1,\cdots,T,
\end{equation}
where $\odot$ denotes the Hadamard product between matrices. Model (\ref{eq2.2}) thus falls within the high-dimensional time-varying VAR model framework which has received increasing attention in recent years. For instance, \cite{DQC17} propose a kernel-weighted $\ell_1$-regularised estimation for a time-varying VAR model; \cite{XCW20} study a high-dimensional VAR model with multiple breaks and estimate smooth time-varying covariance and precision matrices between the break points; \cite{CLLL23} estimate dual network structures via directed Granger causality and undirected partial correlation linkages within the high-dimensional time-varying VAR framework. However, the aforementioned literature often assumes a sparsity condition on the time-varying VAR transition matrices to facilitate the use of the regularised estimation techniques and cannot directly capture possible time-varying network effects. In this paper, we decompose the time-varying transition matrix into two components: ${\mathbf B}_2(\tau_t)$ capturing the momentum effects and ${\mathbf B}_1(\tau_t)\odot\widetilde{\mathbf W}$ capturing the dynamic network spillover effects.

\smallskip

The proposed model (\ref{eq2.1}) contains the homogenous time-varying network VAR model \citep[e.g.,][]{Wu19} as a special case, i.e., 
\[
x_{i,t}=\alpha_{1}^\dagger(\tau_t)\sum_{j\neq i}\widetilde{w}_{ij}x_{j,t-1}+\alpha_{2}^\dagger(\tau_t)x_{i,t-1}+\varepsilon_{i,t}.
\] 
In contrast to the fully heterogenous network VAR model \citep[e.g.,][]{YSM24}, our model achieves substantial dimension reduction, reducing the number of unknown coefficient functions to $K_0^2+K_0$ which is finite (as $K_0$ is assumed to be fixed). In the homogenous or grouped linear network VAR \citep[e.g.,][]{ZPLLW17, CFZ23, ZXF23}, the so-called nodal effect is often incorporated in the model formulation. Hence, we may further extend model (\ref{eq2.1}) to 
\begin{equation}\label{eq2.3}
x_{i,t}=\sum_{j\neq i}\alpha_{g_ig_j}(\tau_t)\widetilde{w}_{ij}x_{j,t-1}+\alpha_{g_i}(\tau_t)x_{i,t-1}+Z_i^{^\intercal}\gamma_{g_i}(\tau_t)+\varepsilon_{i,t},
\end{equation}
where $Z_i$ is a $p$-dimensional vector of node-specific exogenous covariates and $\gamma_{g_i}(\cdot)$ is a vector of smooth coefficient functions.  Letting $\{Z_i\}$ be independent of $\{\varepsilon_{i,t}\}$, the model framework and methodology developed in the sequel can be extended to tackle (\ref{eq2.3}) with slight modification. However, for notational simplicity, we mainly focus on model (\ref{eq2.1}) without the nodal effect throughout the paper. 

%%%%%%%%%%%%%%%%

\subsection{Fundamental assumptions and functional dependence measure}\label{sec2.2}

Let $f^{\prime}(\cdot)$ and $f^{\prime\prime}(\cdot)$ be the first- and second-order derivatives of $f(\cdot)$. We impose the following assumption on model (\ref{eq2.1}).

\begin{assumption}\label{ass:1}

(i)\ For $1\leq g,g^\ast\leq K_0$, $\alpha_{gg^\ast}(\cdot)$ and $\alpha_{g}(\cdot)$ are second-order continuously differentiable functions with
\[
\max_{1\leq g,g^\ast\leq K_0}\sup_{0\leq \tau\leq 1}\left\{\left\vert\alpha_{gg^\ast}^\prime(\tau)\right\vert+\left\vert\alpha_{gg^\ast}^{\prime\prime}(\tau)\right\vert\right\}+\max_{1\leq g\leq K_0}\sup_{0\leq \tau\leq 1}\left\{\left\vert\alpha_{g}^\prime(\tau)\right\vert+\left\vert\alpha_{g}^{\prime\prime}(\tau)\right\vert\right\}\leq c_\alpha,
\]
where $c_\alpha$ is a positive constant. In addition,
\begin{equation}\label{eq2.4}
\max_{1\leq g,g^\ast\leq K_0}\sup_{0\leq \tau\leq 1}\left\vert\alpha_{gg^\ast}(\tau)\right\vert+\max_{1\leq g\leq K_0}\sup_{0\leq \tau\leq 1}\left\vert\alpha_{g}(\tau)\right\vert<1.
\end{equation}

(ii)\ Let $\{\varepsilon_t\}$ be a sequence of i.i.d. random vectors with zero mean, positive definite covariance matrix denoted by ${\boldsymbol\Sigma}_\varepsilon$, and 
\[
\max_{1\leq i\leq N}{\sf E}\left[|\varepsilon_{i,t}|^{q}\right]\leq c_\varepsilon,
\]
where $q>8$ and $c_\varepsilon$ is a positive constant.

\end{assumption}

\begin{remark}\label{re:1}

(i) The smoothness condition on $\alpha_{gg^\ast}(\cdot)$ and $\alpha_{g}(\cdot)$ in Assumption \ref{ass:1}(i) is common for the local linear estimation method and theory \citep[e.g.,][]{FG96}. We may replace it by the condition that $\alpha_{gg^\ast}(\cdot)$ and $\alpha_{g}(\cdot)$ belong to the H\"older class \citep[e.g., Definition 1.2 in][]{T08} when adopting a general local polynomial estimation. The condition (\ref{eq2.4}) in Assumption \ref{ass:1}(i) is a natural extension of the stability assumption for grouped linear network VAR \citep{ZXF23}, ensuring that the underlying time series process is locally stable\footnote{The condition in (\ref{eq2.4}) may be replaced by the following assumption: uniformly over $\tau\in[0,1]$, ${\sf det}({\mathbf I}_N-z{\mathbf B}(\tau))\neq0$ for all $|z|\leq 1$. In fact, (\ref{eq2.4}) is a sufficient condition to guarantee the latter assumption.}. We require a relatively strong moment condition  ($q>8$) in Assumption \ref{ass:1}(ii) to derive the uniform convergence of the kernel-weighted quantities, see Lemmas D.1 and D.2 in Appendix D of the supplement \citep{LPTW24}. 

\smallskip

(ii) For $X_t$ defined in (\ref{eq2.2}), its stationary approximation is given by
\begin{equation}\label{eq2.5}
X_t^\circ(\tau)={\mathbf B}(\tau)X_{t-1}^\circ(\tau)+\varepsilon_t,\ \ 0\leq\tau\leq 1.
\end{equation}
It follows from (\ref{eq2.4}) in Assumption \ref{ass:1}(i) that $\vert{\mathbf B}(\tau)\vert_{\sf O}<1$ uniformly over $0\leq\tau\leq 1$. Hence, we obtain the following Wold representation:
\begin{equation}\label{eq2.6}
X_t^\circ(\tau)=G(\tau,\mathscr{F}_t):=\sum_{j=0}^\infty {\mathbf B}^j(\tau)\varepsilon_{t-j},
\end{equation}
where $\mathscr{F}_t=(\cdots,\varepsilon_{t-1},\varepsilon_{t})$. Assume that 
\begin{equation}\label{eq2.7}
\sup_{0\leq \tau\leq 1} \left\Vert \left\vert X_t^\circ(\tau)\right\vert_\infty\right\Vert_q\leq \theta_{N,q},
\end{equation}
which is weaker than the condition in Example 2.1 of \cite{ZW21} as we allow $\theta_{N,q}$ to diverge as $N$ increases. When $N$ is fixed, the above condition can be simplified to $\sup_{0\leq \tau\leq 1} {\sf E}[\vert X_t^\circ(\tau)\vert_q^q]\leq \theta_{q}$ with $\theta_q$ be a positive constant. Letting $X_t^\circ=X_t^\circ(\tau_t)$, under Assumption \ref{ass:1}(i) and (\ref{eq2.7}), we may show that 
\begin{equation}\label{eq2.8}
\max_{1\leq t\leq T} \left\Vert \left\vert X_t-X_t^\circ\right\vert_\infty\right\Vert_q=O\left(\theta_{N,q}/T\right),
\end{equation}
indicating that $X_t$ may be replaced by $X_t^\circ$ in the asymptotic derivation by restricting the divergence rate of $\theta_{N,q}$. The proof of (\ref{eq2.8}) is provided in Appendix B of the supplement \citep{LPTW24}, which also discusses the connection of the proposed model to the nonlinear functional dependence measure introduced by \cite{Wu05}.  

\end{remark}

%%%%%%%%%%%%%%%%%%%

\section{Estimation of the latent group structure}\label{sec3} 
\renewcommand{\theequation}{3.\arabic{equation}} \setcounter{equation}{0}
\renewcommand{\theprop}{3.\arabic{prop}}\setcounter{prop}{0}

In this section, we introduce the methodology for estimating the latent group membership and number, and present the relevant asymptotic properties.

%%%%%%%%%%%%%%%%%%%

\subsection{Group membership estimation when $K_0$ is pre-specified}\label{sec3.1}

We next introduce the nonparametric estimation and clustering methods and obtain the consistent estimation of the group membership ${\mathscr G}$ when $K_0$ is known a priori. The methodology can be split into the following three stages. 

\medskip

{\bf Stage 1}:\ \ As there is no prior information on the latent group structure, we start with the fully heterogenous time-varying network VAR model:
\begin{equation}\label{eq3.1}
x_{i,t}=\sum_{j\in{\mathscr N}_i}\beta_{ij}(\tau_t)\widetilde{w}_{ij}x_{j,t-1}+\beta_{i}(\tau_t)x_{i,t-1}+\varepsilon_{i,t},
\end{equation}
where  ${\mathscr N}_i=\{j\neq i:\ w_{ij}=1\}$ is the index set of nodes which the $i$-th node follows, $\beta_{ij}(\cdot)=\alpha_{g_ig_j}(\cdot)$ and $\beta_i(\cdot)=\alpha_{g_i}(\cdot)$. Model (\ref{eq3.1}) is similar to that in \cite{YSM24}. Note that $\beta_{ij}(\cdot)$, $j\notin{\mathscr N}_i$, are unidentifiable and thus not estimable. With the smoothness condition in Assumption \ref{ass:1}(i), we adopt the local linear smoothing \citep[e.g.,][]{FG96} to estimate the heterogenous time-varying coefficient functions, only using the sample information from the $i$-th node and its direct neighbors, i.e., $\widetilde{w}_{ij}\neq0$.   

\smallskip

Define
\[
\widetilde{X}_{i,t-1}=\left[\left(\widetilde{w}_{ij}x_{j,t-1}:\ j\in{\mathscr N}_i\right)^{^\intercal}, x_{i,t-1}\right]^{^\intercal},
\] 
which is a random vector with dimension $n_i+1$, where $n_i={\rm card}({\mathscr N}_i)$ is allowed to diverge slowly to infinity and ${\rm card}(\cdot)$ denotes the cardinality of a set. Letting
\[
\beta_{i\bullet}(\tau)=\left[\left(\beta_{ij}(\tau):\ j\in{\mathscr N}_i\right)^{^\intercal}, \beta_i(\tau)\right]^{^\intercal},
\]
with Assumption \ref{ass:1}(i), we have the following Taylor expansion:
\[
\beta_{i\bullet}(\tau_t)\approx \beta_{i\bullet}(\tau)+\beta_{i\bullet}^{\prime}(\tau)(\tau_t-\tau)
\]
when $\tau_t$ falls in a small neighborhood of $\tau$. Define the node-specific local linear weighted objective function:
\begin{equation}\label{eq3.2}
{\cal L}_i(a,b)=\sum_{t=1}^T\left[x_{i,t}-a^{^\intercal}\widetilde X_{i,t-1}-b^{^\intercal}\widetilde X_{i,t-1}(\tau_t-\tau)\right]^2K_h(\tau_t-\tau),
\end{equation}
where $a$ and $b$ are $(n_i+1)$-dimensional vectors, $K_h(\cdot)=\frac{1}{h}K(\cdot/h)$, $K(\cdot)$ is a kernel function and $h$ is a bandwidth. Minimizing ${\cal L}_i(a,b)$ with respect to the vectors $a$ and $b$, we obtain the solution denoted as $\widehat a$ and $\widehat b$, and then the local linear estimate of $\beta_{i\bullet}(\tau)$ as
\begin{equation}\label{eq3.3}
\widehat\beta_{i\bullet}(\tau)=\left[\left(\widehat\beta_{ij}(\tau):\ j\in{\mathscr N}_i\right)^{^\intercal}, \widehat\beta_i(\tau)\right]^{^\intercal}=\widehat a.
\end{equation}
In practice, we obtain the local linear estimates at $\tau_l^\ast$, $l=1,\cdots,L$, a sequence of user-specified equidistant grid points between $0$ and $1$ satisfying $L\rightarrow\infty$ and $L=O(T)$. 

\medskip

{\bf Stage 2}:\ \ Let $\bar{\mathscr N}_N=\{(i,j): 1\leq i\leq N, j\in{\mathscr N}_i\}$. It follows from (\ref{eq2.1}) that there exists a latent homogeneity structure for $\beta_{ij}(\cdot)$, $(i,j)\in\bar{\mathscr N}_N$, and the number of distinct time-varying coefficient functions is at most $K_0^2$. Let $\beta_m^\circ(\cdot)$, $m=1,\cdots,M_0$, denote the true distinct time-varying coefficient functions for network spillover effects, $M_0\leq K_0^2$, and $g_{ij}\in\{1,\cdots,M_0\}$ be the group membership for the index pair $(i,j)\in\bar{\mathscr N}_N$. With $\{\widehat{\beta}_{ij}(\tau_l^\ast):\ 1\leq l\leq L, (i,j)\in\bar{\mathscr N}_N\}$ obtained in Stage 1, combining the clustering algorithm and the ratio criterion with details provided in Appendix A (see also Stage 3 and Section \ref{sec3.2}), we may obtain a consistent estimate of $M_0$ denoted by $\widehat{M}$, and the estimated membership $\widehat{g}_{ij}$. For the $i$-th node, we construct 
\begin{equation}\label{eq3.4} 
\widehat{\beta}_{i\bullet}^\circ(\tau)=\left[\widehat{\beta}_{i1}^\circ(\tau),\cdots,\widehat{\beta}_{i\widehat{M}}^\circ(\tau)\right]^{^\intercal}\quad{\rm with}\quad \widehat{\beta}_{im}^\circ(\tau)=\sum_{j\in{\mathscr N}_i}\widehat{\beta}_{ij}(\tau)\widehat{\omega}_{ij,m},
\end{equation}
where $\tau$ is chosen as the grid points $\tau_{l}^\ast$ defined in Stage 1,
\[
\widehat{\omega}_{ij,m}=\left\{
\begin{array}{ll}
I(\widehat{g}_{ij}=m)/\sum_{j\in{\mathscr N}_i}I(\widehat{g}_{ij}=m),\ &\ {\rm if}\ \sum_{j\in{\mathscr N}_i}I(\widehat{g}_{ij}=m)>0,\\
0,\ &\ {\rm if}\ \sum_{j\in{\mathscr N}_i}I(\widehat{g}_{ij}=m)=0,
\end{array}
\right.
\]
and $I(\cdot)$ denotes the indicator function. 

\medskip

{\bf Stage 3}:\ \ With the estimates $\widehat{\beta}_i(\cdot)$ and $\widehat{\beta}_{i\bullet}^\circ(\cdot)$ defined in Stages 1 and 2, respectively, we may compute the point-wise distance between nodes $i$ and $j$: 
\begin{equation}\label{eq3.5}
\widehat{d}_{ij}(\tau)=\left\vert\widehat{\beta}_i(\tau)-\widehat{\beta}_j(\tau)\right\vert+\left\vert\widehat{\beta}_{i\bullet}^\circ(\tau)-\widehat{\beta}_{j\bullet}^\circ(\tau)\right\vert_2,
\end{equation}
and subsequently define the distance matrix:
\[
\widehat{\mathbf D}=\left\{\widehat{D}_{ij}\right\}_{N\times N},\quad \widehat{D}_{ij}=\frac{1}{L}\sum_{l=1}^L\widehat{d}_{ij}(\tau_l^\ast).
\]
It is clear that the diagonal elements of $\widehat{\mathbf D}$ are zeros. With the distance matrix $\widehat{\mathbf D}$, we may adopt the agglomerative hierarchical clustering algorithm which is commonly used in unsupervised cluster analysis \citep[e.g.,][]{HTF09, ELLS11}. This clustering algorithm has been recently combined with the kernel-based estimation technique to identify the homogeneity/group structure in nonparametric panel regression models. For instance, \cite{Ch19} constructs a similar distance matrix and further estimates the latent group structure in time-varying coefficient panel data models; and \cite{VL20} introduce a bandwidth-free normalized distance measure in the clustering algorithm but assume the panel observations are independent over subjects. The latter assumption may be too restrictive for large-scale network time series data and is thus removed in this paper. Another relevant paper is \cite{Z13} which clusters nonlinear trend functions based on parallelism and allows the number of time series to grow at a slow polynomial rate of $T$. In contrast, the number of nodes $N$ can be much larger than $T$ in this paper, see (\ref{eq3.4}) in Assumption \ref{ass:2}(iii).  

\smallskip

Assuming the true group number $K_0$ is known a priori, we start with $N$ clusters each of which corresponds to one node, search for the smallest off-diagonal entry in $\widehat{\mathbf D}$ (which is the smallest estimated distance between nodes), and merge the two corresponding nodes. Consequently the cluster number reduces from $N$ to $N-1$. Use a linkage technique (such as the single or complete linkage) to calculate the distance between the merged cluster and the remaining ones and update the estimated distance matrix with size $(N-1)\times (N-1)$. Repeat the previous steps with the updated distance matrix, and stop the algorithm when the number of clusters reaches $K_0$. We denote the estimated clusters by $\widehat{\mathscr G}_k$, $k=1, \cdots,K_0$. 

\medskip

Let ${\boldsymbol\Delta}_{i,t}={\sf E}[\widetilde{X}_{i,t}\widetilde{X}_{i,t}^{^\intercal}]$ with $\widetilde{X}_{i,t}$ defined in Stage 1, and write 
\begin{equation}\label{eq3.6} 
\beta_{i\bullet}^\circ(\tau)=\left[\beta_{i1}^\circ(\tau),\cdots,\beta_{iM_0}^\circ(\tau)\right]^{^\intercal}\quad{\rm with}\quad \beta_{im}^\circ(\tau)=\sum_{j\in{\mathscr N}_i}\beta_{ij}(\tau)\omega_{ij,m}
\end{equation}
and
\[
\omega_{ij,m}=\left\{
\begin{array}{ll}
I(g_{ij}=m)/\sum_{j\in{\mathscr N}_i}I(g_{ij}=m),\ &\ {\rm if}\ \sum_{j\in{\mathscr N}_i}I(g_{ij}=m)>0,\\
0,\ &\ {\rm if}\ \sum_{j\in{\mathscr N}_i}I(g_{ij}=m)=0.
\end{array}
\right.
\]
The latter is estimated by $\widehat{\beta}_{i\bullet}^\circ(\tau)$ defined in (\ref{eq3.4}) (up to permutation). The following conditions are required to derive the consistency property of $\widehat{\mathscr G}_k$, $k=1, \cdots,K_0$.   

\smallskip

\begin{assumption}\label{ass:2}

(i)\ The kernel function $K\left(\cdot\right)$ is a symmetric probability density function that is Lipschitz-continuous and has a compact support $\left[-1,1\right]$.

(ii)\ There exist two finite positive constants: $\underline{\lambda}$ and $\overline{\lambda}$, such that
\[
0<\underline{\lambda}\le\min_{1\le i\le N}\min_{0\leq t\leq T-1}\lambda_{\min}({\boldsymbol\Delta}_{i,t})\le\max_{1\le i\le N}\sup_{0\leq t\leq T-1}\lambda_{\max}({\boldsymbol\Delta}_{i,t})\le\overline{\lambda}<\infty.
\]

(iii)\ Let $T$, $N$, and $h$ satisfy $h\rightarrow0$, $Th\rightarrow\infty$ and
\begin{eqnarray}\label{eq3.7}
\frac{N\theta_{N,q}^q}{T^{\frac{q^2-6q-8}{4(q+2)}}\left[h\log(N\vee T)\right]^{q/4}}\to 0
\end{eqnarray}
with $q$ defined in Assumption \ref{ass:1}(ii) and $\theta_{N,q}$ is defined in (\ref{eq2.7}).

(iv) Letting ${\mathscr N}_i(j)=\{k\in{\mathscr N}_i: g_k=j\}$ and $\bar n=\max_{1\leq i\leq N}n_i$, 
\begin{equation}\label{eq3.8}
\min_{1\leq i,j\leq K_0}{\rm card}({\mathscr N}_i(j))\geq1,\quad \bar n=o\left(\sqrt{Th/\log (N\vee T)}\right).
\end{equation}

\end{assumption}

\begin{assumption}\label{ass:3}

Let 
\begin{equation}\label{eq3.9}
\sqrt{\bar n}\left(h^2+\sqrt{\frac{\log (N\vee T)}{Th}}\right)+L^{-1}=o\left(\zeta_{NT}^\dag\wedge\zeta_{NT}^\ddag\right),
\end{equation}
where 
\[
\zeta_{NT}^\dag=\min_{1\leq g_i\neq g_j\leq K_0}\int_0^1\left[\left\vert\alpha_{g_i}(\tau)-\alpha_{g_j}(\tau)\right\vert+\left\vert \beta_{i\bullet}^\circ(\tau)- \beta_{j\bullet}^\circ(\tau)\right\vert_2\right]d\tau
\]
and
\[ 
\zeta_{NT}^\ddag=\min_{1\leq m\neq m^\ast\leq M_0}\int_0^1\left\vert\beta_{m}^\circ(\tau)-\beta_{m^\ast}^\circ(\tau)\right\vert d\tau.
\]

\end{assumption}

\begin{remark}\label{re:2}

(i) Assumption \ref{ass:2}(i) contains some commonly-used conditions imposed on the kernel function \citep[e.g.,][]{FG96}. Assumption \ref{ass:2}(ii) is crucial to ensure that the kernel-weighted random denominator in local linear estimation is non-singular. In fact, the consistency property in Theorem \ref{thm:3.1} below continues to hold by allowing $\underline\lambda$ to slowly converge to zero and strengthening other relevant conditions, e.g., the second restriction in (\ref{eq3.8}) and (\ref{eq3.9}) would be strengthened to 
\[
\bar n=o\left(\underline\lambda\sqrt{Th/\log (N\vee T}\right)\quad {\rm and}\quad \sqrt{\bar n}\left(h^2+\sqrt{\frac{\log (N\vee T)}{Th}}\right)/\underline\lambda+L^{-1}=o\left(\zeta_{NT}^\dag\right),
\]
respectively. Note that $h\rightarrow0$ and $Th\rightarrow\infty$ in Assumption \ref{ass:2}(iii) are regular conditions for kernel-based smoothing, whereas the condition (\ref{eq3.7}) indicates that there is a trade-off between the network size and the required moment condition (i.e., when $q$ increases, $N$ may diverge at a faster rate). The first condition in (\ref{eq3.8}) indicates that node $i$ follows nodes in each of the $K_0$ groups and $\alpha_{g_ig_j}(\cdot)$ is estimable via the local linear method in Stage 1, whereas the second condition in (\ref{eq3.8}) restricts the divergence rate of $n_i$ so that the first-stage local linear estimation of the heterogenous time-varying coefficient functions is uniformly consistent, see Lemma D.3 in the supplement \citep{LPTW24}. 

\smallskip

(ii) Assumption \ref{ass:3} indicates that the minimum distance between groups may converge to zero at a rate slower than a typical nonparametric uniform convergence rate if the grid number $L$ is of order $T$ and $\bar n$ is bounded. The restriction (\ref{eq3.9}) is automatically satisfied if $\zeta_{NT}^\dag$ and $\zeta_{NT}^\ddag$ are strictly larger than a positive constant \citep[e.g.,][]{ZXF23}.

\end{remark}

Theorem \ref{thm:3.1} below establishes the consistency property of the group membership estimation when $K_0$ is pre-specified.

\renewcommand{\thetheorem}{3.\arabic{theorem}}\setcounter{theorem}{0}

\begin{theorem}\label{thm:3.1}

Suppose that Assumptions \ref{ass:1}--\ref{ass:3} hold and $K_0$ is known a priori. Then, as $N,T\rightarrow\infty$ jointly, we have
\begin{equation}\label{eq3.10}
{\sf P}\left(\big\{\widehat{\mathscr G}_k,\ 1\leq k\leq K_0\big\}=\big\{{\mathscr G}_k,\ 1\leq k\leq K_0\big\}\right)\rightarrow1.
\end{equation}

\end{theorem}

\begin{remark}\label{re:3}

The consistency property (\ref{eq3.10}) is similar to the consistency results of group membership estimation in nonparametric panel/longitudinal data models, see Theorem 3.1 in \cite{VL17} and Theorem 4.1(a) in \cite{VL20}. The key step of proving Theorem \ref{thm:3.1} is to show that
\[\max_{1\le k \le K_0}\max_{i,j\in{\mathscr G}_k}\widehat{D}_{ij}<\min_{1\le k\neq l\le K_0}\min_{i\in{\mathscr G}_k,j\in{\mathscr G}_l}\widehat{D}_{ij},\ \ w.p.a.1.\]
This can be proved by using the uniform convergence property of $\widehat d_{ij}(\cdot)$.

\end{remark}

\subsection{Estimation of the group number}\label{sec3.2}

In practice, the true number of groups is unknown and a data-driven criterion is thus required to obtain its consistent estimation. We next introduce an easy-to-implement ratio criterion to estimate $K_0$ consistently. Assuming the group number to be $K$, we may terminate the clustering algorithm in Stage 3 when the cluster number reaches $K$ and obtain the estimated clusters denoted by $\widehat{\mathscr G}_{k|K}$, $k=1,\cdots,K$. With these estimated clusters, we pool the heterogenous time-varying coefficient estimates $\widehat{\beta}_i(\cdot)$ and $\widehat{\beta}_{i\bullet}^\circ(\cdot)$ over $i\in\widehat{\mathscr G}_{k|K}$, and obtain
\begin{equation}\label{eq3.11}
\widehat\beta_{k|K}(\tau)=\frac{1}{{\rm card}\left(\widehat{\mathscr G}_{k|K}\right)}\sum_{i\in\widehat{\mathscr G}_{k|K}}\widehat{\beta}_i(\tau),\quad \widehat\beta_{k|K}^\circ(\tau)=\frac{1}{{\rm card}\left(\widehat{\mathscr G}_{k|K}\right)}\sum_{i\in\widehat{\mathscr G}_{k|K}}\widehat{\beta}_{i\bullet}^\circ(\tau).
\end{equation}
Then, we define the average deviation:
\begin{equation}\label{eq3.12}
\widehat{R}(K)= \frac{1}{KL}\sum_{k=1}^K\sum_{l=1}^L\frac{1}{{\rm card}\left(\widehat{\mathscr G}_{k|K}\right)}\sum_{i\in\widehat{\mathscr G}_{k|K}}\left[\left\vert \widehat{\beta}_i(\tau_l^\ast)-\widehat\beta_{k|K}(\tau_l^\ast)\right\vert+\left\vert \widehat{\beta}_{i\bullet}^\circ(\tau_l^\ast)-\widehat\beta_{k|K}^\circ(\tau_l^\ast)\right\vert_2\right].
\end{equation}
The grouped time-varying network VAR model is either correctly- or over-fitted when $K\geq K_0$, indicating that $\widehat{R}(K)$ converges to zero, and is under-fitted when $K<K_0$. For the latter scenario, at least two groups would be falsely merged, leading to biased estimation of some group-specific time-varying coefficient functions and a relatively large value of $\widehat{R}(K)$. Hence, it is sensible to estimate $K_0$ by
\begin{equation}\label{eq3.13}
\widehat{K}=\argmin_{1\leq K\leq \overline{K}} \frac{\widehat{R}(K)}{\widehat{R}(K-1)},
\end{equation}
where $\overline{K}$ is a pre-specified positive integer larger than $K_0$. In practical implementation, we set $\widehat{R}(1)/\widehat{R}(0)\equiv1$, $\widehat{R}(K)=0$ if it is smaller than $\rho_{NT}$, a user-specified tuning parameter, and define $0/0\equiv1$. A similar ratio criterion is adopted by \cite{YCLL23} to consistently estimate the group number in nonparametric grouped panel quantile regression models. Other applications of the ratio criterion can be found in \cite{LY12} and \cite{LRS20}. 

\smallskip

We require some further conditions to derive the consistency property of $\widehat K$.

\begin{assumption}\label{ass:4}
	
(i)\ There exists a positive constant $c_{\mathscr G}$ such that 
\[\min_{1\le k\le K_0}{\rm card}({\mathscr G}_k)\ge c_{\mathscr G}\cdot N.\]

(ii)\ The tuning parameter $\rho_{NT}$ satisfies that 
\begin{equation}\label{eq3.14}
\rho_{NT}=o\left(\zeta_{NT}^\dag\wedge \zeta_{NT}^\ddag\right),\quad \sqrt{\bar n}\left(h^2+\sqrt{\frac{\log (N\vee T)}{Th}}\right)=o\left(\rho_{NT}\right).
\end{equation}

\end{assumption}

\begin{remark}\label{re:4}

Assumption \ref{ass:4}(i) indicates that the cardinality of ${\mathscr G}_k$ is of the same order over $k$. A similar restriction is also adopted by \cite{Ch19} and \cite{ZXF23}. Assumption \ref{ass:4}(ii) indicates that the order of $\rho_{NT}$ lies between $\sqrt{\bar n}(h^2+\sqrt{\log (N\vee T)/(Th)})$ and $\xi_{NT}^\dag\wedge \xi_{NT}^\ddag$, which is not unreasonable due to Assumption \ref{ass:3}. In particular, when $\xi_{NT}^\dag\wedge \xi_{NT}^\ddag$ is bounded away from zero and $\bar n$ is upper bounded by a positive constant, the conditions in (\ref{eq3.14}) can be simplified to
\[
\rho_{NT}\rightarrow0,\quad h^2+\sqrt{\frac{\log (N\vee T)}{Th}}=o\left(\rho_{NT}\right).
\]

\end{remark}

We establish the consistency property of the ratio criterion in the following theorem.

\begin{theorem}\label{thm:3.2}

Suppose that Assumptions \ref{ass:1}--\ref{ass:4} are satisfied. Then, as $N,T\rightarrow\infty$ jointly, 
\begin{equation}\label{eq3.15}
{\sf P}\left(\widehat K=K_0\right)\rightarrow1.
\end{equation}

\end{theorem}

Finally, we use the estimated group number $\widehat K$ and terminate the clustering algorithm in Section \ref{sec3.1} when the cluster number reaches $\widehat K$. In order to avoid unnecessary notational burden, we still denote the estimated groups as $\widehat{\mathscr G}_k$, $k=1, \cdots,\widehat{K}$. Combining Theorems \ref{thm:3.1} and \ref{thm:3.2}, we readily have the following result.

\renewcommand{\thecorollary}{3.\arabic{corollary}}\setcounter{corollary}{2}

\begin{corollary}\label{cor:3.3}

Suppose that Assumptions \ref{ass:1}--\ref{ass:4} are satisfied. Then, as $N,T\rightarrow\infty$ jointly,
\begin{equation}\label{eq3.16}
{\sf P}\left(\big\{\widehat{\mathscr G}_k,\ 1\leq k\leq \widehat{K}\big\}=\big\{{\mathscr G}_k,\ 1\leq k\leq K_0\big\}\right)\rightarrow1.
\end{equation}

\end{corollary}

%%%%%%%%%%%%%%%%%%%

\section{Post-grouping local linear estimation}\label{sec4} 
\renewcommand{\theequation}{4.\arabic{equation}} \setcounter{equation}{0}

The heterogenous local linear estimation defined in (\ref{eq3.3}) only makes use of the sample information from node $i$ and its direct neighbors, resulting in rather slow convergence rates (see Lemma D.3 in the supplement) and unstable numerical performance if $T$ is not sufficiently large in finite samples. We next aim to address this issue by pooling the sample information over nodes in the same cluster and proposing a post-grouping local linear estimation.

\smallskip

It follows from Corollary \ref{cor:3.3} that, for any $k=1,\cdots,K_0$, there exists $1\leq k^\dag\leq \widehat{K}$ such that ${\mathscr G}_k=\widehat{\mathscr G}_{k^\dag}$ {\em w.p.a.1}. Without loss of generality, we may consider ${\mathscr G}_k=\widehat{\mathscr G}_{k}$ (conditioning on $\widehat{K}=K_0$) throughout this section. For $i\in\widehat{\mathscr G}_k$, define
\[
\check{X}_{i,t-1}=\left[\sum_{j\in\widehat{\mathscr G}_1}\widetilde w_{ij} x_{j,t-1},\cdots,\sum_{j\in\widehat{\mathscr G}_{\widehat{K}}}\widetilde w_{ij} x_{j,t-1}, x_{i,t-1}\right]^{^\intercal},
\] 
which is a random vector with dimension $\widehat{K}+1$. For the $k$-th group, let
\[
\alpha_{k\bullet}(\tau)=\left[\alpha_{k1}(\tau),\cdots,\alpha_{kK_0}(\tau), \alpha_k(\tau)\right]^{^\intercal}
\]
be a vector of true group-specific time-varying coefficient functions to be estimated. For each $k$ and given $\tau\in(0,1)$, we define the following post-grouping local linear objective function: 
\begin{equation}\label{eq4.1}
\sum_{i\in\widehat{\mathscr G}_k}\sum_{t=1}^T\left[x_{i,t}-a^{^\intercal}\check{X}_{i,t-1}-b^{^\intercal}\check{X}_{i,t-1}(\tau_t-\tau)\right]^2K_{h_\dag}(\tau_t-\tau),
\end{equation}
where $h_\dag$ is a bandwidth which may be different from $h$ used in the heterogenous local linear estimation (\ref{eq3.2}) and (\ref{eq3.3}). Minimizing the post-grouping objective function with respect to the vectors $a$ and $b$, we obtain the solutions denoted by $\check a$ and $\check b$, and construct the post-grouping local linear estimation as 
\begin{equation}\label{eq4.2}
\check\alpha_{k\bullet}(\tau)=\left[\check\alpha_{k1}(\tau), \cdots,\check\alpha_{kK_0}(\tau), \check\alpha_{k}(\tau)\right]^{^\intercal}=\check a.
\end{equation}

\smallskip

Let $\sigma_{ij}={\sf E}\left(\varepsilon_{i,t}\varepsilon_{j,t}\right)$ and 
\[
{\boldsymbol\Delta}_{ij}^\diamond(\tau)={\sf E}\left[X_{i,t}^\diamond(\tau) X_{j,t}^{\diamond^\intercal}(\tau)\right],\ \ 1\leq i,j\leq N,
\] 
where 
\[
X_{i,t}^\diamond=\left[\sum_{j\in {\mathscr G}_1}\widetilde w_{ij} x_{j,t}^\circ(\tau),\cdots,\sum_{j\in {\mathscr G}_{K_0}}\widetilde w_{ij} x_{j,t}^\circ(\tau), x_{i,t}^\circ(\tau)\right]^{^\intercal}
\] 
with $x_{i,t}^\circ(\tau)$ being the $i$-th element of $X_{t}^\circ(\tau)$ defined in (\ref{eq2.6}). %We require the following conditions to establish the asymptotic normal distribution theory.

\begin{assumption}\label{ass:5}

(i)\ The bandwidth $h_\dag$ satisfies that $h_\dag\to0$ and $Th_\dag/\log (N\vee T)\to\infty$. In addition, (\ref{eq3.7}) holds when $h$ is replaced by $h_\dag$.

(ii)\ There exists a positive definite matrix ${\boldsymbol\Upsilon}_{{\mathscr G}_k}(\tau)$ such that
\begin{equation}\label{eq4.3}
\frac{1}{{\rm card}({\mathscr G}_k)}\sum_{i,j\in{\mathscr G}_k}\sigma_{ij}{\boldsymbol\Delta}_{ij}^\diamond(\tau)\rightarrow {\boldsymbol\Upsilon}_{{\mathscr G}_k}(\tau)
\end{equation}
as ${\rm card}({\mathscr G}_k)\rightarrow\infty$, and in addition, 
\[
\frac{1}{{\rm card}({\mathscr G}_k)}\sum_{i\in{\mathscr G}_k}{\boldsymbol\Delta}_{i}^\diamond(\tau)\rightarrow \boldsymbol{\Delta}_{{\mathscr G}_k}(\tau),
\]
which is positive definite, where ${\boldsymbol\Delta}_{i}^\diamond(\tau)={\boldsymbol\Delta}_{ii}^\diamond(\tau)$.

\end{assumption}

\begin{remark}\label{re:5}

The bandwidth restriction in Assumption \ref{ass:5}(i) is comparable to that in Assumption \ref{ass:2}(iii). Assumption \ref{ass:5}(ii) allows weak correlation between nodes and can be substantially simplified when $\varepsilon_{i,t}$ are independent over $i$ \citep[e.g.,][]{ZPLLW17}. For example, if $\sigma_{ij}=0$ when $i\neq j$ and $\sigma_{ii}\equiv\sigma^2$, (\ref{eq4.3}) would be simplified to
\[
\frac{1}{{\rm card}({\mathscr G}_k)}\sum_{i,j\in{\mathscr G}_k}\sigma_{ij}{\boldsymbol\Delta}_{ij}^\diamond(\tau)=\frac{\sigma^2}{{\rm card}({\mathscr G}_k)}\sum_{i\in{\mathscr G}_k}{\boldsymbol\Delta}_{i}^\diamond(\tau)\rightarrow \sigma^2 \boldsymbol{\Delta}_{{\mathscr G}_k}(\tau).
\]

\end{remark}

\renewcommand{\thetheorem}{4.\arabic{theorem}}\setcounter{theorem}{0}

\begin{theorem}\label{thm:4.1}

Suppose that Assumptions \ref{ass:1}--\ref{ass:5} are satisfied. For any $\tau\in\left(0,1\right)$, 
\begin{equation}\label{eq4.4}
\sqrt{{\rm card}({\mathscr G}_k) Th_\dag}\left[\check{\alpha}_{k\bullet}(\tau)-\alpha_{k\bullet}(\tau)-\frac{1}{2}h_{\dag}^{2}\mu_2\alpha_{k\bullet}^{\prime\prime}(\tau)\right]\stackrel{d}\longrightarrow {\sf N}\left(\boldsymbol{0}, {\boldsymbol\Omega}_{{\mathscr G}_k}(\tau)\right),
\end{equation}
as $N,T\rightarrow\infty$ jointly, where $
{\boldsymbol\Omega}_{{\mathscr G}_k}(\tau)=\nu_0{\boldsymbol\Delta}_{{\mathscr G}_k}^{-1}(\tau){\boldsymbol\Upsilon}_{{\mathscr G}_k}(\tau){\boldsymbol\Delta}_{{\mathscr G}_k}^{-1}(\tau)$, $\nu_\kappa=\int u^\kappa K^2(u)du$ and $\mu_\kappa=\int u^\kappa K(u)du$ for $k=0,1,\cdots$.

\end{theorem}

\begin{remark}\label{re:6}

Since ${\rm card}({\mathscr G}_k)$ is of order $N$ by Assumption \ref{ass:4}(i), Theorem \ref{thm:4.1} shows that the post-grouping local linear estimation $\check{\alpha}_{k\bullet}(\tau)$ has the point-wise convergence rate $1/\sqrt{NTh_\dag}+h_\dag^2$, which is substantially faster than that for the heterogenous local linear estimation (ignoring the group structure). This is unsurprising since more sample information is used in the post-grouping estimation procedure. Note that \cite{YSM24} use the spline-based estimation for heterogenous functional coefficients, which only achieve the root-$T$ convergence rate. If, in addition,  $\varepsilon_{i,t}$ are independent over $i$, as discussed in Remark \ref{re:5}, we may simplify the asymptotic covariance matrix, i.e., ${\boldsymbol\Omega}_{{\mathscr G}_k}(\tau)=\nu_0\sigma^2{\boldsymbol\Delta}_{{\mathscr G}_k}^{-1}(\tau)$.

\end{remark}

%%%%%%%%%%%%%%%%%%%

\section{Breaks in the group structue}\label{sec5} 
\renewcommand{\theequation}{5.\arabic{equation}} \setcounter{equation}{0}

The model, methodology and theory developed in Sections \ref{sec2}--\ref{sec4} rely on the assumption that the latent group structure is time invariant and the group-specific coefficient functions are smooth over the entire time span. As discussed in the introductory section, this assumption may be restrictive for some empirical applications. Hence, we next make a further extension of the model, methodology and theory, allowing structural breaks in either the group membership, group number or group-specific coefficient functions. Our main interest lies in locating the break point and estimating the group structure before and after the break. We mainly consider the case of a single break for notational brevity and will briefly discuss its extension to the case of multiple breaks later in Remark \ref{re:8}(ii).      

\smallskip

Assume that the break occurs at an unknown time point $t_0$. Let ${\mathscr{G}}^1=\{{\mathscr G}_1^1,\cdots,{\mathscr G}_{K_1}^1\}$ and $g_i^1\in\{1,\cdots,K_1\}$ be the group structure and membership label (for node $i$) before the break, whereas let ${\mathscr{G}}^2=\{{\mathscr G}_1^2,\cdots,{\mathscr G}_{K_2}^2\}$ and $g_i^2\in\{1,\cdots,K_2\}$ be defined similarly for those after the break. Consider the time-varying network VAR model with break in the group structure:
\begin{equation}\label{eq5.1}
x_{i,t}=
\left\{
\begin{array}{ll}
\sum_{j\neq i}\alpha_{g_i^1g_j^1}^1(\tau_t)\widetilde{w}_{ij}x_{j,t-1}+\alpha_{g_i^1}^1(\tau_t)x_{i,t-1}+\varepsilon_{i,t},\ \ &\ \ 1\leq t\leq t_0,\\
\sum_{j\neq i}\alpha_{g_i^2g_j^2}^2(\tau_t)\widetilde{w}_{ij}x_{j,t-1}+\alpha_{g_i^2}^2(\tau_t)x_{i,t-1}+\varepsilon_{i,t},\ \ &\ \ t_0+1\leq t\leq T,\\
\end{array}
\right.
\end{equation}
where $\alpha_{g_i^1g_j^1}^1(\cdot)$ (or $\alpha_{g_i^2g_j^2}^2(\cdot)$) and $\alpha_{g_i^1}^1(\cdot)$ (or $\alpha_{g_i^2}^2(\cdot)$) are the smooth time-varying network spillover and momentum effects before (or after) the break. Model (\ref{eq5.1}) can be seen as an extension of the linear panel model framework (with break in the group structure) in \cite{LOW23} and \cite{WPS23}, taking into account the smooth time-varying feature and network structure.  

\smallskip

Throughout this section, assume that $\bar n$ is bounded. We next introduce a two-stage estimation procedure with break location estimation in Stage 1 and then group estimation in Stage 2.  

\medskip

{\bf Stage 1}:\ \ As the time-varying group structure is latent, similar to (\ref{eq3.1}), we first consider the fully heterogenous time-varying network VAR model with break at $t_0$: 
\begin{equation}\label{eq5.2}
x_{i,t}=\sum_{j\in{\mathscr N}_i}\beta_{ij}^\ddag(\tau_t)\widetilde{w}_{ij}x_{j,t-1}+\beta_{i}^\ddag(\tau_t)x_{i,t-1}+\varepsilon_{i,t},
\end{equation}
where 
\begin{equation}\label{eq5.3}
\beta_{ij}^\ddag(\tau_t)=\left\{
\begin{array}{ll}
\beta_{ij}^1(\tau_t),\ \ &\ \ 1\leq t\leq t_0,\\
\beta_{ij}^2(\tau_t),\ \ &\ \ t_0+1\leq t\leq T,
\end{array}
\right. \quad \beta_{i}^\ddag(\tau_t)=\left\{
\begin{array}{ll}
\beta_{i}^1(\tau_t),\ \ &\ \ 1\leq t\leq t_0,\\
\beta_{i}^2(\tau_t),\ \ &\ \ t_0+1\leq t\leq T.
\end{array}
\right.
\end{equation}
Write 
\[
\beta_{i\bullet}^\ddag(\tau_t)=\left[\left(\beta_{ij}^\ddag(\tau_t):\ j\in{\mathscr N}_i\right)^{^\intercal}, \beta_i^\ddag(\tau_t)\right]^{^\intercal}
\]
as in Section \ref{sec3.1}, and let 
\[
\beta_{i\bullet}^{\ddag,{\sf l}}(\tau)=\lim_{x\uparrow\tau} \beta_{i\bullet}^\ddag(x)\quad {\rm and}\quad \beta_{i\bullet}^{\ddag,{\sf r}}(\tau)=\lim_{x\downarrow\tau} \beta_{i\bullet}^\ddag(x)
\]
denote the left and right limits of $\beta_{i\bullet}^\ddag(\tau)$, respectively. Define
\begin{equation}\label{eq5.4}
\delta_\beta(t)=\max_{1\leq i\leq N} \left\vert \beta_{i\bullet}^{\ddag,{\sf r}}(\tau_t)-\beta_{i\bullet}^{\ddag,{\sf l}}(\tau_t)\right\vert_2.
\end{equation}
It follows from (\ref{eq5.1})--(\ref{eq5.3}) that $\delta_\beta(t)$, $t=1,\cdots,T$, achieve the maximum at $t=t_0$, which motivates the subsequent estimation procedure. Specifically, we estimate $\delta_\beta(t)$ by a one-sided kernel smoothing method\footnote{The extension to the one-sided local polynomial smoothing is straightforward, see \cite{CWW22}.} and then locate the break point by maximizing the estimate of $\delta_\beta(t)$ over $t$.

\smallskip

Let $K^\ddag(\cdot)$ be a one-sided kernel function with a compact support $[0,1]$, say, the one-sided version of the Epanechnikov kernel. Define
\begin{eqnarray}
\widetilde{\boldsymbol\Gamma}_{it}^{\sf l}&=&\frac{1}{Th_\ddag}\sum_{s=1}^T \widetilde{X}_{i,s-1}\widetilde{X}_{i,s-1}^{^\intercal}K^\ddag\left(\frac{\tau_t-\tau_s}{h_\ddag}\right),\quad \widetilde{\boldsymbol\Gamma}_{it}^{\sf r}=\frac{1}{Th_\ddag}\sum_{s=1}^T \widetilde{X}_{i,s-1}\widetilde{X}_{i,s-1}^{^\intercal}K^\ddag\left(\frac{\tau_s-\tau_t}{h_\ddag}\right),\notag\\
\overline{\Gamma}_{it}^{\sf l}&=&\frac{1}{Th_\ddag}\sum_{s=1}^T \widetilde{X}_{i,s-1}x_{i,s}K^\ddag\left(\frac{\tau_t-\tau_s}{h_\ddag}\right),\quad \overline{\Gamma}_{it}^{\sf r}=\frac{1}{Th_\ddag}\sum_{s=1}^T \widetilde{X}_{i,s-1}x_{i,s}K^\ddag\left(\frac{\tau_s-\tau_t}{h_\ddag}\right),\notag
\end{eqnarray} 
where $h_\ddag$ is the bandwidth and $\widetilde{X}_{i,s-1}$ is defined as in Section \ref{sec3.1}. We estimate $\beta_{i\bullet}^{\ddag,{\sf l}}(\tau_t)$ and $\beta_{i\bullet}^{\ddag,{\sf r}}(\tau_t)$ by 
\begin{equation}\label{eq5.5}
\widehat\beta_{i\bullet}^{\ddag,{\sf l}}(\tau_t)=\left(\widetilde{\boldsymbol\Gamma}_{it}^{{\sf l}}\right)^{-1}\overline{\Gamma}_{it}^{\sf l}\quad {\rm and}\quad \widehat\beta_{i\bullet}^{\ddag,{\sf r}}(\tau_t)=\left(\widetilde{\boldsymbol\Gamma}_{it}^{\sf r}\right)^{-1}\overline{\Gamma}_{it}^{\sf r},
\end{equation}
respectively, and then construct 
\begin{equation}\label{eq5.6}
\widehat\delta_\beta(t)=\max_{1\leq i\leq N} \left\vert \widehat\beta_{i\bullet}^{\ddag,{\sf r}}(\tau_t)- \widehat\beta_{i\bullet}^{\ddag,{\sf l}}(\tau_t)\right\vert_2.
\end{equation}
The estimation of $t_0$ is defined as
\begin{equation}\label{eq5.7}
\widehat{t}=\argmax_{t}\widehat\delta_\beta(t).
\end{equation}

\medskip

{\bf Stage 2}:\ \ Let
\[
{\mathscr T}_1=\{1,2,\cdots,\widehat{t}-\lfloor\epsilon_0T\rfloor\}\quad {\rm and}\quad {\mathscr T}_1=\{\widehat{t}+\lfloor\epsilon_0T\rfloor,\cdots,T-1,T\},
\]
where $\epsilon_0$ is an arbitrarily small positive number (say $0.01$). From Theorem \ref{thm:5.1}(i), the group membership and number are time invariant {\em w.p.a.1} over the two time periods ${\mathscr T}_1$ and ${\mathscr T}_2$. Hence, we may adopt the clustering algorithm and ratio criterion developed in Section \ref{sec3} to estimate ${\mathscr G}_k^1$ and $K_1$ (using the network time series sample over ${\mathscr T}_1$) as well as ${\mathscr G}_k^2$ and $K_2$ (using the sample over ${\mathscr T}_2$). We denote the resulting estimates as $\widehat{\mathscr G}_k^1$, $\widehat K_1$, $\widehat{\mathscr G}_k^2$ and $\widehat K_2$, whose consistency property is derived in Theorem \ref{thm:5.1}(ii) below.

\smallskip

The following conditions are required to derive the asymptotic property of the above two-stage estimation method.

\begin{assumption}\label{ass:6}

(i)\ For $1\leq g,g^\ast\leq K_1$, $\alpha_{gg^\ast}^1(\cdot)$ and $\alpha_{g}^1(\cdot)$ have bounded first-order derivatives and satisfy
\[
\max_{1\leq g,g^\ast\leq K_1}\sup_{0\leq \tau\leq 1}\left\vert\alpha_{gg^\ast}^1(\tau)\right\vert+\max_{1\leq g\leq K_1}\sup_{0\leq \tau\leq 1}\left\vert\alpha_{g}^1(\tau)\right\vert<1.
\]
The same conditions hold for $\alpha_{gg^\ast}^2(\cdot)$ and $\alpha_{g}^2(\cdot)$, $1\leq g,g^\ast\leq K_2$.

(ii)\ $K^\ddag(\cdot)$ is positive and Lipschitz-continuous with a compact support $[0,1]$.

 (iii)\ There exist positive constants $c_1\in(0,1)$ and $c_2$ such that $t_0=c_1T$ and $\delta_\beta(t_0)>c_2$.

(iv)\ There exist positive constant $c_3$ and $c_4$ such that 
\[\min_{1\le k\le K_1}{\rm card}({\mathscr G}_k^1)\ge c_3\cdot N\quad {\rm and}\quad \min_{1\le k\le K_2}{\rm card}({\mathscr G}_k^2)\ge c_4\cdot N.\]

\end{assumption}

\begin{remark}\label{re:7}

Assumption \ref{ass:6}(i)(ii) contains some typical conditions on the time-varying coefficient functions and kernel function which are often required when the one-sided kernel smoothing is adopted. In particular, Assumption \ref{ass:6}(i) ensures that the grouped time-varying network VAR process is locally stable over the two time periods separated by the break point. Assumption \ref{ass:6}(iii) indicates that the break location is well separated from the endpoints and the break size is bounded away from zero. The structural break may be due to abrupt changes in the group-specific time-varying coefficients functions or breaks in the group membership or number. More discussion and examples are available in Appendix E of the online supplement \citep{LPTW24}. In fact, by slightly modifying the proof and theory, we may allow $\delta_\beta(t_0)$ to slowly approach zero as in \cite{CWW22}. Assumption \ref{ass:6}(iv) is a natural extension of Assumption \ref{ass:4}(i) to the time-varying group structure.
 
\end{remark}

Let $\beta_{i\bullet}^{\circ1}(\cdot)$ be defined similarly to $\beta_{i\bullet}^\circ(\cdot)$ in (\ref{eq3.6}) but with $\beta_{ij}(\cdot)$ replaced by $\beta_{ij}^1(\cdot)$. As $\beta_m^\circ(\cdot)$ defined in Stage 2 of Section \ref{sec3.1}, we let $\beta_m^{\circ1}(\cdot)$, $m=1,\cdots,M_1$, denote the distinct coefficient functions for time-varying network effects before the break, where $M_1\leq K_1^2$. The definitions of $\beta_{i\bullet}^{\circ2}(\cdot)$ and $\beta_m^{\circ2}(\cdot)$, $m=1,\cdots,M_2$, are analogous. Define
\begin{eqnarray}
\zeta_{NT}^{\dag1}&=&\min_{1\leq g_i^1\neq g_j^1\leq K_1}\int_0^{c_1}\left[\left\vert\alpha_{g_i^1}^1(\tau)-\alpha_{g_j^1}^1(\tau)\right\vert+\left\vert \beta_{i\bullet}^{\circ1}(\tau)- \beta_{j\bullet}^{\circ1}(\tau)\right\vert_2\right]d\tau,\notag\\
\zeta_{NT}^{\dag2}&=&\min_{1\leq g_i^2\neq g_j^2\leq K_2}\int_{c_1}^1\left[\left\vert\alpha_{g_i^2}^2(\tau)-\alpha_{g_j^2}^2(\tau)\right\vert+\left\vert \beta_{i\bullet}^{\circ2}(\tau)- \beta_{j\bullet}^{\circ2}(\tau)\right\vert_2\right]d\tau,\notag\\
\zeta_{NT}^{\ddag1}&=&\min_{1\leq m\neq m^\ast\leq M_1}\int_0^{c_1}\left\vert \beta_{m}^{\circ1}(\tau)-\beta_{m^\ast}^{\circ1}(\tau)\right\vert d\tau,\notag\\
\zeta_{NT}^{\ddag2}&=&\min_{1\leq m\neq m^\ast\leq M_2}\int_{c_1}^1\left\vert \beta_{m}^{\circ2}(\tau)-\beta_{m^\ast}^{\circ2}(\tau)\right\vert d\tau.\notag
\end{eqnarray}
Let ${\mathscr N}_i^1(j)=\{k\in{\mathscr N}_i: g_k^1=j\}$ and ${\mathscr N}_i^2(j)=\{k\in{\mathscr N}_i: g_k^2=j\}$.

\renewcommand{\thetheorem}{5.\arabic{theorem}}\setcounter{theorem}{0}

\begin{theorem}\label{thm:5.1}

Suppose that Assumptions \ref{ass:1}(ii), \ref{ass:2}(ii) and \ref{ass:6}(i)--(iii) are satisfied. 

(i) The break location estimate $\widehat{t}$ has the following approximation order: 
\begin{equation}\label{eq5.8}
\left\vert \frac{\widehat{t}-t_0}{T}\right\vert= O_P\left(\sqrt{\frac{\bar n h_\ddag\log (N\vee T)}{T}}+h_\ddag^2\right).
\end{equation}

(ii) If, in addition, Assumption \ref{ass:6}(iv) is satisfied and (\ref{eq3.8}), (\ref{eq3.9}) and (\ref{eq3.14}) continue to hold when $h$, ${\rm card}({\mathscr N}_i(j))$ and $\zeta_{NT}^\dag \wedge \zeta_{NT}^\ddag$ and replaced by $h_\ddag$, ${\rm card}({\mathscr N}_i^1(j)) \wedge{\rm card}({\mathscr N}_i^2(j))$ and $\min\{\zeta_{NT}^{\dag1}, \zeta_{NT}^{\dag2}, \zeta_{NT}^{\ddag1}, \zeta_{NT}^{\ddag2}\}$, respectively, we have
\begin{eqnarray}
&&{\sf P}\left(\big\{\widehat{\mathscr G}_k^1,\ 1\leq k\leq \widehat{K}_1\big\}=\big\{{\mathscr G}_k^1,\ 1\leq k\leq K_1\big\}\right)\rightarrow1,\notag\\
&&{\sf P}\left(\big\{\widehat{\mathscr G}_k^2,\ 1\leq k\leq \widehat{K}_2\big\}=\big\{{\mathscr G}_k^2,\ 1\leq k\leq K_2\big\}\right)\rightarrow1,\notag\\
&&{\sf P}\left(\widehat K_1=K_1\right)\rightarrow1,\quad {\sf P}\left(\widehat K_2=K_2\right)\rightarrow1,\notag
\end{eqnarray}
as $N,T\rightarrow\infty$ jointly.

\end{theorem}

\begin{remark}\label{re:8}

(i) Theorem \ref{thm:5.1}(i) shows that the scaled break point estimation is consistent. Although the convergence rate in (\ref{eq5.8}) is conservative, it is sufficient to consistently estimate the time-varying group membership and number in Stage 2. The approximation rate may be improved if we replace the one-sided kernel by one-sided local linear smoothing \citep[e.g.,][]{CWW22} in Stage 1. 

\smallskip

(ii) In practice, there are often multiple breaks in the latent group structure, i.e., breaks occur at some unknown but well separated break points $t_1,t_2,\cdots,t_{p}$. In this setting, the methodology and theory continue to hold with minor amendments. For example, we may use the recursive algorithm in \cite{XCW20} to locate the $p$ break points, an idea similar to the binary segmentation commonly used to estimate multiple breaks in parametric models \citep[e.g.,][]{CF12, CF15}.

\smallskip

(iii) In Appendix E of the online supplement \citep{LPTW24}, we discuss a refined break point estimation, making use of the consistently estimated group structures. Under some high-level conditions, we show that the refined estimation of the break location is consistent. Furthermore, we provide a few examples to verify the high-level conditions. 

\end{remark}

%%%%%%%%%%%%%%%%%%%

\section{Numerical studies}\label{sec6} 
\renewcommand{\theequation}{6.\arabic{equation}} \setcounter{equation}{0}

In this section, we conduct both the simulation and empirical studies. Sections \ref{sec6.1} and \ref{sec6.2} assess the finite-sample performance of the developed methodology and verify the main convergence properties via simulation, whereas Section \ref{sec6.3} reports the empirical application to a network time series data set for UK temperature. 

\subsection{Simulation study without break}\label{sec6.1} 

We use the grouped network time-varying VAR model \eqref{eq2.1} for data generation. The entries of the adjacency matrix ${\mathbf W}$ are defined by $w_{ij} = I(u_{ij}\le  \bar{w})$, where $u_{ij}\sim {\sf U}(0,1)$ and $0<\bar{w}<1$, controlling the sparsity level of the network structure. The innovation vectors are generated by $\varepsilon_t\sim {\sf N}({\bf 0}, {\boldsymbol\Sigma}_\varepsilon)$ independently over $t$, where ${\boldsymbol\Sigma}_\varepsilon=\{\sigma_{ij}\}_{N\times N}$ with $\sigma_{ij}=0.1^{|i-j|}$, allowing cross-sectional dependence over components. Consider $K_0=2$ and define the group-specific coefficients as 
\begin{eqnarray*}
\alpha_{g_i} (\tau) &=& \left\{\begin{array}{ll}
0.49 \cos(\pi \tau),\quad & i \in \mathscr{G}_1, \\
-0.2, \quad  & i \in \mathscr{G}_2, 
\end{array} \right. \quad\quad
\alpha_{g_ig_j} (\tau) = \left\{\begin{array}{ll}
0.5-\tau,\quad & \text{ if } i,j\in\mathscr{G}_1, \\
\tau^3-0.5, & \text{otherwise}. \\
\end{array} \right.
\end{eqnarray*}
The group membership is generated as follows: assign each node $i\in \{1,\ldots, N\}$ to $\mathscr{G}_1$ and $\mathscr{G}_2$ with respective probabilities $0.65$ and $0.35$. Note that there exists a further group structure on the time-varying spillover effects $\alpha_{g_ig_j}(\cdot)$ with $M_0=2$, as described in Stage 2 of Section \ref{sec3.1}. In the simulation, we consider two scenarios for generating the group membership: (i) fixed group, i.e., the group membership is only generated once and remains the same over replications; and (ii) random group, i.e., the group membership is randomly generated for each replication. We conduct the simulation over $R=1000$ replications and set $N=100,200$, $T=300,600$, and $\bar{w} = 0.025,0.075$.

\smallskip

We use the Epanechnikov kernel in the local linear smoothing, where the bandwidth is determined by the rule of thumb (\citealp{SW17}): $h=(2.35/\sqrt{12})T^{-1/5}$ for the fully heterogenous local linear estimation (\ref{eq3.2}) and $h_\dag=(2.35/\sqrt{12})[\text{card}(\widehat{\mathscr{G}}_j)T]^{-1/5}$ for the post-grouping estimation (\ref{eq4.1}), where $\widehat{\mathscr{G}}_j$ denotes the estimated group. We notice that the clustering results are insensitive to the bandwidth choice in our simulation. For each simulated data set, we first estimate the group membership and number as in Section \ref{sec3}, and then conduct the post-grouping estimation as in Section \ref{sec4}. To evaluate the group structure estimation accuracy, we adopt the following two measurements:
\begin{eqnarray*}
{\sf AC}(K_0) &=& \frac{1}{R}\sum_{r=1}^R I(\widehat{K}_r=K_0) \quad  \text{and}\quad
{\sf Purity}({\mathscr G}) = \frac{1}{RN}\sum_{r=1}^R \sum_{k=1}^{\widehat{K}_r} \max_{1\le j\le K_0} \left|\widehat{\mathscr{G}}_{k, r} \cap \mathscr{G}_j \right|, 
\end{eqnarray*}
where $\widehat{K}_r$ and $\widehat{\mathscr{G}}_{k, r}$ denote the estimates of the group number and membership in the $r$-th replication. The purity quantity ${\sf Purity}({\mathscr G})$ is a simple and transparent evaluation measure with value close to one when the clustering method is precise. To compare the estimation performance between the pre-grouping local linear estimation and the post-grouping one, we compute the root mean squared errors for the estimated momentum and spillover effects: 
\begin{eqnarray*}
\sf{RMSE}_{M,r}&=& \left\{\frac{1}{NS}\sum_{i=1}^N\sum_{s=1}^S \left|\widetilde{\beta}_{r,i}(\tau_s)- \beta_{i} (\tau_s)\right|^2 \right\}^{1/2},\notag\\
\sf{RMSE}_{S,r}&=& \left\{\frac{1}{NS}\sum_{i=1}^N\sum_{s=1}^S \left|\widetilde{\beta}_{r,i\bullet}^{*}(\tau_s)- \beta_{i\bullet}^\ast(\tau_s)\right|_2^2  \right\}^{1/2},
\end{eqnarray*}
where $ \beta_{i\bullet}^\ast(\cdot)=\left(\beta_{ij}(\cdot):\ j\in{\mathscr N}_i\right)^{^\intercal}$, $\widetilde{\beta}_{r,i}(\cdot)$ and $\widetilde{\beta}_{r,i\bullet}^{*}(\cdot)$ stand for the estimates in the $r$-th replication, $\tau_s = 0.05, 0.1,\cdots, 0.95$ and $S=19$.

\smallskip

The simulation results are summarized in Tables \ref{t1} and \ref{t2}. The numbers in parentheses of Table \ref{t2} are standard deviations of $\sf{RMSE}_{M,r}$ and $\sf{RMSE}_{S,r}$ over 1000 replications. It follows from Table \ref{t1} that both ${\sf AC}(K_0)$ and ${\sf Purity}({\mathscr G})$ converge to one as the time series length $T$ increases from $300$ to $600$, and the results remain stable when the sparsity level $\bar w$ changes from $0.025$ to $0.075$. Table \ref{t2} shows that the post-grouping time-varying coefficient estimation substantially outperforms the pre-grouping one, confirming that the estimation accuracy is significantly improved by making use of the estimated group structure. The standard deviations are generally small, indicating that the nonparametric estimation performance is stable over replications. In addition, both the pre-grouping and post-grouping local linear estimates deteriorate when $\bar w$ increases from $0.025$ to $0.075$.

\begin{table}[tbp]\centering
\setlength{\tabcolsep}{4pt}
\caption{Estimation performance of the group number and membership}\label{t1}
\begin{tabular}{lllrrlrr}
\hline \hline
 &  &  & \multicolumn{2}{c}{Fixed group} &  & \multicolumn{2}{c}{Random group} \\
 \hline
Sparsity & Measurement  & $T\setminus N$ & 100 & 200 &  & 100 & 200 \\
 \hline
$\bar{w}=0.025$ & ${\sf AC}(K_0)$ & 300 & 1.000 & 0.995 &  & 0.999 & 0.997 \\
 &  & 600 & 1.000 & 1.000 &  & 1.000 & 1.000 \\
 & ${\sf Purity}({\mathscr G})$ & 300 & 0.997 & 0.996 &  & 0.996 & 0.995 \\
 &  & 600 & 1.000 & 1.000 &  & 1.000 & 1.000 \\
 &  &  &  &  &  &  &  \\
$\bar{w}=0.075$ & ${\sf AC}(K_0)$ & 300 & 0.999 & 0.989 &  & 0.989 & 0.993 \\
 &  & 600 & 1.000 & 1.000 &  & 1.000 & 1.000 \\
 & ${\sf Purity}({\mathscr G})$ & 300 & 0.988 & 0.958 &  & 0.985 & 0.955 \\
 &  & 600 & 1.000 & 1.000 &  & 1.000 & 0.999 \\
 \hline\hline
\end{tabular}
\end{table}
 
\begin{table}[tbp]\centering\footnotesize
\setlength{\tabcolsep}{4pt}
\caption{Estimation performance of the time-varying momentum and network spillover effects}\label{t2}
\begin{tabular}{lllcccclcccc}
\hline\hline
 &  &  & \multicolumn{4}{c}{Fixed group } &  & \multicolumn{4}{c}{Random group} \\
 \hline
 &  &  & \multicolumn{2}{c}{${\sf RMSE}_{\sf M}$} & \multicolumn{2}{c}{${\sf RMSE}_{\sf S}$} &  & \multicolumn{2}{c}{${\sf RMSE}_{\sf M}$} & \multicolumn{2}{c}{${\sf RMSE}_{\sf S}$} \\
 \hline
 Sparsity & Estimation  & $T\setminus N$ & 100 & 200 & 100 & 200 &  & 100 & 200 & 100 & 200 \\
 \hline
$\bar{w}=0.025$ & {\sf Pre-grouping} & 300 & 0.148 &  \multicolumn{1}{r|}{0.316} & 0.330 & 0.768  & & 0.189 &  \multicolumn{1}{r|}{0.312} & 0.419 & 0.745 \\
 &  &  & (0.023) & \multicolumn{1}{r|}{(0.015)} & (0.011) & (0.012) &  & (0.027) & \multicolumn{1}{r|}{(0.019)}  & (0.038) & (0.037) \\
 &  & 600 & 0.078 & \multicolumn{1}{r|}{0.211} & 0.234 & 0.493 &  & 0.116 & \multicolumn{1}{r|}{0.223} & 0.295 & 0.515 \\
 &  &  & (0.023) & \multicolumn{1}{r|}{(0.014)} & (0.007) & (0.006) &  & (0.026) & \multicolumn{1}{r|}{(0.016)} & (0.025) & (0.025)  \\
 & {\sf Post-grouping} & 300 & 0.035 & \multicolumn{1}{r|}{0.048}  & 0.107 & 0.118 &  & 0.036 & \multicolumn{1}{r|}{0.047} & 0.108 & 0.117  \\
 &  &  & (0.010) & \multicolumn{1}{r|}{(0.007)} & (0.004) & (0.003)  &  & (0.011) & \multicolumn{1}{r|}{(0.008)} & (0.004) & (0.003) \\
 &  & 600 & 0.021 & \multicolumn{1}{r|}{0.024} & 0.077 & 0.081  &  & 0.021 & \multicolumn{1}{r|}{0.026} & 0.078 & 0.082 \\
 &  &  & (0.004) & \multicolumn{1}{r|}{(0.003)}  & (0.002) & (0.002) &  & (0.004) & \multicolumn{1}{r|}{(0.003)} & (0.002) & (0.002) \\
$\bar{w}=0.075$ & {\sf Pre-grouping} & 300 & 0.399 & \multicolumn{1}{r|}{0.751} & 0.967 & 2.268 &  & 0.424 & \multicolumn{1}{r|}{0.768}  & 1.075 & 2.333  \\
 &  &  & (0.023) & \multicolumn{1}{r|}{(0.129)} & (0.015) & (0.020) &  & (0.034) & \multicolumn{1}{r|}{(0.146)} & (0.063) & (0.083)  \\
 &  & 600 & 0.289 & \multicolumn{1}{r|}{0.505} & 0.660 & 1.450  &  & 0.306 & \multicolumn{1}{r|}{0.509} & 0.729 & 1.457 \\
 &  &  & (0.014) & \multicolumn{1}{r|}{(0.048)} & (0.009) & (0.010) &  & (0.019) & \multicolumn{1}{r|}{(0.047)} & (0.038) & (0.042) \\
 & {\sf Post-grouping} & 300 & 0.062 & \multicolumn{1}{r|}{0.102} & 0.125 & 0.156 &  & 0.063 & \multicolumn{1}{r|}{0.104}  & 0.126 & 0.157 \\
 &  &  & (0.013) & \multicolumn{1}{r|}{(0.019)} & (0.004) & (0.003) &  & (0.015) & \multicolumn{1}{r|}{(0.020)} & (0.005) & (0.004)  \\
 &  & 600 & 0.032 & \multicolumn{1}{r|}{0.049} & 0.086 & 0.099 &  & 0.033 & \multicolumn{1}{r|}{0.050} & 0.086 & 0.100 \\
 &  &  & (0.003) & \multicolumn{1}{r|}{(0.004)} & (0.003) & (0.002) &  & (0.004) & \multicolumn{1}{r|}{(0.005)} & (0.003) & (0.003) \\
 \hline \hline
\end{tabular}
\end{table}

\subsection{Simulation study with a break in the group structure}\label{sec6.2}

We next examine the numerical performance of the estimation method with a break in the group structure introduced in Section \ref{sec5}. The break point is set at $t_0=\lfloor T/2\rfloor +1$ when we re-assign each node $i$ to $\mathscr{G}_1$ and $\mathscr{G}_2$ with respective probabilities $0.65$ and $0.35$. This results in a break in the group membership. Before the break time, the data generating process is the same as that in Section \ref{sec6.1}, whereas, after the break, the group-specific time-varying coefficients are defined as 
\[
\alpha_{g_i} (\tau) = \left\{\begin{array}{ll}
-0.49\sin(\pi \tau/2),\quad & i \in \mathscr{G}_1, \\
0.49\sin(\pi \tau/2),\quad & i \in \mathscr{G}_2, \\
\end{array} \right. \quad\quad
\alpha_{g_ig_j} (\tau) = \left\{\begin{array}{ll}
0.49\sin(\pi\tau/2),\quad & \text{ if } i,j \in \mathscr{G}_1,\\
-0.49\sin(\pi\tau/2),\quad & \text{otherwise}.  \\
\end{array} \right.
\]
As in Section \ref{sec6.1}, we consider both the fixed and random groups when generating the group membership (with a break) over $R=1000$ replications. In order to obtain stable finite-sample performance, we slightly increase $T$ from $(300,600)$ to $(400,800)$. The number of nodes remains as $N=100$ and $200$. 

\smallskip

The one-sided version of the Epanechnikov kernel function is adopted in our nonparametric method. We first estimate the break point via (\ref{eq5.7}), compute the (scaled) measurement $\Delta_{t_0} = (\widehat{t}-t_0)/T$, and then report the means and standard deviations (in parentheses) of $\Delta_{t_0}$ in Table \ref{t3}. It is clear that the scaled break point $t_0/T$ can be accurately detected, and its estimation accuracy is not sensitive to the sparsity level of the adjacency matrix ${\mathbf{W}}$. With the estimated break point, we may split the entire time period into the "pre-break" and "post-break" periods and compute their respective ${\sf AC}(K_0)$ and ${\sf Purity}({\mathscr G})$. We further take a simple average of those values over the two periods and report them in Table \ref{appt1}. We note that both the group number and membership are estimated very accurately even when there exists a break in the group structure. We finally compare the estimation performance between the pre-grouping and post-grouping local linear estimation after the break point is detected. When computing ${\sf RMSE}_{\sf M}$ and ${\sf RMSE}_{\sf S}$, we choose $\tau_s = 0.05, 0.1,\ldots, 0.4$ for the ``pre-break" period and $\tau_s = 0.6, 0.65,\ldots, 0.95$ for the ``post-break" period, avoiding possible boundary effect in the estimation. The general pattern in Table \ref{appt2} is very similar to that in Table \ref{t2}, again confirming the significant advantage of the post-grouping estimation.

\begin{table}[tbp]\centering
\caption{Measurements of the (scaled) break point estimation}\label{t3}
\begin{tabular}{llcclcc}
\hline\hline
 &  & \multicolumn{2}{c}{Fixed group} &  & \multicolumn{2}{c}{Random group} \\
 \hline
Sparsity & $T\setminus N$ & 100 & 200 &  & 100 & 200 \\
\hline
$\bar{w}=0.025$ & 400 & -0.002 (0.031) & -0.003  (0.033) &  & -0.001 (0.032) & -0.003 (0.033) \\
 & 800 & -0.001 (0.016) & -0.002 (0.017)  &  & -0.002 (0.016) & 0.000 (0.017) \\
$\bar{w}=0.075$ & 400 & -0.003 (0.033) & -0.002 (0.034)  &  & -0.003 (0.034) & -0.002 (0.034) \\
 & 800 & -0.001 (0.017)  & 0.000 (0.017) &  & 0.000 (0.017) & -0.001 (0.017) \\
 \hline\hline
\end{tabular}
\end{table}

\begin{table}[tbp]\centering
\setlength{\tabcolsep}{4pt}
\caption{Estimation performance of the group number and membership (with a break)}\label{appt1}
\begin{tabular}{llccclcc}
\hline\hline
 &  &  & \multicolumn{2}{c}{Fixed group} &  & \multicolumn{2}{c}{Random group} \\
 \hline
 Sparsity & Measurement  & $T\setminus N$ & 100 & 200 &  & 100 & 200 \\
 \hline
$\bar{w}=0.025$ & ${\sf AC}(K_0)$ & 400 & 1.000 & 1.000 &  & 1.000 & 1.000 \\
 &  & 800 & 1.000 & 1.000 &  & 1.000 & 1.000 \\
 &${\sf Purity}({\mathscr G})$& 400 & 0.999 & 0.999 &  & 0.999 & 0.998 \\
 &  & 800 & 1.000 & 1.000 &  & 1.000 & 1.000 \\
 &  &  &  &  &  &  &  \\
$\bar{w}=0.075$ & ${\sf AC}(K_0)$ & 400 & 0.994 & 0.966 &  & 0.993 & 0.968 \\
 &  & 800 & 1.000 & 1.000 &  & 1.000 & 1.000 \\
 & ${\sf Purity}({\mathscr G})$ & 400 & 0.996 & 0.982 &  & 0.995 & 0.983 \\
 &  & 800 & 1.000 & 1.000 &  & 1.000 & 1.000 \\
 \hline\hline
\end{tabular}
\end{table}
 
\begin{table}[tbp]\centering\footnotesize
\setlength{\tabcolsep}{4pt}
\caption{Estimation performance of the time-varying coefficients with a break in the group structure}\label{appt2}
\begin{tabular}{lllcccclcccc}
\hline\hline
 &  &  & \multicolumn{4}{c}{Fixed group } &  & \multicolumn{4}{c}{Random group} \\
 \hline
 &  &  & \multicolumn{2}{c}{${\sf RMSE}_{\sf M}$} & \multicolumn{2}{c}{${\sf RMSE}_{\sf S}$} &  & \multicolumn{2}{c}{${\sf RMSE}_{\sf M}$} & \multicolumn{2}{c}{${\sf RMSE}_{\sf S}$} \\
 \hline
 Sparsity & Estimation  & $T\setminus N$ & 100 & 200 & 100 & 200 &  & 100 & 200 & 100 & 200 \\
 \hline
$\bar{w}=0.025$ & {\sf Pre-grouping} & 400 &  0.490  &  \multicolumn{1}{r|}{0.562} & 0.553 & 0.785  & & 0.477 &  \multicolumn{1}{r|}{0.545} & 0.569 & 0.765 \\
 &  &  & (0.006) & \multicolumn{1}{r|}{(0.010)} & (0.007) & (0.008) &  &  (0.020) & \multicolumn{1}{r|}{(0.016)}  & (0.027) & (0.028)  \\
 &  & 800 & 0.485 & \multicolumn{1}{r|}{0.490} & 0.521 & 0.622  &  & 0.461 & \multicolumn{1}{r|}{0.494} & 0.515 & 0.627 \\
 &  &  & (0.004) & \multicolumn{1}{r|}{(0.005)} & (0.005) & (0.006)  &  & (0.020) & \multicolumn{1}{r|}{(0.016)} & (0.021) & (0.020)   \\
 & {\sf Post-grouping} & 400 & 0.032 & \multicolumn{1}{r|}{0.040}  & 0.093 & 0.101 &  & 0.034 & \multicolumn{1}{r|}{0.040} & 0.095 & 0.101  \\
 &  &  & (0.006) & \multicolumn{1}{r|}{(0.006)} & (0.003) & (0.002)  &  & (0.009) & \multicolumn{1}{r|}{(0.007)} & (0.003) & (0.002) \\
 &  & 800 & 0.020 & \multicolumn{1}{r|}{0.023} & 0.067 & 0.071   &  & 0.021 & \multicolumn{1}{r|}{0.024} & 0.068 & 0.072\\
 &  &  & (0.003) & \multicolumn{1}{r|}{(0.002)}  & (0.002) & (0.002) &  & (0.004) & \multicolumn{1}{r|}{(0.003)} & (0.002) & (0.002) \\
$\bar{w}=0.075$ & {\sf Pre-grouping} & 400 &  0.613 & \multicolumn{1}{r|}{1.023} & 0.930 & 1.890 &  & 0.636 & \multicolumn{1}{r|}{1.035}  & 0.998 & 1.910   \\
 &  &  & (0.026) & \multicolumn{1}{r|}{(0.070)} & (0.012) & (0.015) &  & (0.027) & \multicolumn{1}{r|}{(0.066)} & (0.046) & (0.058)   \\
 &  & 800 & 0.535 & \multicolumn{1}{r|}{0.730} & 0.731 & 1.260 &  & 0.538  & \multicolumn{1}{r|}{0.743} & 0.759 & 1.291 \\
 &  &  & (0.008) & \multicolumn{1}{r|}{(0.021)} & (0.008) & (0.008) &  & (0.017) & \multicolumn{1}{r|}{(0.027)} & (0.029) & (0.033) \\
 & {\sf Post-grouping} & 400 & 0.046 & \multicolumn{1}{r|}{0.075} & 0.106 & 0.126 &  & 0.050 & \multicolumn{1}{r|}{0.075}  & 0.108 & 0.126  \\
 &  &  & (0.010) & \multicolumn{1}{r|}{(0.016)} & (0.004) & (0.003) &  & (0.011) & \multicolumn{1}{r|}{(0.016)} & (0.004) & (0.004)  \\
 &  & 800 & 0.026 & \multicolumn{1}{r|}{0.038} & 0.074 & 0.083 &  & 0.028 & \multicolumn{1}{r|}{0.038} & 0.074 & 0.083  \\
 &  &  & (0.003) & \multicolumn{1}{r|}{(0.003)} & (0.002) & (0.002) &  & (0.004) & \multicolumn{1}{r|}{(0.005)} & (0.002) & (0.002) \\
 \hline \hline
\end{tabular}
\end{table}

\smallskip

Appendix F in the online supplement \citep{LPTW24} contains extra simulation results: the finite-sample performance of the clustering algorithm introduced in Appendix \ref{appA} on estimating the homogeneity structure for the network spillover effects, and the clustering result by using \cite{ZXF23}'s grouped network VAR model with constant coefficients.

%%%%%%%%%%%%%%%%%%%%

\subsection{An Empirical Study}\label{sec6.3}

\begin{figure}[tbp]\centering
\hspace*{-1cm}\includegraphics[scale=0.25]{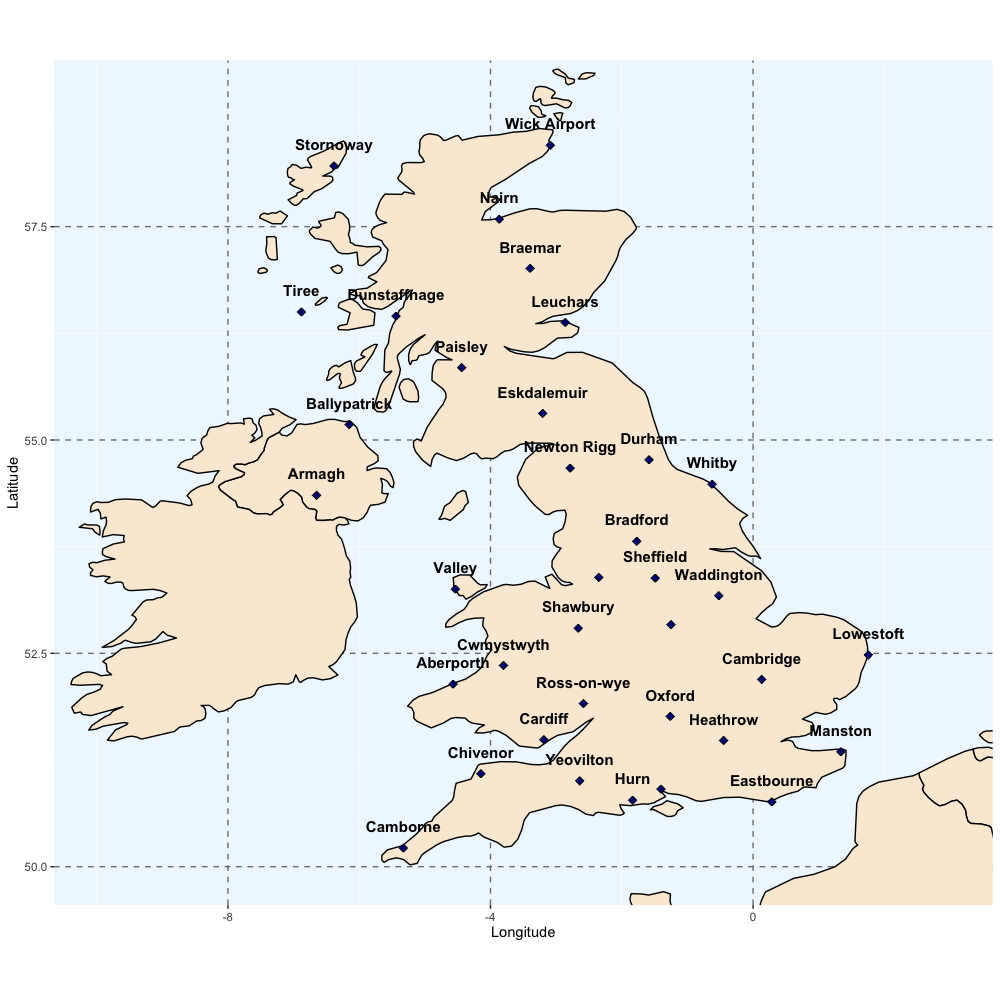}
\caption{\small Geographical locations of the UK weather stations}\label{figstat}
\end{figure}

There has been increasing interest in investigating the spatial pattern of climate data, see, for example, \cite{PSH2009}, \cite{Kendon20}, \cite{Hanlon21} and the references therein. We next apply the proposed model and methodology to analyze a set of UK climate data, exploring the network and latent group structures and allowing for smooth structural changes to account for possible climate changes in the past few decades. The data that we use are collected from the UK Meteorological Office,\footnote{https://www.metoffice.gov.uk/research/climate/maps-and-data/historic-station-data} containing temperature recordings (in Celsius) of 37 weather stations with their geographical locations presented in Figure \ref{figstat}. The original dataset collects minimum and maximum temperatures per month over the period from January 1950 to February 2023. Hence, the time series length is $T=878$. We consider the following two scenarios when building the network VAR model: (i) both the minimum and maximum temperatures are used as elements of $X_t$; (ii) the averaged temperature per month is used as elements of $X_t$. In model (ii), each weather station is treated as a node and $N=37$; whereas in model (i), the minimum and maximum recordings in each weather station are treated as two nodes and $N=74$. The time series observations are standardized to have zero mean and unit standard deviation. The adjacency matrix ${\mathbf W}$ is constructed by following the UK climate region map\footnote{https://www.metoffice.gov.uk/research/climate/maps-and-data/about/regions-map}. Specifically, it is classified into the following five regions: Southern England,  Northern England, Wales, Scotland, and Northern Ireland. When stations $i$ and $j$ are in the same region, we set $w_{ij}=1$, otherwise, $w_{ij}=0$. Consequently, the percentage of non-zero elements of the adjacency matrix is 0.0559 and 0.2267 for the two models. 

\smallskip

The primary interest of this empirical study lies in identifying potential group structure over the weather stations to achieve dimension reduction in the subsequent network VAR model estimation. With the clustering algorithm in Section \ref{sec3}, we obtain two estimated groups, i.e., $\widehat{K}=2$, for both models. Table \ref{table.ep1} reports the estimated group membership which is very similar between the two models. For model (i), all the stations of Group 1 are in Northern Ireland and Wales, while those of Group 2 are in England and Scotland. This indicates that Northern Ireland and Wales often have common patterns in terms of temperature change whereas the weather stations in the island of Great Britain usually have similar temperature recordings. It is noteworthy that, although each weather station in model (i) contains the minimum and maximum temperatures (as two nodes), both nodes are classified into the same group (thus we only report the station name in Table \ref{table.ep1}). For model (ii), three weather stations in the coastal towns (Eastbourne, Tiree and Whitby) move from Group 2 to Group 1. The estimated time-varying coefficient functions are plotted in the online supplement \citep{LPTW24}.

\begin{table}[tbp]\small
\caption{The estimated group membership using the UK temperature time series}\label{table.ep1}
\begin{tabular}{lllllll}
\hline\hline
 Model & Group 1 &  & \multicolumn{4}{c}{Group 2} \\
  \hline\\
Model (i) & \multicolumn{1}{l|}{Aberporth} &  & Bradford & Eskdalemuir & Newton Rigg & Stornoway \\
 & \multicolumn{1}{l|}{Armagh} &  & Braemar & Heathrow & Oxford & Sutton Bonington \\
 & \multicolumn{1}{l|}{Ballypatrick} &  & Camborne & Hurn & Paisley & Tiree \\
 & \multicolumn{1}{l|}{Cardiff} &  & Cambridge & Lerwick & Ringway & Waddington \\
 & \multicolumn{1}{l|}{Cwmystwyth} &  & Chivenor & Leuchars & Ross-on-wye & Whitby \\
 & \multicolumn{1}{l|}{Valley} &  & Dunstaffnage & Lowestoft & Shawbury & Wick Airport \\
 & \multicolumn{1}{l|}{} &  & Durham & Manston & Sheffield & Yeovilton \\
 & \multicolumn{1}{l|}{} &  & Eastbourne & Nairn & \multicolumn{2}{l}{Southampton} \\
 & \multicolumn{1}{l|}{} &  &  &  &  &  \\
Model (ii) & \multicolumn{1}{l|}{Aberporth} &  & Bradford & Eskdalemuir & Nairn & Sheffield \\
 & \multicolumn{1}{l|}{Armagh} &  & Braemar & Heathrow & Newton Rigg & Southampton \\
 & \multicolumn{1}{l|}{Ballypatrick} &  & Camborne & Hurn & Oxford & Stornoway \\
 & \multicolumn{1}{l|}{Cardiff} &  & Cambridge & Lerwick & Paisley & Sutton Bonington \\
 & \multicolumn{1}{l|}{Cwmystwyth} &  & Chivenor & Leuchars & Ringway & Waddington \\
 & \multicolumn{1}{l|}{Valley} &  & Dunstaffnage & Lowestoft & Ross-on-wye & Wick Airport \\
 & \multicolumn{1}{l|}{Eastbourne} &  & Durham & Manston & Shawbury & Yeovilton \\
 & \multicolumn{1}{l|}{Tiree} &  &  &  &  &  \\
 & \multicolumn{1}{l|}{Whitby} &  &  &  &  & \\
 \hline\hline
\end{tabular}
\end{table}

\smallskip

To further show the necessity of accounting for the group structure and smooth structural changes, we compare the out-of-sample prediction performance among the following three methods: fully heterogeneous time-varying network VAR model \citep{YSM24}; the proposed grouped time-varying network VAR model; and the grouped network VAR model with constant coefficients \citep{ZXF23}, denoted by "fully heterogeneous", "grouped+TV" and "grouped+linear", respectively, in Table \ref{ooRMSE}. We leave the last $T_{\sf pre}$ observations out for prediction, where $T_{\sf pre}=12, 24$ and $36$, corresponding to one, two and three years, respectively. For any time point $t_\bullet$ in the prediction (or test) period, we use the observations over $t=1,\cdots, t_\bullet-1$ to get the estimates of time-varying or constant coefficients, which are subsequently used to forecast the value of $X_{t_\bullet}$. We conduct this expanding-window one-step ahead forecasting exercise for all the three models and report their out-of-sample RMSE in Table \ref{ooRMSE}. It follows from Table \ref{ooRMSE} that our proposed model produces the most accurate out-of-sample forecasting results with much smaller RMSE than the other two competing methods.

\begin{table}[tbp]\centering
\caption{Comparison of out-of-sample RMSE}\label{ooRMSE}
\begin{tabular}{lccc}
\hline\hline
Model & $T_{\sf pre}=12$ & $T_{\sf pre}=24$ & $T_{\sf pre}=36$ \\
\hline
  &  &  &  \\
Model (i): fully heterogeneous & 0.939 & 1.048 & 1.030 \\
Model (i): grouped+TV& 0.609 & 0.618 & 0.624 \\
Model (i): grouped+linear & 1.113 & 1.102 & 1.086 \\
  &  &  &  \\
Model (ii): fully heterogeneous & 0.711 & 0.703 & 0.773 \\
Model (ii): grouped+TV & 0.684 & 0.686 & 0.709 \\
Model (ii): grouped+linear  & 1.124 & 1.110 & 1.083\\
\hline\hline
\end{tabular}
\end{table}

%%%%%%%%%%%%%%%%%%%

\section{Conclusion}\label{sec7}

In this paper we have introduced a general nonlinear network VAR model for high-dimensional time series, where the momentum and network spillover effects are allowed to change over time and nodes. To achieve dimension reduction and obtain satisfactory estimation convergence rates, we impose a latent group structure on time-varying coefficients in the heterogenous network VAR model. The unknown group number is determined by an easy-to-implement criterion whereas the group membership is estimated by the agglomerative clustering algorithm with the nonparametrically estimated distance matrix. Theorems \ref{thm:3.1} and \ref{thm:3.2} show that the developed methodology consistently estimates the latent group structure. To further improve the convergence rates of the time-varying coefficient estimation, we have proposed a post-grouping local linear smoothing to estimate the group-specific time-varying momentum and network effects. In addition, we further extend the model, methodology and theory to allow for structural breaks in either the group structure or group-specific coefficient functions. The simulation study demonstrates that (i) the developed method can accurately estimate the latent group structure in finite samples; (ii) the post-grouping local linear estimation significantly outperforms the naive heterogenous estimation which ignores the latent structure; and (iii) the developed two-stage method can accurately locate the break point and estimate the group structure before and after the break. The empirical study of the UK temperature time series shows that there exist two groups over the 37 UK weather stations and our proposed method has better out-of-sample prediction performance than the other two competing methods which ignore either the grouped or time-varying feature in network VAR model building. 

\bigskip

%%%%%%%%%%%%%%%%%%%

\section*{Acknowledgements}

The authors thank the Editor, an Associate Editor and two anonymous referees for their constructive comments, which helped to substantially improve an earlier version of the article. The first author is supported by the Leverhulme Research Fellowship (RF-2023-396).  The second author is supported by the Australian Research Council Discovery Grants (DP210100476).

\bigskip

%%%%%%%%%%%%%%%%%%%

\appendix 
%\addtocounter{section}{0}
\section{Cluster analysis of heterogenous network effects}\label{appA}
\renewcommand{\theequation}{A.\arabic{equation}} \setcounter{equation}{0}

Let $\beta_{ij}(\cdot)=\alpha_{g_ig_j}(\cdot)$ be the heterogenous time-varying network spillover effects defined as in Section \ref{sec3.1}. Due to the homogeneity structure for $\beta_{ij}(\cdot)$, $(i,j)\in\bar{\mathscr N}_N$, there exists a partition of the index pair set $\bar{\mathscr N}_N$, denoted by ${\mathscr G}^\circ=\{{\mathscr G}_1^\circ,\cdots,{\mathscr G}_{M_0}^\circ\}$, such that
\begin{equation}\label{eqA.1}
{\mathscr G}_i^\circ\cap{\mathscr G}_j^\circ=\emptyset,\quad 1\leq i\neq j\leq M_0,\quad \beta_{ij}(\cdot)=\beta_m^\circ(\cdot)\ {\rm for}\ (i,j)\in{\mathscr G}_m^\circ,
\end{equation}
where $M_0$ is a finite positive integer upper bounded by $G_0^2$. Neither the group membership ${\mathscr G}^\circ$ nor the group number $M_0$ is known a priori. Define the distance between index pairs $(i_1,j_1)$ and $(i_2,j_2)$:
\[
\widetilde{D}_{(i_1j_1)(i_2 j_2)}=\frac{1}{L}\sum_{l=1}^L\left\vert \widehat{\beta}_{i_1j_1}(\tau_l^\ast)-\widehat{\beta}_{i_2j_2}(\tau_l^\ast) \right\vert,
\]
where $\widehat{\beta}_{ij}(\cdot)$ and $\tau_l^\ast$ are defined in Section \ref{sec3.1}. With $\widetilde{D}_{(i_1j_1)(i_2 j_2)}$ as the entries, we may further construct an $\bar{N}\times \bar{N}$ distance matrix denoted by $\widetilde{\mathbf D}$, where $\bar{N}=|\bar{\mathscr N}_N|$. Assuming the group number as $M$, we adopt the agglomerative hierarchical clustering algorithm described in Section \ref{sec3.1} with the distance matrix $\widetilde{\mathbf D}$, terminate it when the group number reaches $M$, and denote the resulting group estimates as $\widetilde{\mathscr G}_{1|M}^\circ,\cdots,\widetilde{\mathscr G}_{M|M}^\circ$.

\smallskip

As in Section \ref{sec3.2}, we pool the estimated heterogenous time-varying network spillover effects over $(i,j)\in\widetilde{\mathscr G}_{m|M}^\circ$:
\[
\widetilde\beta_{m|M}(\tau)=\frac{1}{\left\vert\widetilde{\mathscr G}_{m|M}^\circ\right\vert}\sum_{(i,j)\in\widetilde{\mathscr G}_{m|M}^\circ}\widehat{\beta}_{ij}(\tau),
\]
and define the average deviation:
\[
\widetilde{R}(M)= \frac{1}{ML}\sum_{m=1}^M\frac{1}{\left\vert\widetilde{\mathscr G}_{m|M}^\circ\right\vert }\sum_{(i,j)\in\widetilde{\mathscr G}_{m|M}^\circ}\sum_{l=1}^L \left\vert \widehat{\beta}_{ij}(\tau_l^\ast)-\widetilde\beta_{m|M}(\tau_l^\ast)\right\vert.
\]
The group number $M_0$ is estimated by the following ratio criterion:
\begin{equation}\label{eqA.2}
\widehat{M}=\argmin_{1\leq M\leq \overline{M}} \frac{\widetilde{R}(M)}{\widetilde{R}(M-1)},
\end{equation}
where $\overline{M}$ is a pre-specified positive integer. With the above consistent group number number estimate, we again run the agglomerative clustering algorithm, terminate it when the group number reaches $\widehat{M}$, and denote the final estimates of group membership by $\widehat{\mathscr G}_{m}^\circ$, $m=1,\cdots, \widehat M$, from which we obtain the group label estimates $\widehat{g}_{ij}=m$ if $(i,j)\in\widehat{\mathscr G}_m^\circ\cap\bar{\mathscr N}_N$.

\renewcommand{\theproposition}{A.\arabic{proposition}}\setcounter{proposition}{0}

\begin{proposition}\label{prop:A.1}

Suppose that the conditions of Theorems \ref{thm:3.1} and \ref{thm:3.2} hold. Then 
\begin{equation}\label{eqA.3}
{\sf P}\left(\widehat{M}=M_0\right)\rightarrow1,\quad {\sf P}\left(\big\{\widehat{\mathscr G}_m^\circ,\ 1\leq m\leq \widehat{M}\big\}=\big\{{\mathscr G}_m^\circ,\ 1\leq m\leq M_0\big\}\right)\rightarrow1.
\end{equation}

\end{proposition}

\smallskip

The above proposition establishes the consistency properties of the group number estimate $\widehat M$ and the membership estimate $\widehat{\mathscr G}_{m}^\circ$, $m=1,\cdots, \widehat M$. The finite-sample performance of these estimators is reported in Appendix F of the online supplement \citep{LPTW24}.

\bigskip

\small
\newpage
\setcounter{page}{1}

\begin{center}

{\Large Online Supplement to ``Estimation of Grouped Time-Varying Network Vector Autoregression Models"}

\end{center}

\bigskip

This supplemental material contains five appendices: Appendix \ref{appB} gives the proof of (2.8) in Remark 1(ii) and discusses connection of the proposed model to the nonlinear functional dependence measure; Appendix \ref{appC} provides the proofs of main theorems; Appendix \ref{appD} includes the proofs of some technical lemmas; Appendix \ref{appE} discusses a refined estimation of the break location; and Appendix \ref{appF} reports some extra numerical results. Unless explicitly stated differently, all the assumptions, equations, propositions, remarks, theorems and sections mentioned in this supplemental material refer to those presented in the main text. Throughout the supplement, we let $C,C_1,C_2,\cdots$ denote some generic positive constants.

\begin{appendix}
\addtocounter{section}{1}

%%%%%%%%%%%%%%%%%%%

\section{Theoretical justification of Remark 1(ii)}\label{appB}
\renewcommand{\theequation}{B.\arabic{equation}}
\setcounter{equation}{0}

In this appendix, we first provide the proof of (2.8) in Remark 1(ii) and then discuss the nonlinear functional dependence measure introduced by \cite{Wu05}.

\smallskip

\noindent{\bf Proof of (2.8) in Remark 1(ii)}.\ \ The proof is similar to the proof of Lemma A.1 in \cite{ZW21}. It follows from (2.4) in Assumption 1(i) that there exists $0<\chi_0<1$ such that
\begin{eqnarray}
\left\Vert \left\vert X_t-X_t^\circ\right\vert_\infty\right\Vert_q&=&\left\Vert \left\vert {\mathbf B}(\tau_t)\left[X_{t-1}-X_{t-1}^\circ(\tau_t)\right]\right\vert_\infty\right\Vert_q\notag\\
&\leq&\chi_0\left\Vert \left\vert X_{t-1}-X_{t-1}^\circ(\tau_t)\right\vert_\infty\right\Vert_q.\label{eqB.1}
\end{eqnarray}
Letting
\[
\Xi_{t-1,1}=\left\Vert \left\vert X_{t-1}-X_{t-1}^\circ(\tau_{t-1})\right\vert_\infty\right\Vert_q\ \ {\rm and}\ \ 
\Xi_{t-1,2}=\left\Vert \left\vert X_{t-1}^\circ(\tau_{t-1})-X_{t-1}^\circ(\tau_{t})\right\vert_\infty\right\Vert_q,
\]
and using (\ref{eqB.1}) and the triangle inequality, we obtain
\begin{equation}\label{eqB.2}
\left\Vert \left\vert X_t-X_t^\circ\right\vert_\infty\right\Vert_q\leq \chi_0\left(\Xi_{t-1,1}+\Xi_{t-1,2}\right).
\end{equation}
For any $0\leq \tau,\tau^\ast\leq 1$, by the smoothness condition in Assumption 1(i) and (2.7), we may show that 
\begin{eqnarray}
\left\Vert \left\vert X_{t}^\circ(\tau)-X_{t}^\circ(\tau^\ast)\right\vert_\infty\right\Vert_q
&=&\left\Vert \left\vert {\mathbf B}(\tau)X_{t-1}^\circ(\tau)-{\mathbf B}(\tau^\ast) X_{t-1}^\circ(\tau^\ast)\right\vert_\infty\right\Vert_q\notag\\
&\leq&\left\Vert \left\vert {\mathbf B}(\tau)X_{t-1}^\circ(\tau)-{\mathbf B}(\tau^\ast) X_{t-1}^\circ(\tau)\right\vert_\infty\right\Vert_q+\notag\\
&&\left\Vert \left\vert {\mathbf B}(\tau^\ast)X_{t-1}^\circ(\tau)-{\mathbf B}(\tau^\ast) X_{t-1}^\circ(\tau^\ast)\right\vert_\infty\right\Vert_q\notag\\
&\leq& C_1 \left\Vert \left\vert X_{t-1}^\circ(\tau)\right\vert_\infty\right\Vert_q |\tau-\tau^\ast|+\chi_0 \left\Vert \left\vert X_{t-1}^\circ(\tau)-X_{t-1}^\circ(\tau^\ast)\right\vert_\infty\right\Vert_q\notag\\
&\leq&C_1 \theta_{N,q}|\tau-\tau^\ast|+\chi_0 \left\Vert \left\vert X_{t-1}^\circ(\tau)-X_{t-1}^\circ(\tau^\ast)\right\vert_\infty\right\Vert_q,\notag
\end{eqnarray}
indicating that 
\[
\left\Vert \left\vert X_{t}^\circ(\tau)-X_{t}^\circ(\tau^\ast)\right\vert_\infty\right\Vert_q\leq \frac{C_1\theta_{N,q}|\tau-\tau^\ast|}{1-\chi_0}
\]
and thus
\begin{equation}\label{eqB.3}
\Xi_{t-1,2}\leq \frac{C_1}{1-\chi_0}\cdot\frac{\theta_{N,q}}{T}.
\end{equation}
By virtue of (\ref{eqB.2}) and (\ref{eqB.3}), we readily have that 
\begin{equation}\label{eqB.4}
\left\Vert \left\vert X_t-X_t^\circ\right\vert_\infty\right\Vert_q\leq \chi_0\left\Vert \left\vert X_{t-1}-X_{t-1}^\circ\right\vert_\infty\right\Vert_q+ \frac{C_1\chi_0}{1-\chi_0}\cdot\frac{\theta_{N,q}}{T}.
\end{equation}
With (\ref{eqB.4}), using the argument similar to the proof of Lemma 4.5 in \cite{DRW19}, we have
\[
\max_{1\leq t\leq T}\left\Vert \left\vert X_t-X_t^\circ\right\vert_\infty\right\Vert_q\leq  \frac{C_1\chi_0}{(1-\chi_0)^2}\cdot\frac{\theta_{N,q}}{T},
\]
completing the proof of (2.8).\hfill$\Box$

\smallskip

We next connect the grouped time-varying network VAR model to the nonlinear functional dependence measure introduced by \cite{Wu05}, facilitating the development of our main asymptotic theory. Let $\left\{\varepsilon_t^\ast\right\}$ be an independent copy of $\left\{\varepsilon_t\right\}$ and $\mathscr{F}_{t}^{\{l\}}=(\cdots,\varepsilon_{l-1},\varepsilon_l^\ast,\varepsilon_{l+1},\cdots,\varepsilon_t)$ be a coupled version of $\mathscr{F}_t$ replacing $\varepsilon_l$ by $\varepsilon_l^\ast$. Letting $X_{t}^{\circ\{l\}}(\tau)=G(\tau,\mathscr{F}_{t}^{\{l\}})$, as in \cite{ZW21}, we define the node-wise functional dependence measure:
\[
\delta_{i,t,q}=\sup_{\tau\in[0,1]}\left\Vert x_{i,t}^\circ(\tau)-x_{i,t}^{\circ\{0\}}(\tau)\right\Vert_q,
\]
where $x_{i,t}^\circ(\tau)$ and $x_{i,t}^{\circ\{0\}}(\tau)$ are the $i$-th element of $X_t^\circ(\tau)$ and $X_{t}^{\circ\{0\}}(\tau)$, respectively. Furthermore, we construct the node-wise dependence adjusted norm:
\[
\left\Vert x_{i\bullet}\right\Vert_{q,\iota}=\sup_{m\ge0}(m+1)^\iota\Delta_{i,m,q},~~ \Delta_{i,m,q}=\sum_{t=m}^\infty\delta_{i,t,q},
\]
where $\iota\geq0$ depicts the decay rate of the cumulative dependence measure $\Delta_{i,m,q}$. Letting ${\mathbf B}_{i \bullet}^j(\tau)$ be the $i$-th row vector of ${\mathbf B}^j(\tau)$, by (2.4) in Assumption 1(i), $\sup_{\tau\in[0,1]}\vert{\mathbf B}_{i \bullet}^j(\tau)\vert_2$ decays at a geometric rate of $j$ and $\delta_{i,t,q}$ decays at a geometric rate of $t$ for all $i$. For simplicity, we may set $\iota=1$ and write $\Vert x_{i\bullet}\Vert_{q}=\Vert x_{i\bullet}\Vert_{q,1}$. Then we have $\max_{1\leq i\leq N}\Vert x_{i\bullet}\Vert_{q}\le C_2$, where $C_2$ is a positive constant.

\bigskip

%%%%%%%%%%%%%%%%%%%%%%%%%

\section{Proofs of the main asymptotic theorems}\label{appC}
\renewcommand{\theequation}{C.\arabic{equation}}
\setcounter{equation}{0}

We next provide the detailed proofs of the main asymptotic theorems in the main text. 

\smallskip

\subsection{Proof of Theorem 3.1}\label{appC.1} 

To prove (3.10), we only need to show that
\begin{equation}\label{eqC.1}
{\sf P}\left(\max_{1\le k \le K_0}\max_{i,j\in{\mathscr G}_k}\widehat{D}_{ij}<\min_{1\le k\neq l\le K_0}\min_{i\in{\mathscr G}_k,j\in{\mathscr G}_l}\widehat{D}_{ij}\right)\rightarrow1
\end{equation}
as $T$ tends to infinity. Define
\[
D_{ij}=\int_{0}^1d_{ij}(\tau) d\tau\ \ {\rm with}\ \ d_{ij}(\tau)=\left\vert \beta_i(\tau)-\beta_j(\tau)\right\vert+\left\vert\beta_{i\bullet}^\circ(\tau)-\beta_{j\bullet}^\circ(\tau)\right\vert_2.
\]
Notice that $D_{ij}=0$ if $i,j\in{\mathscr G}_k$, and 
\[
\zeta_{NT}^\dag=\min_{1\le k\neq l\le K_0}\min_{i\in{\mathscr G}_k,j\in{\mathscr G}_l}D_{ij}>0.
\] 
Hence, to prove (\ref{eqC.1}), it is sufficient to show that
\begin{equation}\label{eqC.2}
\max_{1\le i\neq j\le N}\big\vert\widehat{D}_{ij}-D_{ij}\big\vert=o_P\left(\zeta_{NT}^\dag\right).
\end{equation}

\smallskip

Letting $\overline{D}_{ij}=\frac{1}{L}\sum_{l=1}^Ld_{ij}(\tau_l^\ast)$, we have
\[
\big\vert\widehat{D}_{ij}-D_{ij}\big\vert\le\big\vert\widehat{D}_{ij}-\overline{D}_{ij}\big\vert+\big\vert\overline{D}_{ij}-D_{ij}\big\vert.
\]
By the definition of the Riemann integral and Assumption 1(i), we readily have that 
\begin{equation}\label{eqC.3}
\max_{1\le i\neq j\le N}\big\vert\overline{D}_{ij}-D_{ij}\big\vert=O\left(1/L\right)=o\left(\zeta_{NT}^\dag\right),
\end{equation}
where $1/L=o(\zeta_{NT}^\dag)$ in Assumption 3 has been used.

\smallskip

By Proposition A.1, without loss of generality, we may assume that $\widehat{M}=M_0$ and $\widehat{\mathscr G}_m^\circ={\mathscr G}_m^\circ$, $m=1,\cdots,M_0$, hold {\em w.p.a.1} in the remaining proof. By the triangle inequality, we have
\begin{eqnarray}
\big\vert\widehat{D}_{ij}-\overline{D}_{ij}\big\vert&=&\bigg\vert\frac{1}{L}\sum_{l=1}^L\left[\widehat{d}_{ij}(\tau_l^\ast)-d_{ij}(\tau_l^\ast)\right]\bigg\vert\notag\\
&\le&\frac{1}{L}\sum_{l=1}^L\bigg\vert\big\vert\widehat{\beta}_i(\tau_l^\ast)-\widehat{\beta}_j(\tau_l^\ast)\big\vert-\big\vert\beta_i(\tau_l^\ast)-\beta_j(\tau_l^\ast)\big\vert\bigg\vert+\notag\\
&&\frac{1}{L}\sum_{l=1}^L\bigg\vert\big\vert\widehat{\beta}_{i\bullet}^\circ(\tau_l^\ast)-\widehat{\beta}_{j\bullet}^\circ(\tau_l^\ast)\big\vert_2-\big\vert\beta_{i\bullet}^\circ(\tau_l^\ast)-\beta_{j\bullet}^\circ(\tau_l^\ast)\big\vert_2\bigg\vert\notag\\
&\leq&\frac{1}{L}\sum_{l=1}^L\big\vert\widehat{\beta}_i(\tau_l^\ast)-\beta_i(\tau_l^\ast)\big\vert+\frac{1}{L}\sum_{l=1}^L\big\vert\widehat{\beta}_j(\tau_l^\ast)-\beta_j(\tau_l^\ast)\big\vert+\notag\\
&&\frac{1}{L}\sum_{l=1}^L \big\vert\widehat{\beta}_{i\bullet}^\circ(\tau_l^\ast)-\beta_{i\bullet}^\circ(\tau_l^\ast)\big\vert_2+\frac{1}{L}\sum_{l=1}^L\big\vert \widehat{\beta}_{j\bullet}^\circ(\tau_l^\ast)-\beta_{j\bullet}^\circ(\tau_l^\ast)\big\vert_2,\notag
\end{eqnarray}
which, together with Lemmas \ref{lem:D.3} and \ref{lem:D.4} and Assumption 3, leads to 
\begin{equation}\label{eqC.4}
\max_{1\leq i\neq j\leq N}\big\vert\widehat{D}_{ij}-\overline{D}_{ij}\big\vert=O_P\left(\sqrt{\frac{\bar n\log (N\vee T)}{Th}}+\sqrt{\bar n}h^2\right)=o_P\left(\zeta_{NT}^\dag\right).
\end{equation}
With (\ref{eqC.3}) and (\ref{eqC.4}), we complete the proof of (\ref{eqC.2}). \hfill$\Box$

\medskip

\subsection{Proof of Theorem 3.2}\label{appC.2} 

Let ${\mathscr E}_{\mathscr G}$ denote the event that $\{\widehat{{\mathscr G}}_1,\cdots,\widehat{{\mathscr G}}_{K_0}\big\}=\big\{{\mathscr G}_1,\cdots,{\mathscr G}_{K_0}\}$. It follows from Theorem 3.1 that ${\sf P}({\mathscr E}_{\mathscr G})\rightarrow1$. Hence, to prove Theorem 3.2, it is sufficient to show that
\begin{equation}\label{eqC.5}
{\sf P}\left(\widehat{K}=K_0\ |\ {\mathscr E}_{\mathscr G}\right)\rightarrow1.
\end{equation}
By the definition of $\widehat{K}$, we only need to show that 
\begin{equation}\label{eqC.6}
{\sf P}\left(\frac{\widehat{R}(K_0)}{\widehat{R}(K_0-1)}=\min_{1\le K\le \overline{K}}\frac{\widehat{R}(K)}{\widehat{R}(K-1)} \ |\ {\mathscr E}_{\mathscr G} \right)\rightarrow1.
\end{equation}
To prove (\ref{eqC.6}), we next consider the following two scenarios: (i) $1\le K\le K_0-1$ and (ii) $K_0+1\le K\le \overline{K}$, corresponding to the under-fitted and over-fitted grouped time-varying network VAR models, respectively. For the case $K_0+1\leq K\leq \overline{K}$, conditional on the event ${\mathscr E}_{\mathscr G}$, by Lemma \ref{lem:D.5}(i), we have $\widehat{R}(K)=o_P(\rho_{NT})$ for $K=K_0+1,\cdots,\overline{K}$. Consequently, we may set $\widehat{R}(K)\equiv0$ {\em w.p.a.1}, and 
\begin{equation}\label{eqC.7}
{\sf P}\left(\frac{\widehat{R}(K)}{\widehat{R}(K-1)} =\frac{0}{0}\equiv1,\ \ K=K_0+1,\cdots,\overline{K}\ \big|\ {\mathscr E}_{\mathscr G}\right)\rightarrow1.
\end{equation}
For the case $1\leq K\leq K_0-1$, conditional on ${\mathscr E}_{\mathscr G}$, by Lemma \ref{lem:D.5}(ii), we have $\widehat{R}(K)\geq \underline{c}\zeta_{NT}^\dag$ {\em w.p.a.1} with $\underline{c}$ being a positive constant strictly larger than zero. 
Hence, there exists a positive constant $\underline{c}_\ast$ such that
\begin{equation}\label{eqC.8}
	{\sf P}\left(\frac{\widehat{R}(K)}{\widehat{R}(K-1)} \geq \underline{c}_\ast,\ \ K=1,\cdots,K_0-1\ \big|\ {\mathscr E}_{\mathscr G}\right)\rightarrow1,
\end{equation}
setting $\frac{\widehat{R}(1)}{\widehat{R}(0)}=1$, and
\begin{equation}\label{eqC.9}
	{\sf P}\left(\frac{\widehat{R}(K_0)}{\widehat{R}(K_0-1)} =0\ \big|\ {\mathscr E}_{\mathscr G}\right)\rightarrow1.
\end{equation}
With (\ref{eqC.7})--(\ref{eqC.9}), we prove (\ref{eqC.6}), completing the proof of Theorem 3.2.\hfill$\Box$

%\medskip

%\noindent{\bf Proof of Corollary 3.3}.\ \ With Theorems 3.1 and 3.2, we can easily prove (3.16).\hfill$\Box$

\medskip

\subsection{Proof of Theorem 4.1}\label{appC.3}

By the consistency properties in Theorem 3.2 and Corollary 3.3, we may prove the asymptotic distribution theory conditional on $\widehat{K}=K_0$ and $\widehat{{\mathscr G}}_k={\mathscr G}_k$, $k=1,\cdots,K_0$. For $k=1,\cdots,K_0$, let
\begin{eqnarray}
{\boldsymbol\Xi}_{kX}(\tau)&=&\frac{1}{{\rm card}({\mathscr G}_k)Th_\dag}\sum_{t=1}^T\sum_{i\in{\mathscr G}_k}\left[
\begin{array}{cc}
X_{i,t-1}^\diamond X_{i,t-1}^{\diamond^\intercal}K_{t0}^{\dagger}(\tau)\ &\ X_{i,t-1}^\diamond X_{i,t-1}^{\diamond^\intercal}K_{t1}^{\dagger}(\tau)\\
X_{i,t-1}^\diamond X_{i,t-1}^{\diamond^\intercal}K_{t1}^{\dagger}(\tau)\ &\ X_{i,t-1}^\diamond X_{i,t-1}^{\diamond^\intercal}K_{t2}^{\dagger}(\tau)
\end{array}
\right],\notag\\
{\boldsymbol\Xi}_{k\alpha}(\tau)&=&\frac{1}{{\rm card}({\mathscr G}_k)Th_\dag}\sum_{t=1}^T\sum_{i\in{\mathscr G}_k}\left[
\begin{array}{c}
X_{i,t-1}^\diamond X_{i,t-1}^{\diamond^\intercal}\left[\alpha_{k\bullet}(\tau_t)-\alpha_{k\bullet}(\tau)-\alpha_{k\bullet}^\prime(\tau)(\tau_t-\tau)\right]K_{t0}^{\dagger}(\tau)
 \\
X_{i,t-1}^\diamond X_{i,t-1}^{\diamond^\intercal}\left[\alpha_{k\bullet}(\tau_t)-\alpha_{k\bullet}(\tau)-\alpha_{k\bullet}^\prime(\tau)(\tau_t-\tau)\right]K_{t1}^{\dagger}(\tau)
\end{array}
\right],\notag\\
{\boldsymbol\Xi}_{k\varepsilon}(\tau)&=&\frac{1}{{\rm card}({\mathscr G}_k)Th_\dag}\sum_{t=1}^T\sum_{i\in{\mathscr G}_k}\left[
\begin{array}{c}
X_{i,t-1}^\diamond\varepsilon_{i,t}K_{t0}^{\dagger}(\tau) \\
X_{i,t-1}^\diamond\varepsilon_{i,t}K_{t1}^{\dagger}(\tau) 
\end{array}
\right],\quad K^\dagger_{t\kappa}(\tau)=\left(\frac{\tau_t-\tau}{h_\dag}\right)^\kappa K\left(\frac{\tau_t-\tau}{h_\dag}\right).\notag
\end{eqnarray}
By the definition of the post-grouping local linear estimation defined in (4.2), we have
\[
\check\alpha_{k\bullet}(\tau)-\alpha_{k\bullet}(\tau)={\mathbf E}_{\bullet}\cdot{\boldsymbol\Xi}_{kX}^{-1}(\tau)\left[{\boldsymbol\Xi}_{k\alpha}(\tau)+{\boldsymbol\Xi}_{k\varepsilon}(\tau)\right]
\]
conditional on $\widehat{K}=K_0$ and $\widehat{{\mathscr G}}_k={\mathscr G}_k$, $k=1,\cdots,K_0$, where ${\mathbf E}_\bullet=\left({\mathbf I}_{K_0+1}, {\mathbf O}_{(K_0+1)\times(K_0+1)}\right)$. In order to prove (4.4), we only need to show that
\begin{eqnarray}
&&{\boldsymbol\Xi}_{kX}(\tau)\stackrel{P}\longrightarrow{\sf diag}\left\{1,\mu_2\right\}\otimes \boldsymbol{\Delta}_{{\mathscr G}_k}(\tau),\label{eqC.10}\\
&&{\boldsymbol\Xi}_{k\alpha}(\tau)=\left[
\begin{array}{c}
\frac{1}{2}h_\dag^2\mu_2\boldsymbol{\Delta}_{{\mathscr G}_k}(\tau)\alpha_{k\bullet}^{\prime\prime}(\tau)\\
{\mathbf 0} \\
\end{array}
\right]+o_P(h_\dag^2),\label{eqC.11}\\
&&\left[{\rm card}({\mathscr G}_k)Th_\dag\right]^{1/2}{\boldsymbol\Xi}_{k\varepsilon}(\tau)\stackrel{d}\longrightarrow{\sf N}\left(\boldsymbol{0},\ {\sf diag}\left\{\nu_0,\nu_2\right\}\otimes{\boldsymbol\Upsilon}_{{\mathscr G}_k}(\tau)\right),\label{eqC.12}
\end{eqnarray}
where $\otimes$ denotes the Kronecker product between matrices. 

\smallskip

By Lemma \ref{lem:D.6} and the approximation result (2.8), we can prove (\ref{eqC.10}). By the smoothness condition in Assumption 1(i), Taylor's expansion of $\alpha_{k\bullet}(\cdot)$ and (\ref{eqC.10}), we can prove (\ref{eqC.11}). We next turn to the proof of (\ref{eqC.12}). Let 
\[
W_{{\mathscr G}_k,t}(\tau)=\left[W_{{\mathscr G}_k,t}^{^\intercal}(\tau,0), W_{{\mathscr G}_k,t}^{^\intercal}(\tau,1)\right]^{^\intercal}
\]
with
\[ 
W_{{\mathscr G}_k,t}(\tau,\kappa)=
\frac{1}{\sqrt{{\rm card}({\mathscr G}_k)}}
\sum_{i\in{\mathscr G}_k}\varepsilon_{i,t}X_{i,t-1}^\diamond K^\dagger_{t\kappa}(\tau),
\]
and write
\[
\sqrt{{\rm card}({\mathscr G}_k)Th_\dag}{\boldsymbol\Xi}_{k\varepsilon}(\tau)=\frac{1}{\sqrt{Th_\dag}}\sum_{t=1}^T W_{{\mathscr G}_k,t}(\tau).
\]
Using Lemma \ref{lem:D.6} and the approximation (2.8), we have 
\begin{eqnarray}
&&\frac{1}{Th_\dag}\sum_{t=1}^T {\sf E}\left[W_{{\mathscr G}_k,t}(\tau,0)W_{{\mathscr G}_k,t}^{^\intercal}(\tau,0) | {\cal F}_{t-1}\right]\notag\\
&=&\frac{1}{{\rm card}({\mathscr G}_k)}\sum_{i,j\in{\mathscr G}_k}\sigma_{ij}\left[\frac{1}{Th_\dag}\sum_{t=1}^T X_{i,t-1}^\diamond X_{i,t-1}^{\diamond^\intercal} K^\ddagger_{t,0}(\tau)\right]\notag\\
&\stackrel{P}\longrightarrow&\frac{\nu_0}{{\rm card}({\mathscr G}_k)}\sum_{i,j\in{\mathscr G}_k}\sigma_{ij}{\boldsymbol\Delta}_{ij}^\diamond(\tau).\notag
\end{eqnarray}
Similarly, we can also prove that
\[
\frac{1}{Th_\dag}\sum_{t=1}^T {\sf E}\left[W_{{\mathscr G}_k,t}(\tau,1)W_{{\mathscr G}_k,t}^{^\intercal}(\tau,1) | {\cal F}_{t-1}\right]\stackrel{P}\longrightarrow \frac{\nu_2}{{\rm card}({\mathscr G}_k)}\sum_{i,j\in{\mathscr G}_k}\sigma_{ij}{\boldsymbol\Delta}_{ij}^\diamond(\tau),
\]
and 
\[
\frac{1}{Th_\dag}\sum_{t=1}^T {\sf E}\left[W_{{\mathscr G}_k,t}(\tau,0)W_{{\mathscr G}_k,t}^{^\intercal}(\tau,1) | {\cal F}_{t-1}\right]=o_P(1).
\]
Hence, we have
\[
\frac{1}{Th_\dag}\sum_{t=1}^T {\sf E}\left[W_{{\mathscr G}_k,t}(\tau)W_{{\mathscr G}_k,t}^{^\intercal}(\tau) | {\cal F}_{t-1}\right]\stackrel{P}\longrightarrow {\sf diag}\left\{\nu_0,\nu_2\right\}\otimes\frac{1}{{\rm card}({\mathscr G}_k)}\sum_{i,j\in{\mathscr G}_k}\sigma_{ij}{\boldsymbol\Delta}_{ij}^\diamond(\tau),
\]
which, together with (4.3) and the martingale central limit theorem \citep[e.g.,][]{HH80}, leads to (\ref{eqC.12}). The proof of Theorem 4.1 is completed. \hfill$\Box$

\medskip

\subsection{Proof of Theorem 5.1}\label{appC.4} 

Let 
\begin{eqnarray}
&&\overline{\Gamma}_{it}^{{\sf l},\varepsilon}=\frac{1}{Th_\ddag}\sum_{s=1}^T \widetilde{X}_{i,s-1}\varepsilon_{i,s}K^\ddag\left(\frac{\tau_t-\tau_s}{h_\ddag}\right),\quad \overline{\Gamma}_{it}^{{\sf r},\varepsilon}=\frac{1}{Th_\ddag}\sum_{s=1}^T \widetilde{X}_{i,s-1}\varepsilon_{i,s}K^\ddag\left(\frac{\tau_s-\tau_t}{h_\ddag}\right),\notag\\
&&\overline\beta_{i\bullet}^{\ddag,{\sf l}}(\tau_t)=\left(\widetilde{\boldsymbol\Gamma}_{it}^{{\sf l}}\right)^{-1}\left(\overline{\Gamma}_{it}^{\sf l}-\overline{\Gamma}_{it}^{{\sf l},\varepsilon}\right),\quad \overline\beta_{i\bullet}^{\ddag,{\sf r}}(\tau_t)=\left(\widetilde{\boldsymbol\Gamma}_{it}^{{\sf r}}\right)^{-1}\left(\overline{\Gamma}_{it}^{\sf r}-\overline{\Gamma}_{it}^{{\sf r},\varepsilon}\right).\notag
\end{eqnarray}
Define
\[
\overline\delta_{\beta,i}(t)=\overline\beta_{i\bullet}^{\ddag,{\sf r}}(\tau_t)- \overline\beta_{i\bullet}^{\ddag,{\sf l}}(\tau_t)\quad {\rm and}\quad \delta_{\beta,i}(t)=\beta_{i\bullet}^{\ddag,{\sf r}}(\tau_t)-\beta_{i\bullet}^{\ddag,{\sf l}}(\tau_t).
\]
By Assumption 2(ii) and 6(i)(ii), we may show that 
\begin{equation}\label{eqC.13}
\overline\delta_{\beta,i}(t)=\widetilde\delta_{\beta,i}(t)+O_P(h_\ddag)
\end{equation}
uniformly over $i=1,\cdots,N$, where
\[
\widetilde\delta_{\beta,i}(t)=\left\{
\begin{array}{ll}
0,\ &\quad t>t_0+Th_\ddag\ \ {\rm or}\ \ t<t_0-Th_\ddag,\\
\left[1-g_i^{\sf l}\left(\frac{t_0-t}{Th_\ddag}\right)\right]\delta_{\beta,i}(t_0),\ &\quad t_0-Th_\ddag\leq t<t_0,\\
\left[1-g_i^{\sf r}\left(\frac{t-t_0}{Th_\ddag}\right)\right]\delta_{\beta,i}(t_0),\ &\quad t_0< t\leq t_0+Th_\ddag,\\
\delta_{\beta,i}(t_0),\ &\quad t= t_0,\\
\end{array}
\right.
\]
and $g_i^{\sf l}(\cdot)$ and $g_i^{\sf r}(\cdot)$ are positive functions satisfying $g_i^{\sf l}(0)=g_i^{\sf r}(0)=0$, $g_i^{\sf l}(1)=g_i^{\sf r}(1)=1$ and
\[
\min_{1\leq i\leq N}\{g_i^{\sf l}(x), g_i^{\sf r}(x)\}\geq \underline{c}_0 x,\quad 0<x<1,
\]
with $\underline{c}_0$ being a positive constant.

\smallskip

Writing 
\[
\overline\delta_{\beta}(t)=\left[\overline\delta_{\beta,1}(t),\cdots,\overline\delta_{\beta,N}(t)\right]^{^\intercal}\quad{\rm and}\quad \widetilde\delta_{\beta}(t)=\left[\widetilde\delta_{\beta,1}(t),\cdots,\widetilde\delta_{\beta,N}(t)\right]^{^\intercal},
\]
by (\ref{eqC.13}) and the triangle inequality, we have
\begin{eqnarray}
\big\vert \overline\delta_{\beta}(t_0)\big\vert_\infty-\big\vert \overline\delta_{\beta}(t)\big\vert_\infty&\geq& \big\vert \widetilde\delta_{\beta}(t_0)\big\vert_\infty-\big\vert \widetilde\delta_{\beta}(t)\big\vert_\infty-O_P(h_\ddag)\notag\\
&\geq& \underline{c}_0\left\vert \frac{t-t_0}{Th_\ddag}\right\vert\cdot \delta_{\beta}(t_0)-O_P(h_\ddag)\label{eqC.14}
\end{eqnarray}
when $t_0-Th_\ddag\leq t\leq t_0+Th_\ddag$, and otherwise
\[
\left\vert \overline\delta_{\beta}(t_0)\right\vert_\infty-\left\vert \overline\delta_{\beta}(t)\right\vert_\infty\geq \left\vert\delta_{\beta}(t_0)\right\vert_\infty-O_P(h_\ddag).
\]
Following the proofs of Lemma \ref{lem:D.1} and (\ref{eqD.27}) in Appendix \ref{appD}, we may show that 
\begin{eqnarray}
&&\max_{1\leq i\leq N}\max_{1\leq t\leq T}\left\vert\left(\widetilde{\boldsymbol\Gamma}_{it}^{{\sf l}}\right)^{-1}\overline{\Gamma}_{it}^{{\sf l},\varepsilon}\right\vert_2=O_P\left(\sqrt{\frac{\bar n\log (N\vee T)}{Th_\ddag}}\right),\label{eqC.15}\\
&&\max_{1\leq i\leq N}\max_{1\leq t\leq T}\left\vert\left(\widetilde{\boldsymbol\Gamma}_{it}^{{\sf r}}\right)^{-1}\overline{\Gamma}_{it}^{{\sf r},\varepsilon}\right\vert_2=O_P\left(\sqrt{\frac{\bar n\log (N\vee T)}{Th_\ddag}}\right).\label{eqC.16}
\end{eqnarray}
By the definition of $\widehat{t}$, we have $\widehat\delta_\beta(\widehat{t})\geq \widehat\delta_\beta(t_0)$, which, together with (\ref{eqC.15}), (\ref{eqC.16}) and the triangle inequality, leads to 
\begin{eqnarray}
\left\vert \overline\delta_{\beta}(t_0)\right\vert_\infty-\left\vert \overline\delta_{\beta}(\widehat t)\right\vert_\infty&\leq& \widehat\delta_\beta(t_0)-\widehat\delta_\beta(\widehat t)+O_P\left(\sqrt{\frac{\bar n\log (N\vee T)}{Th_\ddag}}\right)\notag\\
&\leq&O_P\left(\sqrt{\frac{\bar n\log (N\vee T)}{Th_\ddag}}\right).\label{eqC.17}
\end{eqnarray}
By virtue of (\ref{eqC.14}) and (\ref{eqC.17}), we readily have that 
\[
 \underline{c}_0\left\vert \frac{\widehat{t}-t_0}{Th_\ddag}\right\vert\cdot \delta_{\beta}(t_0)= O_P\left(\sqrt{\frac{\bar n\log (N\vee T)}{Th_\ddag}}+h_\ddag\right),
\]
which completes the proof of (5.8) as $\delta_{\beta}(t_0)$ is bounded away from zero in Assumption 6(iii).

\smallskip

We next turn to the proof of Theorem 5.1(ii). Let 
\[
{\mathscr T}_1^\circ=\{1,2,\cdots,t_0\}\quad {\rm and}\quad {\mathscr T}_2^\circ=\{t_0+1, t_0+2,\cdots,T\}. 
\]
It follows from (5.8) that
\[
{\sf P}({\mathscr T}_1\subset {\mathscr T}_1^\circ)\rightarrow 1\quad {\rm and}\quad {\sf P}({\mathscr T}_2\subset {\mathscr T}_2^\circ)\rightarrow 1,
\]
indicating that the group structure is time invariant {\em w.p.a.1} over the two separate time periods ${\mathscr T}_1$ and ${\mathscr T}_2$. Following the proofs of Theorems 3.1, 3.2 and Corollary 3.3, we can prove Theorem 5.1(ii). Details are omitted to save the space.\hfill$\Box$

\medskip

\subsection{Proof of Proposition A.1}\label{appC.5}

Since the proof is analogous to the arguments used in the proofs of Theorems 3.1 and 3.2, we next only sketch the proof. 

\smallskip

In the first step, we need to prove that 
\begin{equation}\label{eqC.18}
{\sf P}\left(\big\{\widehat{\mathscr G}_m^\circ,\ 1\leq m\leq M_0\big\}=\big\{{\mathscr G}_m^\circ,\ 1\leq m\leq M_0\big\}\right)\rightarrow1.
\end{equation}
For the index pairs $(i_1,j_1)$ and $(i_2, j_2)$ taken from $\bar{\mathscr N}_N$, recall
\[
\widetilde{D}_{(i_1j_1)(i_2 j_2)}=\frac{1}{L}\sum_{l=1}^L\left\vert \widehat{\beta}_{i_1j_1}(\tau_l^\ast)-\widehat{\beta}_{i_2j_2}(\tau_l^\ast) \right\vert,
\]
and define
\[
D_{(i_1j_1)(i_2 j_2)}=\frac{1}{L}\sum_{l=1}^L\left\vert \beta_{i_1j_1}(\tau_l^\ast)-\beta_{i_2j_2}(\tau_l^\ast) \right\vert.
\] 
By (3.9) in Assumption 3, in order to prove (\ref{eqC.18}), it is sufficient to show that  
\begin{equation}\label{eqC.19}
\max_{(i_1,j_1), (i_2, j_2)\in \bar{\mathscr N}_N} \left\vert \widetilde{D}_{(i_1j_1)(i_2 j_2)}-D_{(i_1j_1)(i_2 j_2)} \right\vert=o_P\left(\zeta_{NT}^\ddag\right).
\end{equation}
With Lemma \ref{lem:D.3} and (3.9), following the proof of (\ref{eqC.2}), we can easily prove (\ref{eqC.19}).

\smallskip

In the second step, we need to prove that 
\begin{equation}\label{eqC.20}
{\sf P}\left(\widehat{M}=M_0\ |\ {\mathscr E}_{\mathscr G}^\circ\right)\rightarrow1,
\end{equation}
where ${\mathscr E}_{\mathscr G}^\circ$ denotes the event that $\{\widehat{\mathscr G}_m^\circ,\ 1\leq m\leq M_0\}=\{{\mathscr G}_m^\circ,\ 1\leq m\leq M_0\}$. The proof of (\ref{eqC.20}) is similar to the proof of (\ref{eqC.5}).

\smallskip

In the final step, combining (\ref{eqC.18}) and (\ref{eqC.20}), we readily have that 
\[
{\sf P}\left(\big\{\widehat{\mathscr G}_m^\circ,\ 1\leq m\leq \widehat{M}\big\}=\big\{{\mathscr G}_m^\circ,\ 1\leq m\leq M_0\big\}\right)\rightarrow1,
\]
completing the proof of Proposition A.1.\hfill$\Box$

\bigskip

%%%%%%%%%%%%%%%%%%%

\section{Technical lemmas with proofs}\label{appD}
\renewcommand{\theequation}{D.\arabic{equation}}\setcounter{equation}{0}
\renewcommand{\thelemma}{D.\arabic{lemma}}\setcounter{lemma}{0}

We next prove some technical lemmas which have been used to prove the asymptotic theorems in Appendix \ref{appC}. For any $1\leq i\leq N$, we define 
\begin{eqnarray}
{\boldsymbol\Gamma}_{iX}(\tau)&=&\left[
\begin{array}{cc}
\frac{1}{Th}\sum_{t=1}^T \widetilde{X}_{i,t-1}\widetilde{X}_{i,t-1}^{^\intercal}K_{t0}(\tau)\ &\ \frac{1}{Th}\sum_{t=1}^T \widetilde{X}_{i,t-1}\widetilde{X}_{i,t-1}^{^\intercal}K_{t1}(\tau) \\
\frac{1}{Th}\sum_{t=1}^T \widetilde{X}_{i,t-1}\widetilde{X}_{i,t-1}^{^\intercal}K_{t1}(\tau)\ &\ \frac{1}{Th}\sum_{t=1}^T \widetilde{X}_{i,t-1}\widetilde{X}_{i,t-1}^{^\intercal}K_{t2}(\tau)
\end{array}
\right]\nonumber\\
&=:&\left[
\begin{array}{cc}
{\boldsymbol\Gamma}_{iX}(\tau,0)\ &\ {\boldsymbol\Gamma}_{iX}(\tau,1)\\
{\boldsymbol\Gamma}_{iX}(\tau,1)\ &\ {\boldsymbol\Gamma}_{iX}(\tau,2)
\end{array}
\right],\notag\\
{\boldsymbol\Gamma}_{i\beta}(\tau)&=&\left[
\begin{array}{c}
\frac{1}{Th}\sum_{t=1}^T\widetilde{X}_{i,t-1}\widetilde{X}_{i,t-1}^{^\intercal}\left[\beta_{i\bullet}(\tau_t)-\beta_{i\bullet}(\tau)-\beta_{i\bullet}^\prime(\tau)(\tau_t-\tau)\right]K_{t0}(\tau) \\
\frac{1}{Th}\sum_{t=1}^T\widetilde{X}_{i,t-1}\widetilde{X}_{i,t-1}^{^\intercal}\left[\beta_{i\bullet}(\tau_t)-\beta_{i\bullet}(\tau)-\beta_{i\bullet}^\prime(\tau)(\tau_t-\tau)\right]K_{t1}(\tau)
\end{array}
\right]\nonumber\\
&=:&\left[
\begin{array}{c}
{\boldsymbol\Gamma}_{i\beta}(\tau,0) \\
{\boldsymbol\Gamma}_{i\beta}(\tau,1)
\end{array}
\right],\notag\\
{\boldsymbol\Gamma}_{i\varepsilon}(\tau)&=&\left[
\begin{array}{c}
\frac{1}{Th}\sum_{t=1}^T\widetilde{X}_{i,t-1}\varepsilon_{i,t}K_{t0}(\tau) \\
\frac{1}{Th}\sum_{t=1}^T\widetilde{X}_{i,t-1}\varepsilon_{i,t}K_{t1}(\tau)
\end{array}
\right]
=:\left[
\begin{array}{c}
{\boldsymbol\Gamma}_{i\varepsilon}(\tau,0) \\
{\boldsymbol\Gamma}_{i\varepsilon}(\tau,1)
\end{array}
\right],\notag
\end{eqnarray}
where $K_{tq}(\tau)=\left(\frac{\tau_t-\tau}{h}\right)^qK\left(\frac{\tau_t-\tau}{h}\right)$. From the definition of $\widehat{\beta}_{i\bullet}(\cdot)$ in (3.3), we write 
\begin{equation}\label{eqD.1}
\widehat{\beta}_{i\bullet}(\tau)-\beta_{i\bullet}(\tau)={\mathbf E}_\ast\cdot{\boldsymbol\Gamma}_{iX}^{-1}(\tau)\left[{\boldsymbol\Gamma}_{i\beta}(\tau)+{\boldsymbol\Gamma}_{i\varepsilon}(\tau)\right],
\end{equation}
where ${\mathbf E}_\ast=\left({\mathbf I}_{n_i+1}, {\mathbf O}_{(n_i+1)\times(n_i+1)}\right)$.  As in the previous proofs, we let $C$ denote a generic positive constant whose value may change from one place to another.

\medskip

\begin{lemma}\label{lem:D.1}

Suppose that Assumptions 1 and 2(i)--(iv) and (2.7) in the main text are satisfied. Then we have
\begin{equation}\label{eqD.2}
\max_{1\le i\le N}\sup_{0\leq \tau\leq 1}\left\vert{\boldsymbol\Gamma}_{iX}(\tau)-{\boldsymbol\Gamma}_{iX}^\dag(\tau)\right\vert_{\sf F}=O_P\left(\sqrt{\frac{\bar n^2\log (N\vee T)}{Th}}\right),
\end{equation}
and
\[{\boldsymbol\Gamma}_{iX}^\dag(\tau)=\left[
\begin{array}{cc}
\frac{1}{T}\sum_{t=1}^T{\sf E}\left(\widetilde{X}_{i,t-1}\widetilde{X}_{i,t-1}^{^\intercal}\right)K_{t0}(\tau)\ &\ \frac{1}{T}\sum_{t=1}^T{\sf E}\left(\widetilde{X}_{i,t-1}\widetilde{X}_{i,t-1}^{^\intercal}\right)K_{t1}(\tau)\\
\frac{1}{T}\sum_{t=1}^T{\sf E}\left(\widetilde{X}_{i,t-1}\widetilde{X}_{i,t-1}^{^\intercal}\right)K_{t1}(\tau)\ &\ \frac{1}{T}\sum_{t=1}^T{\sf E}\left(\widetilde{X}_{i,t-1}\widetilde{X}_{i,t-1}^{^\intercal}\right)K_{t2}(\tau)
\end{array}
\right]\]
is positive definite uniformly over $\tau\in[0,1]$.

\end{lemma}

\smallskip

\noindent{\bf Proof of Lemma \ref{lem:D.1}}.\ \ We start with the proof of (\ref{eqD.2}). We only prove the uniform convergence for ${\boldsymbol\Gamma}_{iX}(\cdot,0)$ since the proofs for the other block-matrices ${\boldsymbol\Gamma}_{iX}(\cdot,1)$ and ${\boldsymbol\Gamma}_{iX}(\cdot,2)$ are exactly the same. Let 
{\small\begin{equation}\label{eqD.2.1}
\widehat{\boldsymbol\Delta}_{iT}(\tau)={\boldsymbol\Gamma}_{iX}(\tau,0),\quad {\boldsymbol\Delta}_{i}^\dag(\tau)=\frac{1}{T}\sum_{t=1}^T{\sf E}\left[\widetilde{X}_{i,t-1}\widetilde{X}_{i,t-1}^{^\intercal}\right]K_{t0}(\tau)=\frac{1}{T}\sum_{t=1}^T{\boldsymbol\Delta}_{i,t-1}K_{t0}(\tau),
\end{equation}}
where ${\boldsymbol\Delta}_{i,t}$ is defined as in Assumption 2(ii). Hence, we only show that 
\begin{equation}\label{eqD.3}
\max_{1\le i\le N}\sup_{0\leq \tau\leq 1}\left\vert \widehat{\boldsymbol\Delta}_{iT}(\tau)-{\boldsymbol\Delta}_{i}^\dag(\tau)\right\vert_{\sf F}=O_P\left(\sqrt{\frac{\bar n^2\log (N\vee T)}{Th}}\right).
\end{equation}
As 
\[\left\vert \widehat{\boldsymbol\Delta}_{iT}(\tau)-{\boldsymbol\Delta}_{i}^\dag(\tau)\right\vert_{\sf F}\leq n_i\left\vert \widehat{\boldsymbol\Delta}_{iT}(\tau)-{\boldsymbol\Delta}_{i}^\dag(\tau)\right\vert_\infty\leq \bar n \left\vert \widehat{\boldsymbol\Delta}_{iT}(\tau)-{\boldsymbol\Delta}_{i}^\dag(\tau)\right\vert_\infty,
\]
it is sufficient to show that
\begin{equation}\label{eqD.4}
\max_{1\le i\le N}\sup_{0\leq \tau\leq 1}\left\vert \widehat{\boldsymbol\Delta}_{iT}(\tau)-{\boldsymbol\Delta}_{i}^\dag(\tau)\right\vert_\infty=O_P\left(\sqrt{\frac{\log (N\vee T)}{Th}}\right).
\end{equation}
Let 
\[
\widehat{\boldsymbol\Delta}_{iT}^\circ(\tau)=\frac{1}{T}\sum_{t=1}^T \widetilde{X}_{i,t-1}^\circ\widetilde{X}_{i,t-1}^{\circ^\intercal} K_{t0}(\tau),\quad
{\boldsymbol\Delta}_{i}^\circ(\tau)=\frac{1}{T}\sum_{t=1}^T {\sf E}\left[\widetilde{X}_{i,t-1}^\circ\widetilde{X}_{i,t-1}^{\circ^\intercal}\right] K_{t0}(\tau),
\]
where $\widetilde X_{i,t}^\circ$ is defined similarly to $\widetilde X_{i,t}$ but with elements in $X_t$ replaced by those in $X_t^\circ$. By Lemma \ref{lem:D.2} below, to prove (\ref{eqD.4}), we only need to show that
\begin{equation}\label{eqD.5}
\max_{1\le i\le N}\sup_{0\leq \tau\leq 1}\left\vert \widehat{\boldsymbol\Delta}_{iT}^\circ(\tau)-{\boldsymbol\Delta}_{i}^\circ(\tau)\right\vert_\infty=O_P\left(\sqrt{\frac{\log (N\vee T)}{Th}}\right).
\end{equation}

\smallskip

Let $\widetilde{V}_{i,t}=\widetilde{X}_{i,t-1}^\circ\widetilde{X}_{i,t-1}^{\circ^\intercal}$. We consider covering the closed interval $[0,1]$ by some disjoint sub-intervals ${\mathcal I}_b$ with centres $\tau_b$ and length $\gamma_{NT}=h^2\sqrt{\log(N\vee T)/{Th}}/\xi_{NT}$, $1\le b\le B$, where $\xi_{NT}=T^{(q-2)/[2(q+2)]}\left[h\log(N\vee T)\right]^{1/2}$. The number of sub-intervals, $B$, is upper bounded by $\xi_{NT}/(h^{3/2}\sqrt{\log(N\vee T)/T})$. Observe that
{\small\begin{eqnarray}
\max_{1\le i\le N}\sup_{0\leq \tau\leq 1}\left\vert \widehat{\boldsymbol\Delta}_{iT}^\circ(\tau)-{\boldsymbol\Delta}_{i}^\circ(\tau)\right\vert_\infty&\le&\max_{1\le i\le N}\max_{1\le b\le B}\left\vert\frac{1}{Th}\sum_{t=1}^T\left\{\widetilde{V}_{i,t}-{\sf E}[\widetilde{V}_{i,t}]\right\}K_{t0}(\tau_b)\right\vert_\infty+\notag\\
&&\max_{1\le i\le N}\max_{1\le b\le B}\sup_{\tau\in{\cal I}_b}\left\vert\frac{1}{Th}\sum_{t=1}^T\left\{\widetilde{V}_{i,t}-{\sf E}[\widetilde{V}_{i,t}]\right\}\left[K_{t0}(\tau)-K_{t0}(\tau_b)\right]\right\vert_\infty,\notag
\end{eqnarray}}
We first prove that
\begin{eqnarray}
&&\max_{1\le i\le N}\max_{1\le b\le B}\sup_{\tau\in{\cal I}_b}\left\vert\frac{1}{Th}\sum_{t=1}^T\left\{\widetilde{V}_{i,t}-{\sf E}[\widetilde{V}_{i,t}]\right\}\left[K_{t0}(\tau)-K_{t0}(\tau_b)\right]\right\vert_\infty\notag\\
&&=O_P\left(\sqrt{\frac{\log (N\vee T)}{Th}}\right).\label{eqD.6}
\end{eqnarray}
Let 
\[
\widetilde{V}_{i,t}^\dag=\widetilde{V}_{i,t}I\left\{\vert \widetilde{V}_{i,t}\vert_\infty\le \xi_{NT}\right\}~~\text{and}~~\widetilde{V}_{i,t}^\ddag=\widetilde V_{i,t}-\widetilde{V}_{i,t}^\dag.
\] 
By Assumption 2(i), we readily have that
\[
\max_{1\le b\le B}\sup_{\tau\in{\cal I}_b}\left\vert K_{t0}(\tau)-K_{t0}(\tau_b)\right\vert=O_P\left(\gamma_{NT}h^{-2}\right).
\]
Hence, we may show that
\begin{eqnarray}
&&\max_{1\le i\le N}\max_{1\le b\le B}\sup_{\tau\in{\cal I}_b}\left\vert\frac{1}{Th}\sum_{t=1}^T\left\{\widetilde{V}_{i,t}-{\sf E}[\widetilde{V}_{i,t}]\right\}\left[K_{t0}(\tau)-K_{t0}(\tau_b)\right]\right\vert_\infty\notag\\
&&\le\gamma_{NT}h^{-2}\left\{\max_{1\le i\le N}\frac{1}{T}\sum_{t=1}^T\left\vert \widetilde{V}_{i,t}^\dag-{\sf E}[\widetilde{V}_{i,t}^\dag]\right\vert_\infty+\max_{1\le i\le N} \frac{1}{T}\sum_{t=1}^T\left\vert\widetilde{V}_{i,t}^\ddag-{\sf E}[\widetilde{V}_{i,t}^\ddag]\right\vert_\infty\right\}\notag\\
&&=O_P\left(\sqrt{\frac{\log(N\vee T)}{Th}}\right)+\gamma_{NT}h^{-2}\max_{1\le i\le N} \frac{1}{T}\sum_{t=1}^T\left\vert\widetilde{V}_{i,t}^\ddag-{\sf E}[\widetilde{V}_{i,t}^\ddag]\right\vert_\infty.\label{eqD.7}
\end{eqnarray}
It follows from (2.7) in the main text that
\begin{equation}\label{eqD.8}
\max_{1\leq i\leq N}\max_{1\leq t\leq T} \left\Vert \vert \widetilde{V}_{i,t}\vert_\infty\right\Vert_{q/2}^{q/2}\leq \theta_{N,q}^q,
\end{equation}
which, together with the Markov inequality, indicates that, for any $\eta>0$,
\begin{eqnarray}
&&{\sf P}\left(\max_{1\le i\le N} \frac{1}{T}\sum_{t=1}^T\left\vert\widetilde{V}_{i,t}^\ddag-{\sf E}[\widetilde{V}_{i,t}^\ddag]\right\vert_\infty>\eta \xi_{NT}\right)\le\sum_{i=1}^N\sum_{t=1}^T{\sf P}\left(\left\vert\widetilde{V}_{i,t}\right\vert_\infty>\xi_{NT}\right)\notag\\
&&=O\left(NT\theta_{N,q}^q\xi_{NT}^{-q/2}\right)=o(1),\label{eqD.9}
\end{eqnarray}
where the last equality is due to (3.7) in Assumption 2(iii). With (\ref{eqD.7}) and (\ref{eqD.9}), we complete the proof of (\ref{eqD.5}).

\smallskip

It remains to show that
\begin{equation}\label{eqD.10}
\max_{1\le i\le N}\max_{1\le b\le B}\left\vert\frac{1}{Th}\sum_{t=1}^T\left\{\widetilde{V}_{i,t}-{\sf E}[\widetilde{V}_{i,t}]\right\}K_{t0}(\tau_b)\right\vert_\infty=O_p\left(\sqrt{\frac{\log (N\vee T)}{Th}}\right).
\end{equation}
Letting $\widetilde V_{i,t}(\tau)=\widetilde{V}_{i,t}K_{t0}(\tau)$ for notational simplicity, as $K(\cdot)$ has the support $[-1,1]$,
\[
\sum_{t=1}^T \widetilde V_{i,t}(\tau)=\sum_{t=T_1(\tau)}^{T_2(\tau)} \widetilde V_{i,t}(\tau)=\sum_{t=T_1(\tau)}^{T_2(\tau)}\widetilde{X}_{i,t-1}^\circ\widetilde{X}_{i,t-1}^{\circ^\intercal}K_{t0}(\tau),
\]
where $T_1(\tau)=\lfloor T\tau\rfloor-\lfloor Th\rfloor+1$ and $T_2(\tau)=\lfloor T\tau\rfloor+\lfloor Th\rfloor$. We next adopt some standard techniques in the literature on high-dimensional locally stationary processes \citep[e.g.,][]{ZW21} to prove (\ref{eqD.10}).  Let $M=2\lfloor Th\rfloor, M^\dagger=\lfloor\log M/\log 2\rfloor,u_l=2^l$ for $1\le l\le M^\dagger-1$ and $u_{M^\dagger}=M$. Define
$\widetilde V_{i,t,u}(\tau)={\sf E} [\widetilde V_{i,t}(\tau)\big|\mathscr{F}_{t-u}^{t-1}]$ with $\mathscr{F}_s^t=(\varepsilon_{s},\cdots,\varepsilon_{t})$, and
\[
S_{i,l,T}(\tau)=\sum_{t=T_1(\tau)}^{T_2(\tau)}\left[\widetilde V_{i,t,u_l}(\tau)-\widetilde V_{i,t,u_{l-1}}(\tau)\right],\ \ l=1,\cdots,M^\dagger.
\]
Then, for any $1\leq i\leq N$ and $1\leq b\leq B$, we may decompose
\begin{eqnarray}
\frac{1}{Th}\sum_{t=1}^T\left\{\widetilde{V}_{i,t}-{\sf E}\left[\widetilde{V}_{i,t}\right]\right\}K_{t0}(\tau_b)&=&\sum_{t=T_1(\tau_b)}^{T_2(\tau_b)} \left\{\widetilde V_{i,t}(\tau_b)-{\sf E}\left[\widetilde V_{i,t}(\tau_b)\right]\right\}\notag\\
&=&\frac{1}{Th}\left(\sum_{t=T_1(\tau_b)}^{T_2(\tau_b)}\left[\widetilde V_{i,t}(\tau_b)-\widetilde V_{i,t,M}(\tau_b)\right]+\sum_{l=2}^{M^\dagger} S_{i,l,T}(\tau_b)+\right.\notag\\
&&\left.\sum_{t=T_1(\tau_b)}^{T_2(\tau_b)}\left\{\widetilde V_{i,t,2}(\tau_b)-{\sf E}\left[\widetilde V_{i,t,2}(\tau_b)\right]\right\}\right).\label{eqD.11}
\end{eqnarray}

\smallskip

We next separately tackle the three terms on the right side of (\ref{eqD.11}). Since 
\[
\widetilde V_{i,t}(\tau)-\widetilde V_{i,t,M}(\tau)=\sum_{k=M+1}^\infty\left[\widetilde V_{i,t,k}(\tau)-\widetilde V_{i,t,k-1}(\tau)\right],\] 
we have
\begin{eqnarray}
&&\left\Vert\Bigg\vert\sum_{t=T_1(\tau_b)}^{T_2(\tau_b)}\left[\widetilde V_{i,t}(\tau_b)-\widetilde V_{i,t,M}(\tau_b)\right]\Bigg\vert_\infty\right\Vert_{q/2}\notag\\
&=&\left\Vert\Bigg\vert\sum_{t=T_1(\tau_b)}^{T_2(\tau_b)}\sum_{k=M+1}^\infty\left[\widetilde V_{i,t,k}(\tau_b)-\widetilde V_{i,t,k-1}(\tau_b)\right]\Bigg\vert_\infty\right\Vert_{q/2}\notag\\
&\le&C\sum_{k=M+1}^\infty\left\Vert\Bigg\vert\sum_{t=T_1(\tau_b)}^{T_2(\tau_b)}\left[\widetilde V_{i,t,k}(\tau_b)-\widetilde V_{i,t,k-1}(\tau_b)\right]\Bigg\vert_\infty\right\Vert_{q/2}.\notag
\end{eqnarray}
Note that, for any $1\leq i\leq N$, $1\leq b\leq B$ and $k\ge M+1$, ${\sf vec}(\widetilde V_{i,t,k}(\tau_b)-\widetilde V_{i,t,k-1}(\tau_b)),~T_1(\tau)\le t\le T_2(\tau)$, is a sequence of martingale difference vectors. By Lemma D.3 of \cite{ZW21}, we have 
\begin{eqnarray}
&&\sum_{k=M+1}^\infty\left\Vert\Bigg\vert\sum_{t=T_1(\tau_b)}^{T_2(\tau_b)}\left(\widetilde V_{i,t,k}(\tau_b)-\widetilde V_{i,t,k-1}(\tau_b)\right)\Bigg\vert_\infty\right\Vert_{q/2}\notag\\
&=&\sum_{k=M+1}^\infty\left\Vert\Bigg\vert\sum_{t=T_1(\tau_b)}^{T_2(\tau_b)}{\sf vec}\left(\widetilde V_{i,t,k}(\tau_b)-\widetilde V_{i,t,k-1}(\tau_b)\right)\Bigg\vert_\infty\right\Vert_{q/2}\notag\\
&\le&C\sum_{k=M+1}^\infty \sqrt{\sum_{t=T_1(\tau_b)}^{T_2(\tau_b)}\left\Vert\bigg\vert{\sf vec}\left( \widetilde V_{i,t,k}(\tau_b)-\widetilde V_{i,t,k-1}(\tau_b)\right)\bigg\vert_{\infty}\right\Vert_{q/2}^2}\notag\\
&=&C\sum_{k=M+1}^\infty \sqrt{\sum_{t=T_1(\tau_b)}^{T_2(\tau_b)}\left\Vert\bigg\vert \widetilde V_{i,t,k}(\tau_b)-\widetilde V_{i,t,k-1}(\tau_b)\bigg\vert_{\infty}\right\Vert_{q/2}^2}.\notag
\end{eqnarray}
With the triangle inequality, 
\begin{eqnarray}
&&\left\Vert\left\vert \widetilde V_{i,t,k}(\tau_b)-\widetilde V_{i,t,k-1}(\tau_b)\right\vert_{\infty}\right\Vert_{q/2}\notag\\
&=&\left\Vert\bigg\vert{\sf E}\left(\widetilde V_{i,t}(\tau_b)-\widetilde V_{i,t}^{\{t-k\}}(\tau_b)\big|\mathscr{F}_{t-k}^{t-1}\right)\bigg\vert_\infty\right\Vert_{q/2}\notag\\
&\le&\left\Vert\bigg\vert\left(\widetilde{X}_{i,t-1}^\circ-\widetilde{X}_{i,t-1}^{\circ\{t-k\}}\right)\widetilde{X}_{i,t-1}^{^\intercal}K_{t0}(\tau_b)\bigg\vert_\infty\right\Vert_{q/2}+\notag\\
&&\left\Vert\bigg\vert \widetilde{X}_{i,t-1}^{\circ\{t-k\}}\left(\widetilde{X}_{i,t-1}^\circ-\widetilde{X}_{i,t-1}^{\circ\{t-k\}}\right)^{^\intercal}K_{t0}(\tau_b)\bigg\vert_\infty\right\Vert_{q/2},\notag
\end{eqnarray}
where $\widetilde V_{i,t}^{\{l\}}(\tau)=\widetilde{X}_{i,t-1}^{\circ\{l\}}\widetilde{X}_{i,t-1}^{\circ\{l\}^{^\intercal}}K_{t0}(\tau)$ and 
\[
\widetilde{X}_{i,t}^{\circ\{l\}}=\left[\left(\widetilde{w}_{ij}x_{j,t}^{\circ\{l\}}:\ j\in{\mathscr N}_i\right)^{^\intercal},\ x_{i,t}^{\circ\{l\}}\right]^{^\intercal}
\] 
with $x_{i,t}^{\circ\{l\}}$ being the $i$-th element of $X_{t}^{\circ\{l\}}=X_{t}^{\circ\{l\}}(\tau_t)$ defined in Appendix \ref{appB}. Following the discussion in Appendix \ref{appB} and noting that $\sum_{j\in{\mathscr N}_i}\widetilde w_{ij}=1$, we have
\[
\left\Vert\left\vert\widetilde{X}_{i,t-1}^\circ-\widetilde{X}_{i,t-1}^{\circ\{t-k\}}\right\vert_\infty\right\Vert_{q}\leq \delta_{i,k-1,q}+\sum_{j\in{\mathscr N}_i}\widetilde w_{ij}\delta_{j,k-1,q},
\] 
and 
\begin{eqnarray}
\left\Vert\left\vert\widetilde{X}_{i,t-1}^\circ K_{t0}(\tau_b)\right\vert_\infty\right\Vert_q&\le&\sum_{k=0}^\infty\left\Vert\bigg\vert \left[{\sf E}(\widetilde{X}_{i,t-1}^\circ\big|\mathscr{F}_{t-k-1})-{\sf E}(\widetilde{X}_{i,t-1}^\circ\big|\mathscr{F}_{t-k-2})\right]K_{t0}(\tau_b)\bigg\vert_\infty\right\Vert_{q}\notag\\
&\le&C\sum_{k=0}^\infty\left(\delta_{i,k,q}+\sum_{j\in{\mathscr N}_i}\widetilde w_{ij}\delta_{j,k,q}\right)\notag\\
&\leq&C\max_{1\leq i\leq N}\left\Vert x_{i\bullet}\right\Vert_{q,\iota}\notag
\end{eqnarray}
which is bounded, where $\mathscr{F}_t=(\cdots,\varepsilon_{t-1},\varepsilon_{t})$ as in Section 2 of the main text. As discussed in Appendix \ref{appB}, we set $\iota=1$ in the rest of the proof. By the Jensen and Cauchy-Schwarz inequalities, we have 
\begin{eqnarray}
&&\left\Vert\left\vert\left(\widetilde{X}_{i,t-1}^\circ-\widetilde{X}_{i,t-1}^{\circ\{t-k\}}\right)\widetilde{X}_{i,t-1}^{\circ^\intercal}K_{t0}(\tau_b)\right\vert_\infty\right\Vert_{q/2}\notag\\
&\leq&\left\Vert\left\vert\widetilde{X}_{i,t-1}^\circ-\widetilde{X}_{i,t-1}^{\circ\{t-k\}}\right\vert_\infty\right\Vert_q\cdot\left\Vert\left\vert\widetilde{X}_{i,t-1}^\circ K_{t0}(\tau_b)\right\vert_\infty\right\Vert_q\notag\\
&\le& C\left(\delta_{i,k-1,q}+\sum_{j\in{\mathscr N}_i}\widetilde w_{ij}\delta_{j,k-1,q}\right),\notag
\end{eqnarray}
and similarly,
\[
\left\Vert\left\vert \widetilde{X}_{i,t-1}^{\circ\{t-k\}}\left(\widetilde{X}_{i,t-1}^\circ-\widetilde{X}_{i,t-1}^{\circ\{t-k\}}\right)^{^\intercal}K_{t0}(\tau_b)\right\vert_\infty\right\Vert_{q/2}\leq C\left(\delta_{i,k-1,q}+\sum_{j\in{\mathscr N}_i}\widetilde w_{ij}\delta_{j,k-1,q}\right).
\]
Since $M=2\lfloor Th\rfloor$, we have 
\begin{eqnarray}
&&\sum_{k=M+1}^\infty \sqrt{\sum_{t=T_1(\tau_b)}^{T_2(\tau_b)}\left\Vert\left\vert\notag \widetilde V_{i,t,k}(\tau_b)-\widetilde V_{i,t,k-1}(\tau_b)\right\vert_{\infty}\right\Vert_{q/2}^2}\notag\\
&\le&C(Th)^{1/2}\sum_{k=M+1}^\infty\left(\delta_{i,k-1,q}+\sum_{j\in{\mathscr N}_i}\widetilde w_{ij}\delta_{j,k-1,q}\right)\notag\\
&=&C(Th)^{1/2}\left(\Delta_{i,M,q}+\sum_{j\in{\mathscr N}_i}\widetilde w_{ij}\Delta_{j,M,q}\right)\notag\\
&\le&C(Th)^{-1/2}\max_{1\leq i\leq N}\left\Vert x_{i\bullet}\right\Vert_{q,1},\label{eqD.12}
\end{eqnarray}
where the last inequality is due to the fact that 
\[
\max_{1\leq i\leq N}\Delta_{i,M,q}\leq M^{-1}\max_{1\leq i\leq N}\left\Vert x_{i\bullet}\right\Vert_{q,1}
\]
using the definition of $\left\Vert x_{i\bullet}\right\Vert_{q,\iota}$ and setting $\iota=1$. Then, by (\ref{eqD.12}) and the Markov inequality, we have for any $z>0$,
\begin{equation}\label{eqD.13}
{\sf P}\left(\left\vert\sum_{t=T_1(\tau_b)}^{T_2(\tau_b)}\left[\widetilde V_{i,t}(\tau_b)-\widetilde V_{i,t,M}(\tau_b)\right]\right\vert_\infty>z\right)\le \frac{C}{z^{q/2}(Th)^{q/4}}\max_{1\leq i\leq N}\left\Vert x_{i\bullet}\right\Vert_{q,1}^{q/2},
\end{equation}
where $C$ is a positive constant independent of $1\leq i\leq N$ and $1\leq b\leq B$.

\smallskip

We next turn to the second term on the right side of (\ref{eqD.11}). For each $2\le l\le M^\dagger$, define
\[
U_{i,k,l}(\tau)=\sum_{t=(k-1)u_l+T_1(\tau)}^{(ku_l+T_1(\tau))\wedge T_2(\tau)}\left[\widetilde V_{i,t,u_l}(\tau)-\widetilde V_{i,t,u_{l-1}}(\tau)\right],\ \ 1\le k\le \lceil M/u_l\rceil,
\]
and
\[
S_{i,l,T}^{\rm e}(\tau)=\sum_{k \text{ is even}}U_{i,k,l}(\tau), ~~ S_{i,l,T}^{\rm o}(\tau)=\sum_{k \text{ is odd}}U_{i,k,l}(\tau).
\]
Let $\lambda_l=(l-1)^{-2}/(\pi^2/3)$ if $2\le l\le M^\dagger/2$ and $\lambda_l=(M^\dagger+1-l)^{-2}/(\pi^2/3)$ if $M^\dagger/2<l<M^\dagger$. It is easy to verify that $\sum_{l=2}^{M^\dagger}\lambda_l\le1$. Since $U_{i,k_1,l}(\tau)$ and $U_{i,k_2,l}(\tau)$ are independent for $\vert k_1-k_2\vert>1$, by Lemma D.4 of \cite{ZW21}, we have, for any $z>0$,
\begin{eqnarray}
{\sf P}\left(\left\vert S_{i,l,T}^{\rm e}(\tau_b)\right\vert_\infty\ge\lambda_lz\right)&\le& C\left((\lambda_lz)^{-q/2}\sum_{k\text{ is even}}{\sf E}\left[\vert U_{i,k,l}(\tau_b)\vert_\infty^{q/2}\right]+\right.\notag\\
&&\left.\exp\left\{-\frac{(\lambda_lz)^2}{C\sum\limits_{k\text{ is even}}{\sf E}\left[\vert U_{i,k,l}(\tau_b)\vert_\infty^2\right]}\right\}\right),\notag
\end{eqnarray}
where $2\le l\le M^\dagger$. Similarly to the proof of (\ref{eqD.12}), we may show that 
\[
\Vert \vert U_{i,j,l}(\tau)\vert_\infty\Vert_{q/2}\leq C\cdot u_l^{-1/2}\max_{1\leq i\leq N}\left\Vert x_{i\bullet}\right\Vert_{q,1}
\]
and
\[
\Vert\vert U_{i,j,l}(\tau)\vert_\infty\Vert_2\leq C\cdot u_l^{-1/2}\max_{1\leq i\leq N}\left\Vert x_{i\bullet}\right\Vert_{4,1}.
\]
Similar results also hold for $\vert S_{i,l,T}^{\rm o}(\tau)\vert_\infty$ with details omitted to save space. A combination of the above arguments yields that
\begin{eqnarray}
{\sf P}\left(\left\vert\sum_{l=2}^{M^\dagger}S_{i,l,T}(\tau_b)\right\vert_\infty\ge2z\right)
&\le&\sum_{l=2}^{M^\dagger}{\sf P}\left(\vert S_{i,l,T}(\tau_b)\vert_{\infty}\ge2\lambda_l z\right)\notag\\
&\le&\sum_{l=2}^{M^\dagger}{\sf P}\left(\vert S_{i,l,T}^{\rm e}(\tau_b)\vert_\infty\ge\lambda_lz\right)+\sum_{l=2}^{M^\dagger}{\sf P}\left(\vert S_{i,l,T}^{\rm o}(\tau_b)\vert_\infty\ge\lambda_lz\right)\notag\\
&\le&C\left\{\frac{Th}{z^{q/2}}\cdot\sum_{l=2}^{M^\dagger}u_l^{-(q/4+1)}\lambda_l^{-q/2}\left(\max_{1\leq i\leq N}\left\Vert x_{i\bullet}\right\Vert_{q,1}^{q/2}\right)+\right.\notag\\
&&\left.\sum_{l=2}^{M^\dagger}\exp\left(-\frac{\lambda_l^2u_l^{2}z^2}{C(Th)\max\limits_{1\leq i\leq N}\left\Vert x_{i\bullet}\right\Vert_{4,1}^2}\right)\right\}.\notag
\end{eqnarray}
Making use of the fact that $\sum_{l=2}^{M^\dagger}\lambda_l^{-q/2}u_l^{-(q/4+1)}$ is bounded and $\min_{l\ge 1}\lambda_l^2u_l^{2}>0$, we have
\begin{eqnarray}
{\sf P}\left(\left\vert\sum_{l=2}^{M^\dagger}S_{i,l,T}(\tau_b)\right\vert_\infty\ge2z\right)&\le& C\left(\frac{Th}{z^{q/2}}\left(\max_{1\leq i\leq N}\left\Vert x_{i\bullet}\right\Vert_{q,1}^{q/2}\right)+\right.\notag\\
&&\left.\exp\left\{-\frac{z^2}{C(Th)}\left(\max\limits_{1\leq i\leq N}\left\Vert x_{i\bullet}\right\Vert_{4,1}^2\right)^{-1}\right\}\right),\label{eqD.14}
\end{eqnarray}
where $C$ is a positive constant independent of $1\leq i\leq N$ and $1\leq b\leq B$.

\smallskip

We next tackle the third term on the right side of (\ref{eqD.11}). Note that
\[
\widetilde V_{i,t,2}(\tau)-{\sf E}\left[\widetilde V_{i,t,2}(\tau)\right]=\sum_{s=-\infty}^t\left\{{\sf E}\left[\widetilde V_{i,t,2}(\tau)\big|\mathscr{F}_{s-1}\right]-{\sf E}\left[\widetilde V_{i,t,2}(\tau)\big|\mathscr{F}_{s-2}\right]\right\}.
\]
For any $s$, by the triangle and Jensen inequalities, we have
{\small\begin{eqnarray}
&&\left\Vert\bigg\vert{\sf E}\left[\widetilde V_{i,t,2}(\tau_b)\big|\mathscr{F}_{s-1}\right]-{\sf E}\left[\widetilde V_{i,t,2}(\tau_b)\big|\mathscr{F}_{s-2}\right]\bigg\vert_\infty\right\Vert_{q/2}\notag\\
&\le&\left\Vert\left\vert\left(\widetilde{X}_{i,t-1}^\circ-\widetilde{X}_{i,t-1}^{\circ\{s-1\}}\right)\widetilde{X}_{i,t-1}^{\circ^\intercal}K_{t0}(\tau_b)\right\vert_\infty\right\Vert_{q/2}+\left\Vert\left\vert \widetilde{X}_{i,t-1}^{\circ\{s-1\}}\left(\widetilde{X}_{i,t-1}^\circ-\widetilde{X}_{i,t-1}^{\circ\{s-1\}}\right)^{^\intercal}K_{t0}(\tau_b)\right\vert_\infty\right\Vert_{q/2}\notag\\
&\le&C\left(\delta_{i,t-s,q}+\sum_{j\in{\mathscr N}_i}\widetilde w_{ij}\delta_{j,t-s,q}\right).\notag
\end{eqnarray}}
Consequently,
\[
\left\Vert\left\vert \widetilde V_{i,t,2}(\tau_b)-{\sf E}\left[\widetilde V_{i,t,2}(\tau_b)\right]\right\vert_\infty\right\Vert_{q/2}\le C\sum_{s=-\infty}^t \left(\delta_{i,t-s,q}+\sum_{j\in{\mathscr N}_i}\widetilde w_{ij}\delta_{j,t-s,q}\right)\le C\max_{1\leq i\leq N}\Delta_{i,0,q},
\]
and
\begin{equation}\label{eqD.15}
\sum_{t=T_1(\tau_b)}^{T_2(\tau_b)}\left\Vert\left\vert \widetilde V_{i,t,2}(\tau_b)-{\sf E}\left[\widetilde V_{i,t,2}(\tau_b)\right]\right\vert_\infty\right\Vert_{q/2}^{q/2}\le  C(Th)\max_{1\leq i\leq N}\left\Vert x_{i\bullet}\right\Vert_{q,\alpha}^{q/2}.
\end{equation}
Note that $\widetilde V_{i,t_1,2}(\tau)$ and $\widetilde V_{i,t_2,2}(\tau)$ are independent when $\vert t_1-t_2\vert>2$. Following the proof of (\ref{eqD.14}) and using (\ref{eqD.15}), we have
\begin{eqnarray}
&&{\sf P}\left(\bigg\vert\sum_{t=T_1(\tau_b)}^{T_2(\tau_b)}\left\{\widetilde V_{i,t,2}(\tau_b)-{\sf E}\left[\widetilde V_{i,t,2}(\tau_b)\right]\right\}\bigg\vert_\infty\ge z\right)\notag\\
&\le&C\left(\frac{Th}{z^{q/2}}\left(\max_{1\leq i\leq N}\left\Vert x_{i\bullet}\right\Vert_{q,1}^{q/2}\right)+\exp\left\{-\frac{z^2}{C(Th)}\left(\max\limits_{1\leq i\leq N}\left\Vert x_{i\bullet}\right\Vert_{4,1}^{2}\right)^{-1}\right\}\right),\label{eqD.16}
\end{eqnarray}
where $C$ is a positive constant independent of $1\leq i\leq N$ and $1\leq b\leq B$.

\smallskip

By virtue of (\ref{eqD.11}), (\ref{eqD.13}), (\ref{eqD.14}) and (\ref{eqD.16}), we readily have that
\begin{eqnarray}
&&{\sf P}\left(\left\vert \sum_{t=1}^T\left\{\widetilde{V}_{i,t}-{\sf E}[\widetilde{V}_{i,t}]\right\}K_{t0}(\tau_b)\right\vert_\infty>z\right)\notag\\
&\le& C\left(\frac{Th}{z^{q/2}}\left(\max_{1\leq i\leq N}\left\Vert x_{i\bullet}\right\Vert_{q,1}^{q/2}\right)+\exp\left\{-\frac{z^2}{C(Th)}\left(\max\limits_{1\leq i\leq N}\left\Vert x_{i\bullet}\right\Vert_{4,1}^{2}\right)^{-1}\right\}\right).\label{eqD.17}
\end{eqnarray}
As discussed in Appendix \ref{appB}, $\left\Vert x_{i\bullet}\right\Vert_{q,1}+\left\Vert x_{i\bullet}\right\Vert_{4,1}$ is bounded uniformly over $i$. Setting $z=\eta\sqrt{Th\log (N\vee T)}$ in (\ref{eqD.17}), we may show that, for any $\eta>0$,
\begin{eqnarray}
&&{\sf P}\left(\max_{1\le i\le N}\max_{1\le b\le B}\left\vert\frac{1}{Th}\sum_{t=1}^T\left\{\widetilde{V}_{i,t}-{\sf E}[\widetilde{V}_{i,t}]\right\}K_{t0}(\tau_b)\right\vert_\infty>\eta\sqrt{\frac{\log (N\vee T)}{Th}}\right)\notag\\
&\le&\sum_{i=1}^N\sum_{b=1}^B{\sf P}\left( \left\vert \sum_{t=1}^T\left\{\widetilde{V}_{i,t}-{\sf E}[\widetilde{V}_{i,t}]\right\}K_{t0}(\tau_b)\right\vert_\infty>\eta\sqrt{Th\log (N\vee T)}\right)\notag\\
&=&O\left(NB\exp\left\{-C\eta^2\log (N\vee T)\right\}\right)+O\left(\frac{NB}{\eta^{q/2}(Th)^{q/4-1}(\log (N\vee T))^{q/4}}\right).\label{eqD.18}
\end{eqnarray}
Noting that $B$ diverge at a certain polynomial rate of $T$ and letting $\eta$ be sufficiently large, the first order on the right side of (\ref{eqD.18}) converges to zero. By (3.7) in Assumption 2(iii), we may show that the second order also converges to zero. The proof of (\ref{eqD.10}) is completed. 

\smallskip

By Assumption 2(i)(ii), it is easy to show that ${\boldsymbol\Gamma}_{iX}^\dag(\tau)$ is positive definite uniformly over $0\leq \tau\leq 1$. The proof of Lemma \ref{lem:D.1} is completed.\hfill$\Box$

\medskip

\begin{lemma}\label{lem:D.2}

Suppose that Assumptions 1 and 2(i)(iii) and (2.7) in the main text are satisfied. Let 
\[
\widehat{\boldsymbol\Delta}_{iT}^\circ(\tau)=\frac{1}{T}\sum_{t=1}^T \widetilde{X}_{i,t-1}^\circ\widetilde{X}_{i,t-1}^{\circ^\intercal} K_{t0}(\tau),\quad
{\boldsymbol\Delta}_{i}^\circ(\tau)=\frac{1}{T}\sum_{t=1}^T {\sf E}\left[\widetilde{X}_{i,t-1}^\circ\widetilde{X}_{i,t-1}^{\circ^\intercal}\right] K_{t0}(\tau),
\]
where $\widetilde X_{i,t}^\circ$ is defined similarly to $\widetilde X_{i,t}$ in Section 3.1 but with elements in $X_t$ replaced by those in $X_t^\circ$. Then we have
\begin{equation}\label{eqD.19}
\max_{1\le i\le N}\sup_{0\leq \tau\leq 1}\left\vert \widehat{\boldsymbol\Delta}_{iT}(\tau)-\widehat{\boldsymbol\Delta}_{iT}^\circ(\tau)\right\vert_{\infty}=o_P\left(\sqrt{\frac{\log (N\vee T)}{Th}}\right)
\end{equation}
and
\begin{equation}\label{eqD.20}
\max_{1\le i\le N}\sup_{0\leq \tau \leq 1}\left\vert {\boldsymbol\Delta}_{i}^\dag(\tau)-{\boldsymbol\Delta}_{i}^\circ(\tau)\right\vert_{\infty}=o\left(\sqrt{\frac{\log (N\vee T)}{Th}}\right),
\end{equation}
where $\widehat{\boldsymbol\Delta}_{iT}(\tau)$ and ${\boldsymbol\Delta}_{i}^\dag(\tau)$ are defined in (\ref{eqD.2.1}).

\end{lemma}

\smallskip

\noindent{\bf Proof of Lemma \ref{lem:D.2}}.\ \ With (2.7), by slightly modifying the proof of (2.8) in Appendix \ref{appB}, we may show that
\begin{equation}\label{eqD.21}
\max_{0\leq t\leq T-1} \left\Vert \left\vert X_t-X_t^\circ\right\vert_\infty\right\Vert_q=O\left(\theta_{N,q}/T\right).
\end{equation}
By (2.7), (\ref{eqD.21}), the classic $c_r$- and Cauchy-Schwarz inequalities, we have
{\small\begin{eqnarray}
&&\max_{1\leq i\leq N}\max_{1\leq t\leq T}\left\Vert \left\vert\widetilde{X}_{i,t-1}\widetilde{X}_{i,t-1}^{^\intercal} -\widetilde{X}_{i,t-1}^\circ\widetilde{X}_{i,t-1}^{\circ^\intercal}\right\vert_\infty \right\Vert_{q/2}^{q/2}\notag\\
%&\leq&C\max_{1\leq i\leq N}\max_{1\leq t\leq T}\left\{\left\Vert \left\vert(\widetilde{X}_{i,t-1}-\widetilde{X}_{i,t-1}^\circ)\widetilde{X}_{i,t-1}^{^\intercal}\right\vert_\infty \right\Vert_{q/2}^{q/2}+\left\Vert \left\vert\widetilde{X}_{i,t-1}^{\circ}(\widetilde{X}_{i,t-1}-\widetilde{X}_{i,t-1}^\circ)^{^\intercal}\right\vert_\infty \right\Vert_{q/2}^{q/2}\right\}\notag\\
&\leq&C\max_{1\leq i\leq N}\max_{1\leq t\leq T}\left\Vert \left\vert\widetilde{X}_{i,t-1}-\widetilde{X}_{i,t-1}^\circ\right\vert_\infty \right\Vert_{q}^{q/2}
\left[\left\Vert\left\vert \widetilde{X}_{i,t-1}\right\vert_\infty\right\Vert_q^{q/2}+\left\Vert\left\vert \widetilde{X}_{i,t-1}^\circ\right\vert_\infty\right\Vert_q^{q/2}\right]\notag\\
&\leq&C\max_{0\leq t\leq T-1}\left\Vert \left\vert X_{t}-X_{t}^\circ\right\vert_\infty \right\Vert_{q}^{q/2}
\left[\left\Vert\left\vert X_{t}\right\vert_\infty\right\Vert_q^{q/2}+\left\Vert\left\vert X_{t}^\circ\right\vert_\infty\right\Vert_q^{q/2}\right]\notag\\
&=&O\left(\theta_{N,q}^qT^{-q/2}\right).\label{eqD.22}
\end{eqnarray}}
With (\ref{eqD.22}), the Bonferroni and Markov inequalities and (3.7) in Assumption 2(iii), we may show that, for any $\eta>0$,
\begin{eqnarray}
&&{\sf P}\left(\max_{1\leq i\leq N}\max_{1\leq t\leq T} \left\vert\widetilde{X}_{i,t-1}\widetilde{X}_{i,t-1}^{^\intercal} -\widetilde{X}_{i,t-1}^\circ\widetilde{X}_{i,t-1}^{\circ^\intercal}\right\vert_\infty>\eta \sqrt{\frac{\log (N\vee T)}{Th}}\right)\notag\\
&\leq&\sum_{i=1}^N\sum_{t=1}^T{\sf P}\left(\left\vert\widetilde{X}_{i,t-1}\widetilde{X}_{i,t-1}^{^\intercal} -\widetilde{X}_{i,t-1}^\circ\widetilde{X}_{i,t-1}^{\circ^\intercal}\right\vert_\infty>\eta \sqrt{\frac{\log (N\vee T)}{Th}}\right)\notag\\
&\leq&\left(\eta^2 \frac{\log (N\vee T)}{Th}\right)^{-q/4}\sum_{i=1}^N\sum_{t=1}^T \left\Vert \left\vert\widetilde{X}_{i,t-1}\widetilde{X}_{i,t-1}^{^\intercal} -\widetilde{X}_{i,t-1}^\circ\widetilde{X}_{i,t-1}^{\circ^\intercal}\right\vert_\infty \right\Vert_{q/2}^{q/2}\notag\\
&\leq&C\eta^{-q/2}\cdot\frac{N\theta_{N,q}^qh^{q/4}}{T^{q/4-1}[\log (N\vee T)]^{q/4}}=o(1),\notag
\end{eqnarray}
indicating that
\begin{equation}\label{eqD.23}
\max_{1\leq i\leq N}\max_{1\leq t\leq T} \left\vert\widetilde{X}_{i,t-1}\widetilde{X}_{i,t-1}^{^\intercal} -\widetilde{X}_{i,t-1}^\circ\widetilde{X}_{i,t-1}^{\circ^\intercal}\right\vert_\infty=o_P\left(\sqrt{\frac{\log (N\vee T)}{Th}}\right).
\end{equation}
By (\ref{eqD.23}) and Assumption 2(i), we may show that 
\begin{eqnarray}
&&\max_{1\le i\le N}\sup_{0\leq \tau\leq 1}\left\vert \widehat{\boldsymbol\Delta}_{iT}(\tau)-\widehat{\boldsymbol\Delta}_{iT}^\circ(\tau)\right\vert_{\infty}\notag\\
&\leq&\max_{1\le i\le N}\sup_{0\leq \tau\leq 1}\frac{1}{T}\sum_{t=1}^T \left\vert\widetilde{X}_{i,t-1}\widetilde{X}_{i,t-1}^{^\intercal} -\widetilde{X}_{i,t-1}^\circ\widetilde{X}_{i,t-1}^{\circ^\intercal}\right\vert_\infty K_{t0}(\tau)\notag\\
&\leq&\max_{1\leq i\leq N}\max_{1\leq s\leq T} \left\vert\widetilde{X}_{i,s-1}\widetilde{X}_{i,s-1}^{^\intercal} -\widetilde{X}_{i,s-1}^\circ\widetilde{X}_{i,s-1}^{\circ^\intercal}\right\vert_\infty \sup_{0\leq \tau\leq 1}\frac{1}{T}\sum_{t=1}^T K_{t0}(\tau)\notag\\
&=&o_P\left(\sqrt{\frac{\log (N\vee T)}{Th}}\right),\notag
\end{eqnarray}
completing the proof of (\ref{eqD.19}). With (\ref{eqD.23}) and Assumption 2(i), we can also prove (\ref{eqD.20}). \hfill$\Box$

\medskip

In the following lemma, we derive the uniform consistency of the local linear estimation of the heterogenous time-varying coefficient functions defined in (3.3).

\begin{lemma}\label{lem:D.3}

Suppose that Assumptions 1 and 2(i)--(iv) and (2.7) in the main text are satisfied. Letting $\widehat\beta_{i\bullet}(\cdot)$ be defined in (3.3), we have
\begin{equation}\label{eqD.24}
\max_{1\le i\le N}\max_{1\leq l\leq L}\left\vert \widehat\beta_{i\bullet}(\tau_l^\ast)-\beta_{i\bullet}(\tau_l^\ast)\right\vert_2=O_P\left(\sqrt{\bar n}\left(\sqrt{\frac{\log (N\vee T)}{Th}}+h^2\right)\right),
\end{equation}
where the number of grid points $L$ satisfies $L=O(T)$.

\end{lemma}

\smallskip

\noindent{\bf Proof of Lemma \ref{lem:D.3}}.\ \ By Lemma \ref{lem:D.1} and Assumptions 1(i) and 2(ii), we may show that 
\begin{equation}\label{eqD.25}
\max_{1\le i\le N}\max_{1\leq l\leq L}\lambda_{\max}\left({\boldsymbol\Gamma}_{iX}^{-1}(\tau_l^\ast)\right)=O_P(1)
\end{equation}
and
\begin{equation}\label{eqD.26}
\max_{1\le i\le N}\max_{1\leq l\leq L}\left\vert {\boldsymbol\Gamma}_{iX}^{-1}(\tau_l^\ast){\boldsymbol\Gamma}_{i\beta}(\tau_l^\ast)\right\vert_2=O_P\left(\sqrt{\bar n}h^2\right).
\end{equation}
In view of (\ref{eqD.25}) and (\ref{eqD.26}), in order to prove (\ref{eqD.24}), we only need to show that 
\[
\max_{1\le i\le N}\max_{1\leq l\leq L}\left\vert {\boldsymbol\Gamma}_{i\varepsilon}(\tau_l^\ast)\right\vert_2=O_P\left(\sqrt{\frac{\bar n\log (N\vee T)}{Th}}\right).
\]
As in the proof of Lemma \ref{lem:D.1}, to save the space, we only provide the proof of
\begin{equation}\label{eqD.27}
\max_{1\le i\le N}\max_{1\leq l\leq L}\left\vert\frac{1}{Th}\sum_{t=1}^T\widetilde{X}_{i,t-1}\varepsilon_{i,t}K_{t0}(\tau_l^\ast)\right\vert_2=O_P\left(\sqrt{\frac{\bar n\log (N\vee T)}{Th}}\right).
\end{equation}

\smallskip

We next use the truncation technique and the concentration inequality for martingale to prove (\ref{eqD.27}). Let $\xi_{NT}=T^{(q-2)/[2(q+2)]}\left[h\log(N\vee T)\right]^{1/2}$ be defined as in the proof of Lemma \ref{lem:D.1}, $\widetilde W_{i,t}=\widetilde{X}_{i,t-1}\varepsilon_{i,t}$, 
\[
\widetilde{W}_{i,t}^{\dag}=\widetilde{X}_{i,t-1}\varepsilon_{i,t}I\left\{\vert\widetilde{X}_{i,t-1}\vert_\infty\le \xi_{NT}^{1/2},\ \vert\varepsilon_{i,t}\vert\le\xi_{NT}^{1/2}\right\}~~\text{and}~~\widetilde{W}_{i,t}^{\ddag}=\widetilde W_{i,t}-\widetilde{W}_{i,t}^{\dag}.
\] 
Note that
\begin{eqnarray}
&&\max_{1\le i\le N}\max_{1\leq l\leq L}\left\vert\frac{1}{Th}\sum_{t=1}^T\widetilde{X}_{i,t-1}\varepsilon_{i,t}K_{t0}(\tau_l^\ast)\right\vert_2\notag\\
&\leq&\max_{1\le i\le N}\max_{1\leq l\leq L}\left\vert\frac{1}{Th}\sum_{t=1}^T\left\{\widetilde{W}_{i,t}^{\dag}-{\sf E}\left[\widetilde{W}_{i,t}^{\dag}\right]\right\}K_{t0}(\tau_l^\ast)\right\vert_2+\notag\\
&&\max_{1\le i\le N}\max_{1\leq l\leq L}\left\vert\frac{1}{Th}\sum_{t=1}^T\left\{\widetilde{W}_{i,t}^{\ddag}-{\sf E}\left[\widetilde{W}_{i,t}^{\ddag}\right]\right\}K_{t0}(\tau_l^\ast)\right\vert_2.\notag
\end{eqnarray}
For any $\eta>0$, by the Markov inequality, Assumptions 1(ii) and 2(iii), and noting that  
\[
\max_{1\leq i\leq N}\max_{0\leq t\leq T-1} \left\Vert \vert \widetilde{X}_{i,t}\vert_\infty\right\Vert_{q}^{q}\leq \theta_{N,q}^q,
\]
we have
\begin{eqnarray}
&&{\sf P}\left(\max_{1\le i\le N}\max_{1\leq l\leq L}\left\vert\frac{1}{Th}\sum_{t=1}^T\left\{\widetilde{W}_{i,t}^{\ddag}-{\sf E}\left[\widetilde{W}_{i,t}^{\ddag}\right]\right\}K_{t0}(\tau_l^\ast)\right\vert_2>\eta \sqrt{\frac{\bar n\log (N\vee T)}{Th}}\right)\label{eqD.28}\\
&\leq&{\sf P}\left(\max_{1\le i\le N}\max_{0\leq t\leq T-1}\vert\widetilde{X}_{i,t}\vert_\infty> \xi_{NT}^{1/2}\right)+{\sf P}\left(\max_{1\le i\le N}\max_{1\leq t\leq T}\vert\varepsilon_{i,t}\vert> \xi_{NT}^{1/2}\right)\notag\\
&\leq&\sum_{i=1}^N\sum_{t=0}^{T-1}{\sf P}\left(\vert\widetilde{X}_{i,t}\vert_\infty> \xi_{NT}^{1/2}\right)+\sum_{i=1}^N\sum_{t=1}^T{\sf P}\left(\vert\varepsilon_{i,t}\vert> \xi_{NT}^{1/2}\right)\notag\\
&=&O\left(NT\theta_{N,q}^q\xi_{NT}^{-q/2}+NT\xi_{NT}^{-q/2}\right)=o(1).\notag
\end{eqnarray}
By the concentration inequality for martingales \citep[e.g.,][]{Fr75}, we have
{\small\begin{eqnarray}
&&{\sf P}\left(\max_{1\le i\le N}\max_{1\leq l\leq L}\left\vert\frac{1}{Th}\sum_{t=1}^T\left\{\widetilde{W}_{i,t}^{\dag}-{\sf E}\left[\widetilde{W}_{i,t}^{\dag}\right]\right\}K_{t0}(\tau_l^\ast)\right\vert_2>\eta \sqrt{\frac{\bar n\log (N\vee T)}{Th}}\right)\notag\\
&\leq&\sum_{i=1}^N\sum_{l=1}^L{\sf P}\left(\left\vert\frac{1}{Th}\sum_{t=1}^T\left\{\widetilde{W}_{i,t}^{\dag}-{\sf E}\left[\widetilde{W}_{i,t}^{\dag}\right]\right\}K_{t0}(\tau_l^\ast)\right\vert_2>\eta \sqrt{\frac{\bar n\log ( N\vee T)}{Th}}\right)\notag\\
&=&O\left(NL\exp\left\{-\frac{\eta^2(Th)\log(N\vee T)}{CTh}\right\}\right)=o(1),\label{eqD.29}
\end{eqnarray}}
by letting $\eta$ be sufficiently large. With (\ref{eqD.28}) and (\ref{eqD.29}), we complete the proof of (\ref{eqD.27}).\hfill$\Box$

\medskip

\begin{lemma}\label{lem:D.4}

Suppose that the conditions of Lemma \ref{lem:D.3} are satisfied and
\begin{equation}\label{eqD.30}
{\sf P}\left(\widehat{M}=M_0\quad {\rm and}\quad \widehat{\mathscr G}_m^\circ={\mathscr G}_m^\circ,\ 1\leq m\leq M_0\right)\rightarrow1.
\end{equation}
Letting $\widehat\beta_{i\bullet}^\circ(\cdot)$ be defined in (3.4) of the main text, we have 
\begin{equation}\label{eqD.31}
\max_{1\le i\le N}\max_{1\leq l\leq L}\left\vert \widehat\beta_{i\bullet}^\circ(\tau_l^\ast)-\beta_{i\bullet}^\circ(\tau_l^\ast)\right\vert_2=O_P\left(\sqrt{\bar n}\left(\sqrt{\frac{\log (N\vee T)}{Th}}+h^2\right)\right).
\end{equation}

\end{lemma}

\smallskip

\noindent{\bf Proof of Lemma \ref{lem:D.4}}.\ Define
\[
{\mathscr E}_g=\left\{\widehat{g}_{ij}=g_{ij}, (i,j)\in\bar{\mathscr N}_N\right\}\cap\big\{\widehat M=M_0\big\},
\]
where $\widehat{g}_{ij}$ and $\bar{\mathscr N}_N$ are defined in Appendix A, and $g_{ij}$ is defined in Section 3.1. By (\ref{eqD.30}), we readily have that ${\sf P}({\mathscr E}_g)\rightarrow 1$. Hence, it is sufficient to prove (\ref{eqD.31}) conditional on the event ${\mathscr E}_g$. 
Let $\widehat{\omega}_{ij,m}$ and $\omega_{ij,m}$ be defined as in Section 3.1. Note that 
\begin{equation}\label{eqD.32}
\widehat{\beta}_{im}^\circ(\tau)=\sum_{j\in{\mathscr N}_i}\widehat{\beta}_{ij}(\tau)\widehat{\omega}_{ij,m}=\sum_{j\in{\mathscr N}_i}\widehat{\beta}_{ij}(\tau) \omega_{ij,m}=:\overline{\beta}_{im}^\circ(\tau)
\end{equation}
and
\begin{equation}\label{eqD.33}
\overline{\beta}_{i\bullet}^\circ(\tau)-\beta_{i\bullet}^\circ(\tau)=\left\{\sum_{j\in{\mathscr N}_i}\left[\widehat{\beta}_{ij}(\tau)-\beta_{ij}(\tau)\right] \omega_{ij,1},\cdots, \sum_{j\in{\mathscr N}_i}\left[\widehat{\beta}_{ij}(\tau)-\beta_{ij}(\tau)\right] \omega_{ij,M_0}\right\}^{^\intercal}
\end{equation}
conditional on ${\mathscr E}_g$, where $\overline{\beta}_{i\bullet}^\circ(\tau)=\left[\overline{\beta}_{i1}^\circ(\tau),\cdots,\overline{\beta}_{iM_0}^\circ(\tau)\right]^{^\intercal}$. Combining (\ref{eqD.32}), (\ref{eqD.33}) and Lemma \ref{lem:D.3}, we can complete the proof of (\ref{eqD.31}) conditional on ${\mathscr E}_g$.\hfill$\Box$

\medskip

The following lemma is crucial to prove the consistency property of the ratio criterion.

\begin{lemma}\label{lem:D.5}

Suppose that the conditions of Theorem 3.2 are satisfied. Define ${\mathscr E}_{\mathscr G}$ as the event that $\{\widehat{{\mathscr G}}_1,\cdots,\widehat{{\mathscr G}}_{K_0}\big\}=\big\{{\mathscr G}_1,\cdots,{\mathscr G}_{K_0}\}$. Then, conditional on ${\mathscr E}_{\mathscr G}$, we have (i) for $K=K_0, K_0+1,\cdots,\overline{K}$, $\widehat{R}(K)=o_P(\rho_{NT})$, where $\rho_{NT}$ is defined in Section 3.2; and (ii) for $K=1,\cdots,K_0-1$, $\widehat{R}(K)\geq \underline{c}\zeta_{NT}^\dag$ w.p.a.1, where $\zeta_{NT}^\dag$ is defined in Assumption 3 and $\underline{c}$ is a positive constant strictly larger than zero.   
 
\end{lemma} 

%\smallskip

\noindent{\bf Proof of Lemma \ref{lem:D.5}}.\ \ For the case $K_0\leq K\leq \overline{K}$, conditional on the event ${\mathscr E}_{\mathscr G}$, the grouped time-varying network VAR model is either correctly- or over-fitted. Some of ${\mathscr G}_k$, $k=1,\cdots,K_0$, are further split into smaller groups when $K>K_0$. Without loss of generality, we only consider the case of $K=K_0+1$ (conditional on ${\mathscr E}_{\mathscr G}$) and assume that ${\mathscr G}_{K_0}$ is split into ${\mathscr G}_{K_0}^\dag$ and ${\mathscr G}_{K_0}^\ddag$. It follows from the latent group assumption in Section 2.1 that there exist $\alpha_{k\bullet}^\circ(\cdot)$, $k=1,\cdots, K_0$, such that $\beta_{i\bullet}^\circ(\cdot)=\alpha_{k\bullet}^\circ(\cdot)$ when $g_i=k$. By Assumption 4(i) as well as Lemmas \ref{lem:D.3} and \ref{lem:D.4}, we may show that
\begin{eqnarray}
\widehat{R}(K_0+1)&=&\frac{1}{L(K_0+1)}\sum_{k=1}^{K_0-1}\frac{1}{{\rm card}({\mathscr G}_k)}\sum_{i\in{\mathscr G}_{k}}\sum_{l=1}^L\left[\left\vert \beta_i(\tau_l^\ast)-\alpha_{k}(\tau_l^\ast)\right\vert+\left\vert \beta_{i\bullet}^\circ(\tau_l^\ast)-\alpha_{k\bullet}^\circ(\tau_l^\ast)\right\vert_2\right]+\notag\\
&&\frac{1}{L(K_0+1){\rm card}({\mathscr G}_{K_0}^\dag)}\sum_{i\in{\mathscr G}_{K_0}^\dag}\sum_{l=1}^L\left[\left\vert \beta_i(\tau_l^\ast)-\alpha_{K_0}(\tau_l^\ast)\right\vert+\left\vert \beta_{i\bullet}^\circ(\tau_l^\ast)-\alpha_{K_0\bullet}^\circ(\tau_l^\ast)\right\vert_2\right]+\notag\\
&&\frac{1}{L(K_0+1){\rm card}({\mathscr G}_{K_0}^\ddag)}\sum_{i\in{\mathscr G}_{K_0}^\ddag}\sum_{l=1}^L\left[\left\vert \beta_i(\tau_l^\ast)-\alpha_{K_0}(\tau_l^\ast)\right\vert+\left\vert \beta_{i\bullet}^\circ(\tau_l^\ast)-\alpha_{K_0\bullet}^\circ(\tau_l^\ast)\right\vert_2\right]+\notag\\
&&O_P\left(\sqrt{\bar n}\left(\sqrt{\frac{\log (N\vee T)}{Th}}+h^2\right)\right)\notag\\
&=&O_P\left(\sqrt{\bar n}\left(\sqrt{\frac{\log (N\vee T)}{Th}}+h^2\right)\right)=o_P(\rho_{NT})\notag
\end{eqnarray}
conditional on ${\mathscr E}_{\mathscr G}$. The proof is similar for $\widehat{R}(K)$, $K=K_0, K_0+2,\cdots,\overline{K}$. 

\smallskip

For the case $1\leq K\leq K_0-1$, conditional on the event ${\mathscr E}_{\mathscr G}$, the model is under-fitted and at least two groups are falsely merged. Without loss of generality, we only consider the case of $K=K_0-1$ (conditional on ${\mathscr E}_{\mathscr G}$) and assume that ${\mathscr G}_{K_0-1}$ and ${\mathscr G}_{K_0}$ are merged. Define
\[
\alpha_{\ast}(\tau)=\frac{1}{{\rm card}({\mathscr G}_{K_0-1}\cup{\mathscr G}_{K_0})}\left[\left\vert {\mathscr G}_{K_0-1}\right\vert \alpha_{K_0-1}(\tau)+\left\vert{\mathscr G}_{K_0}\right\vert \alpha_{K_0}(\tau)\right]
\]
and 
\[
\alpha_{\ast}^\circ(\tau)=\frac{1}{{\rm card}({\mathscr G}_{K_0-1}\cup{\mathscr G}_{K_0})}\left[\left\vert {\mathscr G}_{K_0-1}\right\vert \alpha_{K_0-1\bullet}^\circ(\tau)+\left\vert{\mathscr G}_{K_0}\right\vert \alpha_{K_0\bullet}^\circ(\tau)\right].
\]
By Assumptions 3 and 4, and Lemmas \ref{lem:D.3} and \ref{lem:D.4}, we can prove that
{\footnotesize\begin{eqnarray}
\widehat{R}(K_0-1)&=& \frac{1}{L(K_0-1)}\sum_{k=1}^{K_0-2}\frac{1}{{\rm card}({\mathscr G}_k)}\sum_{i\in{\mathscr G}_{k}}\sum_{l=1}^L\left[\left\vert \beta_i(\tau_l^\ast)-\alpha_{k}(\tau_l^\ast)\right\vert+\left\vert \beta_{i\bullet}^\circ(\tau_l^\ast)-\alpha_{k\bullet}^\circ(\tau_l^\ast)\right\vert_2\right]+\notag\\
&&\frac{1}{L(K_0-1){\rm card}({\mathscr G}_{K_0-1}\cup{\mathscr G}_{K_0})}\sum_{i\in {\mathscr G}_{K_0-1}\cup{\mathscr G}_{K_0}}\sum_{l=1}^L\left[\left\vert \beta_i(\tau_l^\ast)-\alpha_{\ast}(\tau_l^\ast)\right\vert+\left\vert \beta_{i\bullet}^\circ(\tau_l^\ast)-\alpha_{\ast}^\circ(\tau_l^\ast)\right\vert_2\right]\notag\\
&&+O_P\left(\sqrt{\bar n}\left(\sqrt{\frac{\log N\vee T}{Th}}+h^2\right)\right)\notag\\
&=&2\frac{{\rm card}({\mathscr G}_{K_0-1}){\rm card}({\mathscr G}_{K_0})}{L(K_0-1){\rm card}({\mathscr G}_{K_0-1}\cup{\mathscr G}_{K_0})}\sum_{l=1}^L\left[\left\vert \alpha_{K_0}(\tau_l^\ast)-\alpha_{K_0-1}(\tau_l^\ast)\right\vert+\left\vert \alpha_{K_0\bullet}^\circ(\tau_l^\ast)-\alpha_{K_0-1\bullet}^\circ(\tau_l^\ast)\right\vert_2\right]\notag\\
&&+O_P\left(\sqrt{\bar n}\left(\sqrt{\frac{\log N\vee T}{Th}}+h^2\right)\right)\notag\\
&\geq&\underline{c}\xi_{NT}^\dag\quad w.p.a.1,\notag
\end{eqnarray}}
conditional on ${\mathscr E}_{\mathscr G}$, where $\underline{c}$ is a positive constant. The same result also holds for $\widehat{R}(K)$, $K=1,\cdots,K_0-2$. The proof of Lemma \ref{lem:D.5} is completed. \hfill$\Box$

\medskip

The following lemma is useful to prove the limit distribution theory of the post-grouping local linear estimation.

\begin{lemma}\label{lem:D.6}

Suppose that Assumptions 1, 2(i) and 5(i) in the main text are satisfied. Let

\[
\check{\boldsymbol\Delta}_{ij,\kappa}(\tau)=\frac{1}{Th_\dag}\sum_{t=1}^TX_{i,t-1}^\diamond X_{i,t-1}^{\diamond^\intercal}K_{t\kappa}^{\dagger}(\tau),\quad K^\dagger_{t\kappa}(\tau)=\left(\frac{\tau_t-\tau}{h_\dag}\right)^\kappa K\left(\frac{\tau_t-\tau}{h_\dag}\right),
\]
where $X_{i,t}^\diamond$ is defined in Section 4.1. Then we have
\begin{equation}\label{eqD.34}
\max_{1\le i,j\le N}\left\vert \check{\boldsymbol\Delta}_{ij,\kappa}(\tau)-{\sf E}\left[ \check{\boldsymbol\Delta}_{ij,\kappa}(\tau)\right]\right\vert_{\sf F}=O_P\left(\sqrt{\frac{\log (N\vee T)}{Th}}\right)
\end{equation}
for $\kappa=0,1,2$ and $\tau\in[0,1]$.

\end{lemma}

\noindent{\bf Proof of Lemma \ref{lem:D.6}}.\ The proof is similar to the proof of Lemma \ref{lem:D.1}. Details are omitted here to save space. \hfill $\Box$

\bigskip

%%%%%%%%%%%%%%%%%%

\section{Refined estimation of the break point}\label{appE}
\renewcommand{\theequation}{E.\arabic{equation}}\setcounter{equation}{0}

\medskip

\subsection{Refined estimation methodology}\label{appE:1}

In Section 5 of the main text, we construct an estimation of the break location $t_0$, which is shown to be consistent with scaling, i.e.,
\[
\left\vert \frac{\widehat t-t_0}{T}\right\vert=o_P(1),
\] 
see Theorem 5.1(i). Although this consistency property is sufficient to consistently estimate the group membership and number in the subsequent stage, there is a natural question on whether the precision of the break location estimation can be improved. We next aim to tackle this issue by introducing a refined break point estimation which achieves improvement of the estimation accuracy by making use of the estimated group structure. 

\smallskip

As in Section 4, we define
\begin{eqnarray}
\check{X}_{i,t}^1=\left[\sum_{j\in\widehat{\mathscr G}_1^1}\widetilde w_{ij} x_{j,t},\cdots,\sum_{j\in\widehat{\mathscr G}_{\widehat{K}_1}^{1}}\widetilde w_{ij} x_{j,t}, x_{i,t}\right]^{^\intercal},\notag\\
\check{X}_{i,t}^2=\left[\sum_{j\in\widehat{\mathscr G}_1^2}\widetilde w_{ij} x_{j,t},\cdots,\sum_{j\in\widehat{\mathscr G}_{\widehat{K}_2}^{2}}\widetilde w_{ij} x_{j,t}, x_{i,t}\right]^{^\intercal}.\notag
\end{eqnarray}
For any $t$, we use the cross-node regression to estimate the group-specific coefficient functions:
\begin{eqnarray}
\check\alpha_{k\bullet}^1(\tau_t)&=&\left[\sum_{i\in\widehat{\mathscr G}_k^1}\check{X}_{i,t-1}^1\left(\check{X}_{i,t-1}^{1}\right)^{^{\intercal}}\right]^{-1}\sum_{i\in\widehat{\mathscr G}_k^1}\check{X}_{i,t-1}^1x_{i,t},\quad k=1,\cdots,\widehat{K}_1,\notag\\
\check\alpha_{k\bullet}^2(\tau_t)&=&\left[\sum_{i\in\widehat{\mathscr G}_k^2}\check{X}_{i,t-1}^2\left(\check{X}_{i,t-1}^{2}\right)^{^{\intercal}}\right]^{-1}\sum_{i\in\widehat{\mathscr G}_k^2}\check{X}_{i,t-1}^2x_{i,t}.\quad k=1,\cdots,\widehat{K}_2.\notag
\end{eqnarray}
As in Section 4, without loss of generality, we assume 
\begin{eqnarray}
&&\widehat{\mathscr G}_k^1={\mathscr G}_k^1,\quad 1\leq k\leq K_1,\quad {\rm and}\quad \widehat{K}_1=K_1,\notag\\
&&\widehat{\mathscr G}_k^2={\mathscr G}_k^2,\quad 1\leq k\leq K_2,\quad {\rm and}\quad \widehat{K}_2=K_2.\notag
\end{eqnarray}
We expect $\check\alpha_{k\bullet}^1(\tau_t)$ to be a consistent estimate of $
\alpha_{k\bullet}^1(\tau_t):=\left[\alpha_{k1}^1(\tau_t),\cdots,\alpha_{kK_1}^1(\tau_t), \alpha_k^1(\tau_t)\right]^{^\intercal}$ when $t\leq t_0$ but this consistency becomes invalid for at least one $k$ when $t>t_0$. Similarly, we expect $\check\alpha_{k\bullet}^2(\tau_t)$ to be a consistent estimate of $
\alpha_{k\bullet}^2(\tau_t):=\left[\alpha_{k1}^2(\tau_t),\cdots,\alpha_{kK_2}^2(\tau_t), \alpha_k^2(\tau_t)\right]^{^\intercal}$ when $t> t_0$ but the consistency becomes invalid for at least one $k$ when $t\leq t_0$. Define
\begin{equation}\label{eqE.1}
\check\delta_\alpha(t)=\max_{1\leq k\leq \widehat{K}_1}\left\vert \check\alpha_{k\bullet}^1(\tau_{t+1})-\check\alpha_{k\bullet}^1(\tau_{t})\right\vert_2+\max_{1\leq k\leq \widehat{K}_2}\left\vert \check\alpha_{k\bullet}^2(\tau_{t+1})-\check\alpha_{k\bullet}^2(\tau_{t})\right\vert_2.
\end{equation}
The refined break location estimation is then obtained by
\begin{equation}\label{eqE.2}
\check{t}=\argmax_{t} \check\delta_\alpha(t).
\end{equation}

\subsection{Assumptions and consistency}\label{appE:2}

We next provide some high-level conditions and then derive the consistency property of $\check t$. We start with the introduction of some notation. Let 
\begin{eqnarray}
X_{i,t}^{1,\diamond}&=&\left[\sum_{g_j^1=1}\widetilde w_{ij} x_{j,t},\cdots,\sum_{g_j^1=K_1}\widetilde w_{ij} x_{j,t}, x_{i,t}\right]^{^\intercal},\notag\\
X_{i,t}^{2,\diamond}&=&\left[\sum_{g_j^2=1}\widetilde w_{ij} x_{j,t},\cdots,\sum_{g_j^2=K_2}\widetilde w_{ij} x_{j,t}, x_{i,t}\right]^{^\intercal},\notag\\
{\boldsymbol\Delta}_{ij,t}^{1,\diamond}&=&{\sf E}\left[X_{i,t}^{1,\diamond}\left(X_{j,t}^{1,\diamond}\right)^{^\intercal}\right],\quad {\boldsymbol\Delta}_{i,t}^{1,\diamond}={\boldsymbol\Delta}_{ii,t}^{1,\diamond},\notag\\
{\boldsymbol\Delta}_{ij,t}^{2,\diamond}&=&{\sf E}\left[X_{i,t}^{2,\diamond}\left(X_{j,t}^{2,\diamond}\right)^{^\intercal}\right],\quad {\boldsymbol\Delta}_{i,t}^{2,\diamond}={\boldsymbol\Delta}_{ii,t}^{2,\diamond}.\notag
\end{eqnarray}

\renewcommand{\theassumption}{E.\arabic{assumption}}\setcounter{assumption}{0}

\begin{assumption}\label{ass:E.1}

(i)\ Let 
\[
{\boldsymbol\Delta}_{{\mathscr G}_k^1,t}=\frac{1}{{\rm card}({\mathscr G}_k^1)}\sum_{g_i^1=k}{\boldsymbol\Delta}_{i,t}^{1,\diamond}\quad \quad {\boldsymbol\Delta}_{{\mathscr G}_k^2,t}=\frac{1}{{\rm card}({\mathscr G}_k^2)}\sum_{g_i^2=k}{\boldsymbol\Delta}_{i,t}^{2,\diamond}
\]
be positive definite uniformly over $k$ and $t$.

(ii)\ There exist $\Delta_{{\mathscr G}_k^1,t}^\alpha$, $k=1,\cdots,K_1$, and $\Delta_{{\mathscr G}_k^2,t}^\alpha$, $k=1,\cdots,K_2$, such that 
\begin{eqnarray}
&&\max_{1\leq k\leq K_1}\max_{t_0<t\leq T}\left\vert \frac{1}{{\rm card}({\mathscr G}_k^1)}\sum_{g_i^1=k}X_{i,t}^{1,\diamond}\left[\sum_{j\neq i}\alpha_{g_i^2g_j^2}^2(\tau_t)\widetilde{w}_{ij}x_{j,t-1}+\alpha_{g_i^2}^2(\tau_t)x_{i,t-1}\right]-\Delta_{{\mathscr G}_k^1,t}^\alpha\right\vert_2=o_P(1),\notag\\
&&\max_{1\leq k\leq K_2}\max_{1<t\leq t_0}\left\vert \frac{1}{{\rm card}({\mathscr G}_k^2)}\sum_{g_i^2=k}X_{i,t}^{2,\diamond}\left[\sum_{j\neq i}\alpha_{g_i^1g_j^1}^1(\tau_t)\widetilde{w}_{ij}x_{j,t-1}+\alpha_{g_i^1}^1(\tau_t)x_{i,t-1}\right]-\Delta_{{\mathscr G}_k^2,t}^\alpha\right\vert_2=o_P(1).\notag
\end{eqnarray}

\end{assumption}

\begin{assumption}\label{ass:E.2}

For any group ${\mathscr G}$ with cardinality sufficiently large,  
\begin{eqnarray}
&&\max_{1\leq t\leq T}{\sf E}\left(\left\vert\sum_{i\in {\mathscr G}}\left[X_{i,t}^{1,\diamond}\left(X_{i,t}^{1,\diamond}\right)^{^\intercal}-{\boldsymbol\Delta}_{i,t}^{1,\diamond}\right]\right\vert_{\sf F}^\iota+\left\vert\sum_{i\in {\mathscr G}}\left[X_{i,t}^{2,\diamond}\left(X_{i,t}^{2,\diamond}\right)^{^\intercal}-{\boldsymbol\Delta}_{i,t}^{2,\diamond}\right]\right\vert_{\sf F}^\iota\right)=O\left([{\rm card}({\mathscr G})]^{\iota/2}\right),\notag\\
&&\max_{1\leq t\leq T}{\sf E}\left(\left\vert\sum_{i\in {\mathscr G}} X_{i,t-1}^{1,\diamond}\varepsilon_{i,t}\right\vert_{2}^\iota+\left\vert\sum_{i\in {\mathscr G}} X_{i,t-1}^{2,\diamond}\varepsilon_{i,t}\right\vert_{2}^\iota\right)=O\left([{\rm card}({\mathscr G})]^{\iota/2}\right),\notag
\end{eqnarray}
where $\iota$ is a positive constant larger than $2$. In addition, $T/N^{\iota/2}\rightarrow0$.

\end{assumption}

\begin{assumption}\label{ass:E.3}

Let
\[
\delta_\alpha:=\max_{1\leq k\leq K_1}\left\vert \alpha_{k\bullet}^\dag(\tau_{t_0+1})-\alpha_{k\bullet}^1(\tau_{t_0})\right\vert_2+\max_{1\leq k\leq K_2}\left\vert \alpha_{k\bullet}^2(\tau_{t_0+1})-\alpha_{k\bullet}^\ddag(\tau_{t_0})\right\vert_2\geq c_5>0
\]
where 
\begin{eqnarray}
\alpha_{k\bullet}^\dag(\tau_{t})&=&{\boldsymbol\Delta}_{{\mathscr G}_k^1,t}^{-1}\Delta_{{\mathscr G}_k^1,t}^\alpha,\quad t_0<t\leq T,\notag\\
\alpha_{k\bullet}^\ddag(\tau_{t})&=&{\boldsymbol\Delta}_{{\mathscr G}_k^2,t}^{-1}\Delta_{{\mathscr G}_k^2,t}^\alpha,\quad 1< t\leq t_0.\notag
\end{eqnarray}

\end{assumption}

\smallskip

Assumption \ref{ass:E.1}(i) is similar to Assumption 5(ii) in the main text. Meanwhile, a combination of the conditions in Assumption \ref{ass:E.1}(i)(ii) ensures that the limits of $\check\alpha_{k\bullet}^1(\cdot)$ and $\check\alpha_{k\bullet}^2(\cdot)$ are well defined even when groups are misclassified, i.e., $\alpha_{k\bullet}^\dag(\cdot)$ and $\alpha_{k\bullet}^\ddag(\cdot)$ in Assumption \ref{ass:E.3} are well defined. Assumption \ref{ass:E.2} imposes some high-level moment conditions to restrict weak correlation between nodes and indicates that $T$ can be either smaller or larger than $N$ (depending on the value of $\iota$). Assumption \ref{ass:E.3} indicates that the break size at $t_0$ needs to be bounded away from zero. This condition can be satisfied when there are breaks in the group-specific coefficient functions, group membership or number. Some examples are provided in Appendix \ref{appE:3} to verify this condition.

\smallskip

\renewcommand{\theproposition}{E.\arabic{proposition}}\setcounter{proposition}{0}

\begin{proposition}\label{prop:E.1}

Suppose that the conditions of Theorem 5.1 and Assumptions \ref{ass:E.1}--\ref{ass:E.3} are satisfied. Then we have ${\sf P}(\check t=t_0)\rightarrow1$. 

\end{proposition}

\medskip

\noindent{\bf Proof of Proposition \ref{prop:E.1}}.\ \ Let ${\mathscr E}_{\mathscr G}^1$ denote the event that $\widehat{\mathscr G}_k^1={\mathscr G}_k^1$, $1\leq k\leq K_1$, and $\widehat{K}_1=K_1$, and let ${\mathscr E}_{\mathscr G}^2$ denote the event that $\widehat{\mathscr G}_k^2={\mathscr G}_k^2$, $1\leq k\leq K_2$, and $\widehat{K}_2=K_2$. For $1\leq t\leq t_0$, conditional on ${\mathscr E}_{\mathscr G}^1\cap {\mathscr E}_{\mathscr G}^2$, the grouped time-varying network VAR model is correctly fitted. Hence, we have
\begin{eqnarray}
\check\alpha_{k\bullet}^1(\tau_t)&=&\left[\sum_{i\in\widehat{\mathscr G}_k^1}\check{X}_{i,t-1}^1\left(\check{X}_{i,t-1}^{1}\right)^{^{\intercal}}\right]^{-1}\sum_{i\in\widehat{\mathscr G}_k^1}\check{X}_{i,t-1}^1x_{i,t}\notag\\
&=&\left[\sum_{i\in {\mathscr G}_k^1}X_{i,t-1}^{1,\diamond}\left(X_{i,t-1}^{1,\diamond}\right)^{^{\intercal}}\right]^{-1}\sum_{i\in {\mathscr G}_k^1}X_{i,t-1}^{1,\diamond}x_{i,t}\notag\\
&=&\alpha_{k\bullet}^1(\tau_t)+\left[\sum_{i\in {\mathscr G}_k^1}X_{i,t-1}^{1,\diamond}\left(X_{i,t-1}^{1,\diamond}\right)^{^{\intercal}}\right]^{-1}\sum_{i\in {\mathscr G}_k^1}X_{i,t-1}^{1,\diamond}\varepsilon_{i,t}\label{eqE.3}
\end{eqnarray}
conditional on ${\mathscr E}_{\mathscr G}^1\cap {\mathscr E}_{\mathscr G}^2$. By Assumptions \ref{ass:E.2} and 6(iv) and the Bonferroni and Markov inequalities, we may show that, for any $\eta>0$ 
\begin{eqnarray}
&&{\sf P}\left(\max_{1\leq t\leq t_0}\left\vert\frac{1}{{\rm card}({\mathscr G}_k^1)}\sum_{i\in {\mathscr G}_k^1}X_{i,t-1}^{1,\diamond}\varepsilon_{i,t}\right\vert_2>\eta\right)\notag\\
&\leq&\sum_{t=1}^{t_0}{\sf P}\left(\left\vert\frac{1}{{\rm card}({\mathscr G}_k^1)}\sum_{i\in {\mathscr G}_k^1}X_{i,t-1}^{1,\diamond}\varepsilon_{i,t}\right\vert_2>\eta\right)\notag\\
&\leq&\sum_{t=1}^{t_0}[{\rm card}({\mathscr G}_k^1)]^{-\iota} {\sf E}\left\vert \sum_{i\in {\mathscr G}_k^1}X_{i,t-1}^{1,\diamond}\varepsilon_{i,t}\right\vert_2^\iota\notag\\
&=&O\left(TN^{-\iota/2}\right)=o(1), \notag
\end{eqnarray}
which leads to
\begin{equation}\label{eqE.4}
\max_{1\leq t\leq t_0}\left\vert\frac{1}{{\rm card}({\mathscr G}_k^1)}\sum_{i\in {\mathscr G}_k^1}X_{i,t-1}^{1,\diamond}\varepsilon_{i,t}\right\vert_2=o_P(1).
\end{equation}
Similarly, we can also prove that 
\begin{equation}\label{eqE.5}
\max_{1\leq t\leq t_0}\left\vert\frac{1}{{\rm card}({\mathscr G}_k^1)}\sum_{i\in {\mathscr G}}\left[X_{i,t}^{1,\diamond}\left(X_{i,t}^{1,\diamond}\right)^{^\intercal}-{\boldsymbol\Delta}_{i,t}^{1,\diamond}\right]\right\vert_{\sf F}=o_P(1).
\end{equation}
With (\ref{eqE.3})--(\ref{eqE.5}), as $K_1$ is fixed, we readily have that
\begin{equation}\label{eqE.6}
\max_{1\leq k\leq K_1}\max_{1\leq t\leq t_0}\left\vert\check\alpha_{k\bullet}^1(\tau_t)-\alpha_{k\bullet}^1(\tau_t)\right\vert_2=o_P(1).
\end{equation}
Similarly, by Assumptions \ref{ass:E.1} and \ref{ass:E.2}, we also have 
\begin{eqnarray}
&&\max_{1\leq k\leq K_1}\max_{t_0+1\leq t\leq T}\left\vert\check\alpha_{k\bullet}^1(\tau_t)-\alpha_{k\bullet}^\dag(\tau_t)\right\vert_2=o_P(1),\label{eqE.7}\\
&&\max_{1\leq k\leq K_1}\max_{t_0+1\leq t\leq T}\left\vert\check\alpha_{k\bullet}^2(\tau_t)-\alpha_{k\bullet}^2(\tau_t)\right\vert_2=o_P(1),\label{eqE.8}\\
&&\max_{1\leq k\leq K_1}\max_{1\leq t\leq t_0}\left\vert\check\alpha_{k\bullet}^2(\tau_t)-\alpha_{k\bullet}^\ddag(\tau_t)\right\vert_2=o_P(1).\label{eqE.9}
\end{eqnarray}
Note that $\alpha^1_{k\bullet}(\cdot)$ and $\alpha^\ddag_{k\bullet}(\cdot)$ are continuous over $[0,\tau_{t_0}]$ whereas $\alpha^2_{k\bullet}(\cdot)$ and $\alpha^\dag_{k\bullet}(\cdot)$ are continuous over $[\tau_{t_0},1]$. By virtue of (\ref{eqE.6})--(\ref{eqE.9}), we have $\check\delta_\alpha(t)=o_P(1)$ when $t\neq t_0$ and $\check\delta_\alpha(t_0)-\delta_\alpha(t_0)=o_P(1)$. Finally, by Assumption \ref{ass:E.3}, we may prove that ${\sf P}(\check t=t_0)\rightarrow1$.\hfill$\Box$

\subsection{Verification of Assumption \ref{ass:E.3}}\label{appE:3}

We next provide a few examples to verify the condition $\delta_\alpha(t_0)>c_5$ in Assumption \ref{ass:E.3}.

\medskip

\noindent{\bf Example E.1}.\ \ Suppose that the group structure is time invariant, i.e., ${\mathscr G}^1={\mathscr G}^2$ and $K_1=K_2=K_0$, but there exists a $k_0\in\{1,2,\cdots,K_0\}$ such that $\alpha_{k_0\bullet}(\cdot)$ defined in Section 4 is discontinuous at $\tau_{t_0}$ satisfying
\[\left\vert \alpha_{k_0\bullet}(\tau_{t_0+1})-\alpha_{k_0\bullet}(\tau_{t_0})\right\vert_2\geq c_6>0,\]
where $\alpha_{k_0\bullet}^{\sf r}(\cdot)$ and $\alpha_{k_0\bullet}^{\sf l}(\cdot)$ denote the right and left limits of $\alpha_{k_0\bullet}(\cdot)$, respectively. In this case, we may verify that
\[
\alpha_{k\bullet}^{\dag}(\tau_{t})=\alpha_{k\bullet}(\tau_{t}),\quad \alpha_{k\bullet}^{\ddag}(\tau_{t})=\alpha_{k\bullet}(\tau_{t}).
\]
Hence, it is easy to show that 
\[
\delta_\alpha\geq \left\vert \alpha_{k_0\bullet}(\tau_{t_0+1})-\alpha_{k_0\bullet}(\tau_{t_0})\right\vert_2\geq c_6,
\]
verifying Assumption \ref{ass:E.3}. Similarly, the condition of $\delta_\beta(t_0)>c_2$ in Assumption 6(iii) is also satisfied.

\medskip

In the following examples, for notational simplicity, we consider a special time-varying network VAR model with break in either the group membership or number: 
\begin{equation}\label{eqE.10}
x_{i,t}=\left\{
\begin{array}{ll}
\gamma_{g_i^1}(\tau_t)\sum_{j\neq i}\widetilde w_{ij}x_{j,t-1}+\alpha_{g_i^1}(\tau_t)x_{i,t-1}+\varepsilon_{i,t},\ &\ 1\leq t\leq t_0,\\
\gamma_{g_i^2}(\tau_t)\sum_{j\neq i}\widetilde w_{ij}x_{j,t-1}+\alpha_{g_i^2}(\tau_t)x_{i,t-1}+\varepsilon_{i,t},\ &\ t_0+1\leq t\leq T,
\end{array}
\right.
\end{equation}
where $\gamma_{g_i^1}(\cdot)$ and $\gamma_{g_i^2}(\cdot)$ denote the time-varying network effects invariant over nodes which the $i$-th node follows. This model is considered in an earlier version of our paper with no break in the group structure.

\medskip

\noindent{\bf Example E.2}.\ \ Suppose that model (\ref{eqE.10}) holds, the group number is time invariant, $K_1=K_2=2$, but there is a break in the group membership at the time point $t_0$. Before the break point, $g_i^1=1$ for $1\leq i\leq \lfloor N/2\rfloor$ and $g_i^1=2$ for $\lfloor N/2\rfloor +1\leq i\leq N$; after the break point, $g_i^2=1$ for $1\leq i\leq \lfloor N/4\rfloor$ and $g_i^2=2$ for $\lfloor N/4\rfloor +1\leq i\leq N$. The group-specific coefficient functions $\gamma_1(\cdot)$, $\gamma_2(\cdot)$, $\alpha_1(\cdot)$ and $\alpha_2(\cdot)$ are continuous over $[0,1]$, and 
\begin{equation}\label{eqE.11}
\left\vert \gamma_2(\tau_{t_0})-\gamma_1(\tau_{t_0})\right\vert+\left\vert \alpha_2(\tau_{t_0})-\alpha_1(\tau_{t_0})\right\vert\geq c_7>0.
\end{equation}
Note that
\[
\alpha_{1\bullet}(\tau_t)=\left[\gamma_1(\tau_t),\alpha_1(\tau_t)\right]^{^\intercal},\quad 
\alpha_{2\bullet}(\tau_t)=\left[\gamma_2(\tau_t),\alpha_2(\tau_t)\right]^{^\intercal}.
\]
With some elementary calculations, we may show that 
\begin{eqnarray}
&&\alpha_{1\bullet}^\dag(\tau_t)={\mathbf W}_1^\dag 
\left[
\begin{array}{c}
\gamma_1(\tau_t)\\
\alpha_1(\tau_t)
\end{array}
\right]
+{\mathbf W}_2^\dag 
\left[
\begin{array}{c}
\gamma_2(\tau_t)\\
\alpha_2(\tau_t)
\end{array}
\right],\quad 
\alpha_{2\bullet}^\dag(\tau_t)=\left[\gamma_2(\tau_t),\alpha_2(\tau_t)\right]^{^\intercal},\notag\\
&&\alpha_{1\bullet}^\ddag(\tau_t)=\left[\gamma_1(\tau_t),\alpha_1(\tau_t)\right]^{^\intercal},\quad \alpha_{2\bullet}^\ddag(\tau_t)={\mathbf W}_1^\ddag 
\left[
\begin{array}{c}
\gamma_1(\tau_t)\\
\alpha_1(\tau_t)
\end{array}
\right]
+{\mathbf W}_2^\ddag 
\left[
\begin{array}{c}
\gamma_2(\tau_t)\\
\alpha_2(\tau_t)
\end{array}
\right],\notag
\end{eqnarray}
where ${\mathbf W}_1^\dag$, ${\mathbf W}_2^\dag$, ${\mathbf W}_1^\ddag$ and ${\mathbf W}_2^\ddag$ are all positive definite satisfying ${\mathbf W}_1^\dag+{\mathbf W}_2^\dag={\mathbf I}_2$ and ${\mathbf W}_1^\ddag+{\mathbf W}_2^\ddag={\mathbf I}_2$. Under the condition (\ref{eqE.11}), it is easy to verify Assumption \ref{ass:E.3}. The verification of $\delta_\beta(t_0)>c_2$ in Assumption 6(iii) is similar.

\medskip

It is worth pointing out that Assumption \ref{ass:E.3} may be invalid when the break in the group membership is sparse. In Example E.2, consider $g_i^1=1$ for $1\leq i\leq \lfloor N/2\rfloor$ and $g_i^1=2$ for $\lfloor N/2\rfloor +1\leq i\leq N$ before the break point; $g_i^2=1$ for $1\leq i\leq \lfloor N/2\rfloor-s$ and $g_i^2=2$ for $\lfloor N/2\rfloor-s+1\leq i\leq N$ after the break point, where $s$ is a small and fixed positive integer. In this case, $\delta_\alpha$ would converge to zero, indicating that Assumption \ref{ass:E.3} is violated. 

\medskip

\noindent{\bf Example E.3}.\ \ Suppose that model (\ref{eqE.10}) holds. Consider a break in the group number: $K_1=2$ and $K_2=1$. Before the break time $t_0$, $g_i^1=1$ for $1\leq i\leq \lfloor N/2\rfloor$ and $g_i^1=2$ for $\lfloor N/2\rfloor +1\leq i\leq N$; after the break point, the two groups merge, i.e., $g_i^2\equiv1$. Using the arguments in Example E.2, we can similarly verify Assumption \ref{ass:E.3} as well as $\delta_\beta(t_0)>c_2$ in Assumption 6(iii).

\section{Extra numerical results}\label{appF}
\renewcommand{\thetable}{F.\arabic{table}}
\renewcommand{\thefigure}{F.\arabic{figure}}

\subsection{Results of the clustering algorithm in Appendix A}\label{appF.1}

We next report the finite-sample performance of the clustering algorithm introduced in Appendix A to determine the homogeneity structure on the time-varying network spillover effects. We start with the data generating process provided in Section 6.1. Note that $M_0=2$ and 
\[
{\mathscr G}_1^\circ=\left\{(i,j)\in\bar{\mathscr N}_N:\ i\in{\mathscr G}_1, j\in{\mathscr G}_1\right\},\quad {\mathscr G}_2^\circ=\bar{\mathscr N}_N/ {\mathscr G}_1^\circ.
\]
To evaluate the homogeneity structure estimation accuracy, as in Section 6.1, we compute
\begin{eqnarray*}
{\sf AC}(M_0) &=& \frac{1}{R}\sum_{r=1}^R I(\widehat{M}_r=M_0) \quad  \text{and}\quad
{\sf Purity}({\mathscr G}^\circ) = \frac{1}{RN}\sum_{r=1}^R \sum_{m=1}^{\widehat{M}_r} \max_{1\le j\le M_0} \left|\widehat{\mathscr{G}}_{m,r}^\circ \cap \mathscr{G}_j^\circ \right|, 
\end{eqnarray*}
where $\widehat{M}_r$ and $\widehat{\mathscr{G}}_{m, r}^\circ$ are the estimates of the group number and membership in the $r$-th replication. The relevant results are summarized in Table \ref{apt1}. In general, the performance of the group number and membership estimates improve as $T$ increases from $300$ to $600$ whereas the membership estimation is sensitive to the network sparsity level with ${\sf Purity}({\mathscr G})$ decreasing significantly as $\bar w$ increases from $0.025$ to $0.075$.

\begin{table}[tbp]\centering
\setlength{\tabcolsep}{4pt}
\caption{Estimation performance of the group number $M_0$ and membership ${\cal G}^\circ$}\label{apt1}
\begin{tabular}{lllrrlrr}
\hline\hline
 &  &  & \multicolumn{2}{c}{Fixed group} &  & \multicolumn{2}{c}{Random group} \\
 \hline
Sparsity & Measurement & $T\setminus N$ & 100 & 200 &  & 100 & 200 \\
 \hline
$\bar{w}=0.025$ & ${\sf AC}(M_0)$  & 300 & 0.890 & 0.972 &  & 0.916 & 0.978 \\
 &  & 600 & 1.000 & 0.999 &  & 0.974 & 0.998 \\
 & ${\sf Purity}({\mathscr G}^\circ)$ & 300 & 0.939 & 0.761 &  & 0.901 & 0.759 \\
 &  & 600 & 0.982 & 0.874 &  & 0.961 & 0.860 \\
 &  &  &  &  &  &  &  \\
$\bar{w}=0.075$ & ${\sf AC}(M_0)$  & 300 & 0.904 & 0.702 &  & 0.856 & 0.697 \\
 &  & 600 & 0.999 & 0.829 &  & 0.988 & 0.857 \\
 & ${\sf Purity}({\mathscr G}^\circ)$ & 300 & 0.649 & 0.471 &  & 0.622 & 0.467 \\
 &  & 600 & 0.782 & 0.547 &  & 0.762 & 0.556 \\
 \hline\hline
\end{tabular}
\end{table}
 
 \smallskip

In addition, we report the clustering results when there exists a break in the group membership. The data generating process is described in Section 6.2. Table \ref{apt3} summarizes the estimation performance for the group number $M_0$ and membership ${\mathscr G}^\circ$. The findings drawn from Table \ref{apt3} are similar to those from Table \ref{apt1}.

\begin{table}[tbp]\centering
\setlength{\tabcolsep}{4pt}
\caption{Estimation performance of the group number $M_0$ and membership ${\cal G}^\circ$ (with a break)}\label{apt3}
\begin{tabular}{llrrrlrr}
\hline\hline
 &  &  & \multicolumn{2}{c}{Fixed group} &  & \multicolumn{2}{c}{Random group} \\
 \hline
Sparsity & Measurement  & $T\setminus N$ & 100 & 200 &  & 100 & 200 \\
\hline
$\bar{w}=0.025$ & ${\sf AC}(M_0)$ & 400 & 1.000 & 0.995 &  & 1.000 & 0.996 \\
 &  & 800 & 1.000 & 0.998 &  & 1.000 & 0.999 \\
 & ${\sf Purity}({\mathscr G}^\circ)$ & 400 & 0.961 & 0.811 &  & 0.930 & 0.808 \\
 &  & 800 & 0.989 & 0.903 &  & 0.975 & 0.895 \\
 &  &  &  &  &  &  &  \\
$\bar{w}=0.075$ &${\sf AC}(M_0)$ & 400 & 0.989 & 0.933 &  & 0.975 & 0.938 \\
 &  & 800 & 1.000 & 0.980 &  & 0.995 & 0.979 \\
 & ${\sf Purity}({\mathscr G}^\circ)$ & 400 & 0.729 & 0.578 &  & 0.712 & 0.577 \\
 &  & 800 & 0.827 & 0.658 &  & 0.812 & 0.654 \\
 \hline\hline
\end{tabular}
\end{table}

\subsection{Clustering results for grouped network VAR with constant coefficients}\label{appF.2}

We next report the additional simulation result by treating the data generating process in Section 6.1 as \cite{ZXF23}'s grouped network VAR with constant coefficients and adopting their clustering algorithm. The estimation results for the group structure are shown in Table \ref{tablecom}. 
By comparing the results with those in Table 1, we note that failure to account for smooth structural changes in network VAR would substantially affect the estimation accuracy of the latent group structure for network time series, resulting in lower values of ${\sf AC}(K_0)$ and ${\sf Purity}({\mathscr G})$.

\begin{table}[h]\centering
\setlength{\tabcolsep}{4pt}
\caption{Estimation performance of the group structure using \cite{ZXF23}'s model and algorithm}\label{tablecom}
\begin{tabular}{lllrrlrr}
\hline\hline
 &  &  & \multicolumn{2}{c}{Fixed group} &  & \multicolumn{2}{c}{Random group} \\
 \hline
 Sparsity & Measurement & $T\setminus N$ & 100 & 200 &  & 100 & 200 \\
 \hline
$\bar{w}=0.025$ & ${\sf AC}(K_0)$ & 300 & 0.844 & 0.851 &  & 0.827 & 0.830 \\
 &  & 600 &  0.670 & 0.714 &  & 0.557 & 0.743 \\
 & ${\sf Purity}({\mathscr G})$ & 300 & 0.555 & 0.577 &  & 0.561 & 0.582 \\
 &  & 600 &  0.538 & 0.681 &  & 0.551 & 0.675 \\
$\bar{w}=0.075$ & ${\sf AC}(K_0)$ & 300 &  0.802 & 0.773 &  &0.796 & 0.823 \\
 &  & 600 & 0.763 & 0.705 &  &  0.719 & 0.733 \\
 & ${\sf Purity}({\mathscr G})$ & 300 & 0.616 & 0.644 &  & 0.604 & 0.663 \\
 &  & 600 &  0.711 & 0.750 &  &  0.685 & 0.757 \\
 \hline\hline
\end{tabular}
\end{table}

\subsection{Extra empirical result}

\begin{figure}[tbp]\centering
\includegraphics[scale=0.20]{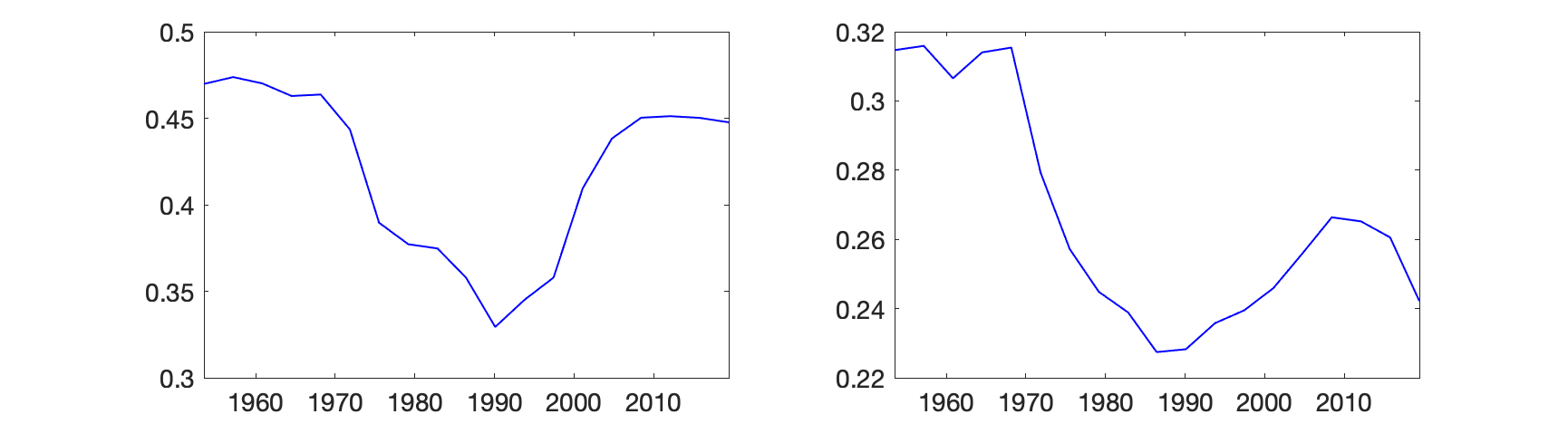}
\caption{\small Estimated time-varying momentum effects in model (i)}\label{fig1}
\end{figure}

\begin{figure}[tbp]\centering
\includegraphics[scale=0.20]{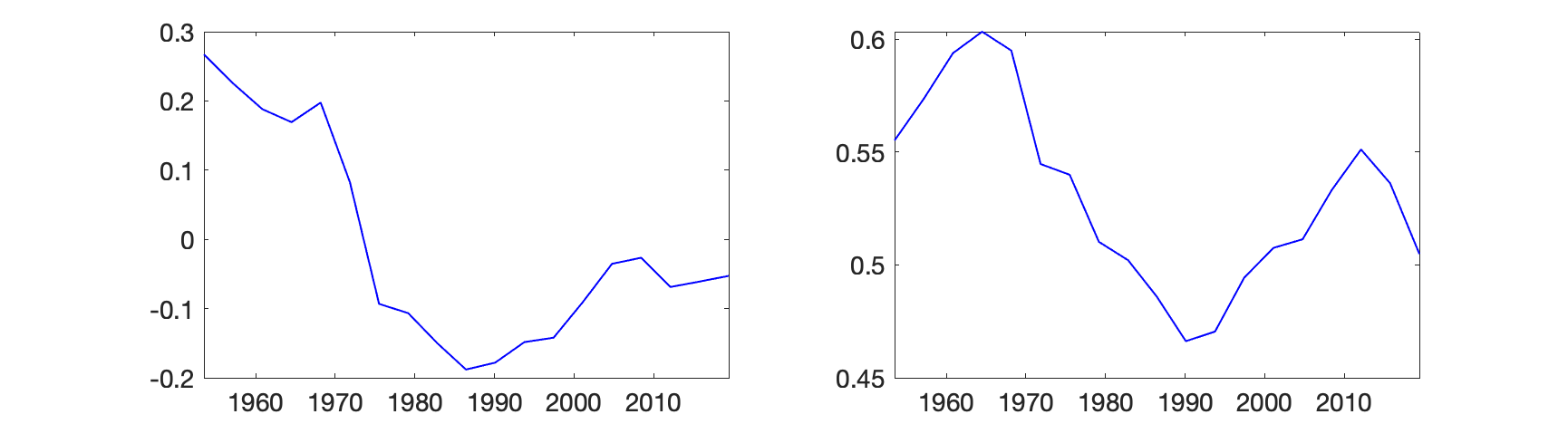}
\caption{\small Estimated time-varying momentum effects in model (ii)}\label{fig2}
\end{figure}

In addition, we plot the estimated time-varying momentum effects from both models in Figures \ref{fig1} and \ref{fig2}, respectively. In each figure, the left plot is for Group 1, while the right one is for Group 2. Although both figures present similar patterns of movement across groups, the Y-axis covers different ranges.

\end{appendix}

\end{document}